\documentclass[11pt,a4paper,usenames,dvipsnames]{article} 

\usepackage{adjustbox}
\usepackage{amsfonts}
\usepackage{amssymb}
\usepackage{amsmath}
\usepackage{amsthm}
\usepackage{authblk}
\usepackage{array}
\usepackage{bbm}
\usepackage{booktabs}
\usepackage{caption}
\usepackage{subcaption}
\usepackage{dsfont}
\usepackage{enumerate}
\usepackage{framed,multirow}
\usepackage{graphics}
\usepackage{graphicx}
\usepackage{hhline}
\usepackage{hyperref}
\usepackage{inputenc}
\usepackage{latexsym}
\usepackage{rotating}
\usepackage{mathrsfs}  
\usepackage{tikz}
\usepackage[normalem]{ulem}

\usepackage{url}
\usepackage{colortbl}
\usepackage{hyperref}
\usepackage{url}

\definecolor{bronze}{rgb}{0.8, 0.5, 0.2}
\definecolor{gold}{rgb}{0.83, 0.69, 0.22}
\definecolor{silver}{rgb}{0.44, 0.5, 0.56}

\title{SHREC 2022: Fitting and recognition of simple geometric primitives on point clouds}

\author[1, \thanks{Track organizer}]{Chiara Romanengo}
\author[1, $^*$, \thanks{Corresponding author}]{Andrea Raffo}
\author[1]{Silvia Biasotti}
\author[1]{Bianca Falcidieno}
\author[2]{Vlassis Fotis}
\author[2]{Ioannis Romanelis}
\author[2]{Eleftheria Psatha}
\author[2]{Konstantinos Moustakas}
\author[3]{Ivan Sipiran}
\author[4,5,6]{Quang-Thuc Nguyen}
\author[4,6]{Chi-Bien Chu}
\author[4,6]{Khoi-Nguyen Nguyen-Ngoc}
\author[4,6]{Dinh-Khoi Vo}
\author[4,6]{Tuan-An To}
\author[4,6]{Nham-Tan Nguyen}
\author[4,6]{Nhat-Quynh Le-Pham}
\author[4,5,6]{Hai-Dang Nguyen}
\author[4,5,6]{Minh-Triet Tran}
\author[7]{Yifan Qie}
\author[7]{Nabil Anwer}

\affil[1]{Istituto di Matematica Applicata e Tecnologie Informatiche  ``E. Magenes", Consiglio Nazionale delle Ricerche, Via de Marini 6, 16149 Genova, Italy}
\affil[2]{Department of Electrical and Computer Engineering, University of Patras, Patras, 26504, Greece}
\affil[3]{Department of Computer Science, University of Chile, Beauchef 851, Santiago, Chile}
\affil[4]{University of Science, VNU-HCM, Ho Chi Minh City, Vietnam}
\affil[5]{John von Neumann Institute, VNU-HCM, Ho Chi Minh City, Vietnam}
\affil[6]{Vietnam National University, Ho Chi Minh City, Vietnam}
\affil[7]{Automated Production Research Laboratory (LURPA), ENS Paris-Saclay, Université Paris-Saclay, 91190 Gif-sur-Yvette, France}

\date{}                     %% if you don't need date to appear
\newcolumntype{a}{>{\columncolor{blue!12}}m}
\newcolumntype{z}{>{\columncolor{teal!25}}m}

\begin{document}
\maketitle

\begin{abstract}
This paper presents the methods that have participated in the SHREC 2022 track on the fitting and recognition of simple geometric primitives on point clouds. As simple primitives we mean the classical surface primitives derived from constructive solid geometry, i.e., planes, spheres, cylinders, cones and tori.
The aim of the track is to evaluate the quality of automatic algorithms for fitting and recognising geometric primitives on point clouds. Specifically, the goal is to identify, for each point cloud, its primitive type and some geometric descriptors. For this purpose, we created a synthetic dataset, divided into a training set and a test set, containing segments perturbed with different kinds of point cloud artifacts.
Among the six participants to this track, two  are based on direct methods, while  four are either fully based on deep learning or combine direct and neural approaches. The performance of the methods is evaluated using various classification and approximation measures.  
\end{abstract}

\section{Introduction}
\label{sec:intro}
Fitting and recognition are of paramount importance in multiple application domains, such as reverse engineering and data compression. In most real-world scenarios, acquisition methods tend to produce data affected by noise and potentially by other imperfections (e.g., non-uniform or low sampling density, misalignment and missing data). The literature on methods for fitting and recognition of simple primitives is quite consolidated \cite{Kaiser2019} and recently we have seen the growth of data-driven methods, see for instance \cite{datasetGuibas_CVPR,Raffo:2020, ParseNet, Yan2021HPNetDP, SAPORTA2022289} resulting in a growing demand for appropriate datasets for training and their evaluation. It is therefore worth analyzing and comparing the outcome of simple primitive fitting methods on a controlled dataset with different kinds of perturbations and primitive variations and in terms of different evaluation measures. 

\subsection{Related datasets}
Given the strong interest in the problem of fitting geometric primitives, several datasets are becoming available but, generally, they are not specifically designed for simple geometric primitives, possibly affected by perturbations, missing points and local variations. A first collection of one million CAD models targeted to data-driven problems was provided by \cite{Koch_2019_CVPR}, in which each object is associated to a ground truth for different quantities, as for example the indices of the patch segmentation and the analytic representation of surfaces and curves. A lot of surface types are considered and the models are represented as triangle meshes. Then, it is worth mentioning \cite{Fit4CAD}, a benchmark of point clouds representing CAD objects aimed at evaluating methods for detecting simple geometric primitives. Despite having some features similar to our purposes, such as sub-sampled clouds and missing parts, it is not big enough and was designed to test model segmentation methods. A dataset of mechanical components is proposed by \cite{datasetGuibas_CVPR}, which generated the point samples from some models of the American National Standards Institute (ANSI) provided by TraceParts. In this collection, four types of primitives are considered (plane, sphere, cylinder, cone). Another synthetic dataset is presented in \cite{Sharma_2018_CVPR}, but it proposes only cylinders, spheres and cubes as surface types. These existing datasets are not suitable for our purposes since our intent is to analyze the different methods for fitting geometric primitives on point clouds that present different types of specific artifacts. Indeed, in our dataset all point clouds have been artificially generated from exact surfaces and then optionally perturbed by simulating uniform or Gaussian noise of different intensity, undersampling, missing parts, small deformations, or a combination of such point cloud artifacts. In addition, we are not interested in the whole segmentation process, but we are focused on a sub-issue of it, that is the problem of classification and recognition of segments.

\subsection{Contribution}

The aim of this SHREC track is to evaluate the quality of automatic algorithms for fitting and recognising geometric primitives in point clouds. The goal is to identify, for each point cloud, its primitive type (i.e., planes, spheres, cylinders, cones and tori) and some geometric descriptors. 

To this aim, a dataset of 3D segments, corresponding to parts of primitives and represented as point clouds, divided in training set and test set was generated. For each segment in the training set, its primitive type and an analytical representation of it were made available to the track participants. For each point cloud in the test set, the goal of the contest is twofold: to recognise the primitive type from the data and to provide an analytical representation of the primitive recognised.
The relevance or non-relevance of the simple primitives has been evaluated on the basis of a ground truth that derived from the knowledge of the primitive that originated that segment. Finally, the performance of the methods is discussed using classification, recognition and approximation measures that quantify the adherence to the original primitives and the goodness of fit. 

\subsection{Organization}
The remainder of this paper is organized as follows.
Section \ref{sec:benchmark} describes the characteristics of the proposed benchmark: the dataset, the ground truth, and the quality measures chosen to evaluate classification, approximation and recognition of primitives. Section \ref{sec:description_of_methods} presents a description of the six automatic algorithms that participate to this SHREC track: two of them are fully-based on neural networks, two are direct methods, while the remaining ones use deep learning in an initial classification step and proceed by applying direct methods. Finally, Section \ref{sec:analysis} provides an accurate comparative analysis of the results obtained by the six methods.

\section{The benchmark}\label{sec:benchmark}
The benchmark is available at \url{https://github.com/chiararomanengo/SHREC2022.git}.

\subsection{Dataset and ground truth}\label{sec:dataset_description}
The dataset is composed of $46,925$ three-dimensional segments represented as point clouds; it is divided into a training set and a test set that contain, respectively, $46,000$ and $925$ point clouds. Each point cloud is provided in a TXT file listing one triplet per line. 

\begin{figure}[h!]
\centering
\resizebox{0.70\textwidth }{!}{	
\begin{tabular}{c|>{\centering}m{2.0cm}| >{\centering}m{2.0cm}|>{\centering}m{2.0cm}|
>{\centering}m{2.0cm}|>{\centering}m{2.0cm}| c}
%First row
\hhline{~-----~} 
& \cellcolor{BlueViolet!20}T1
& \cellcolor{BlueViolet!20}T2 & \cellcolor{BlueViolet!20}T3 & \cellcolor{BlueViolet!20}T4 & 
\cellcolor{BlueViolet!20}T5 & \\ 
\hhline{------~}

% CLEAN DATA
\multicolumn{1}{|c|}{\cellcolor{BlueViolet!20} \begin{turn}{90}$\text{A0}$\end{turn}} 
& 
\includegraphics[scale=0.18, trim={3.5cm 0.5cm 2.5cm 1cm}, clip]{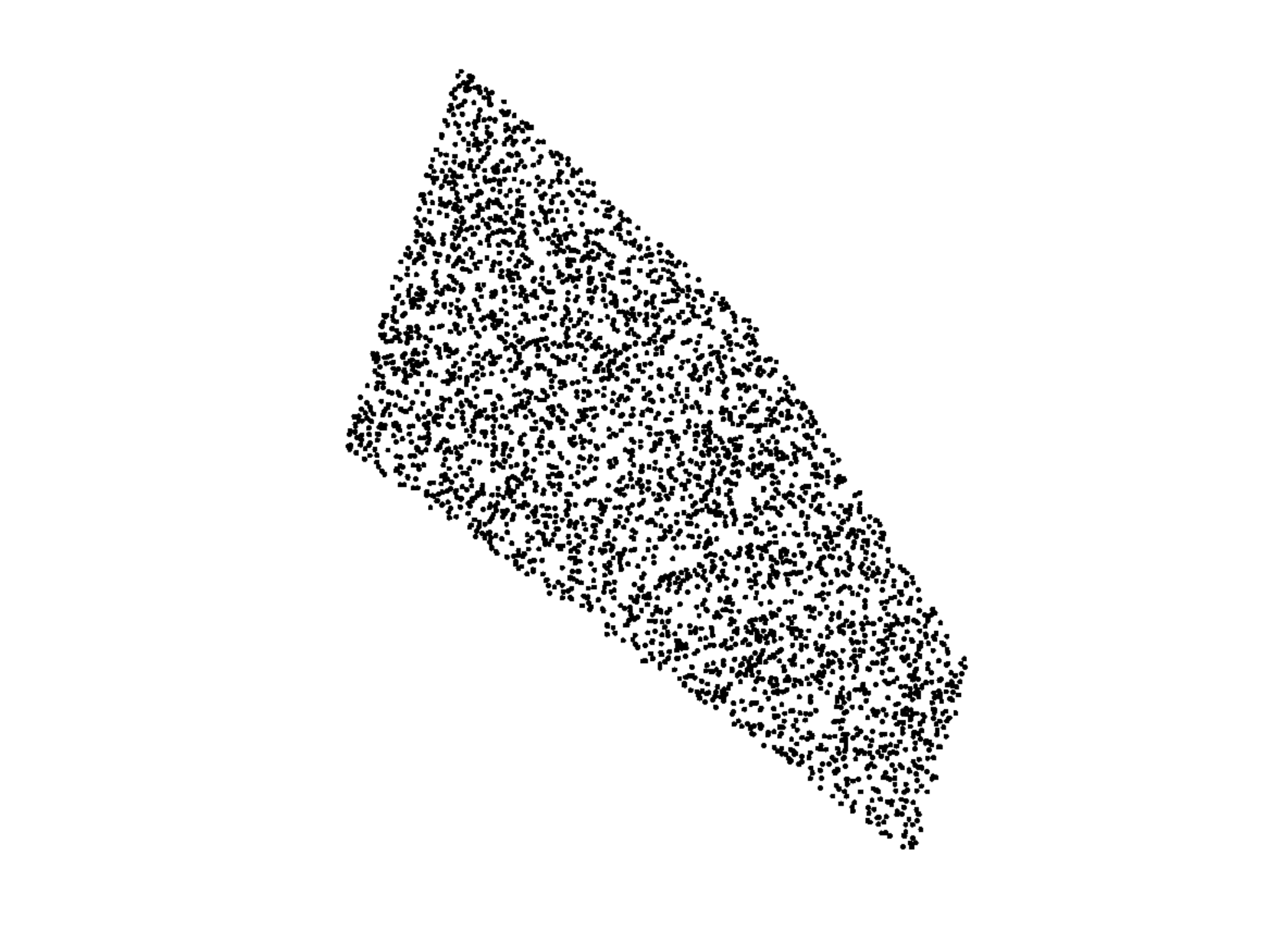}
&
\includegraphics[scale=0.25, trim={5.5cm 2.5cm 5cm 2cm}, clip]{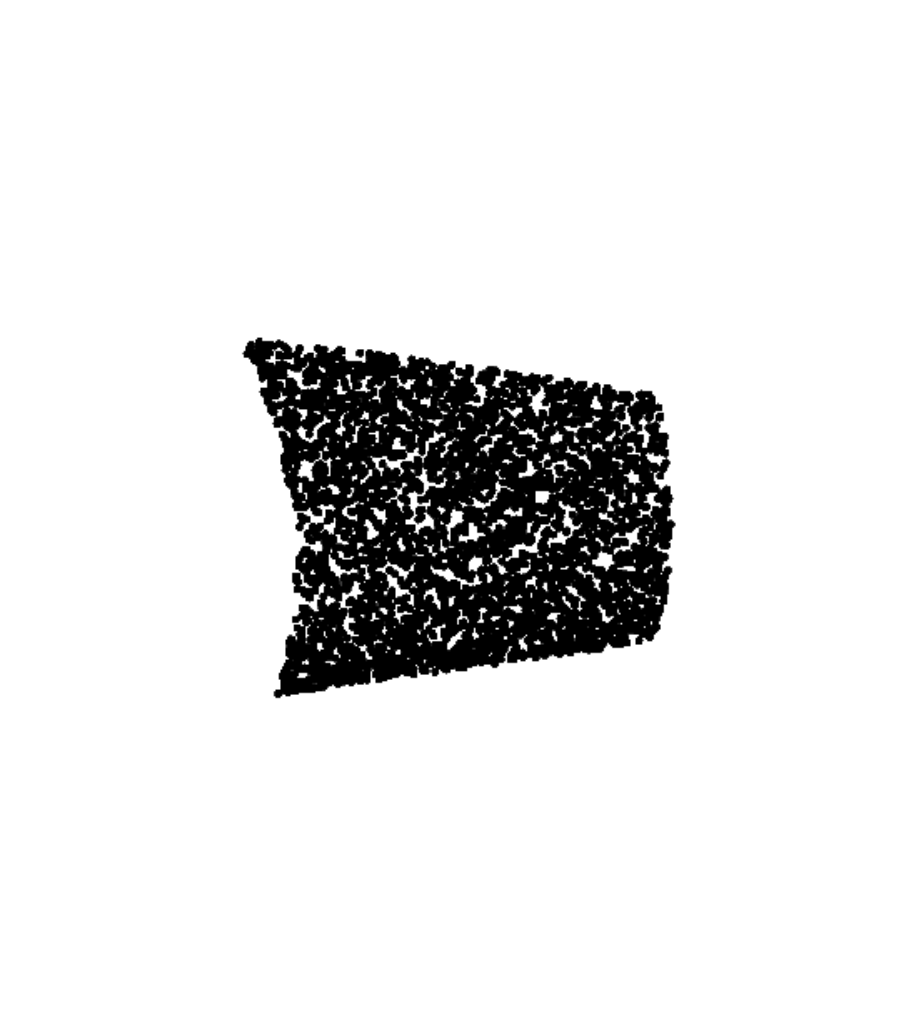}
& 
\includegraphics[scale=0.20, trim={5.5cm 2.5cm 5cm 2cm}, clip]{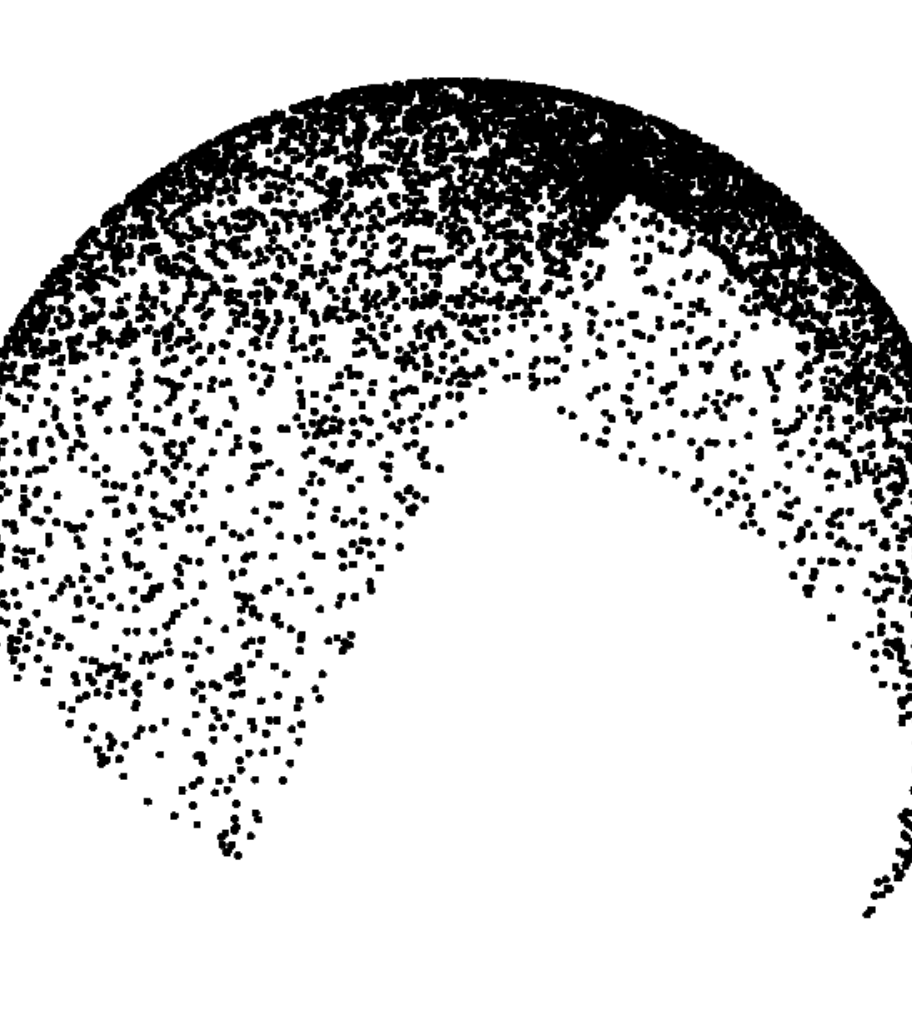}
& 
\includegraphics[scale=0.20, trim={5cm 2.5cm 5cm 2.5cm}, clip]{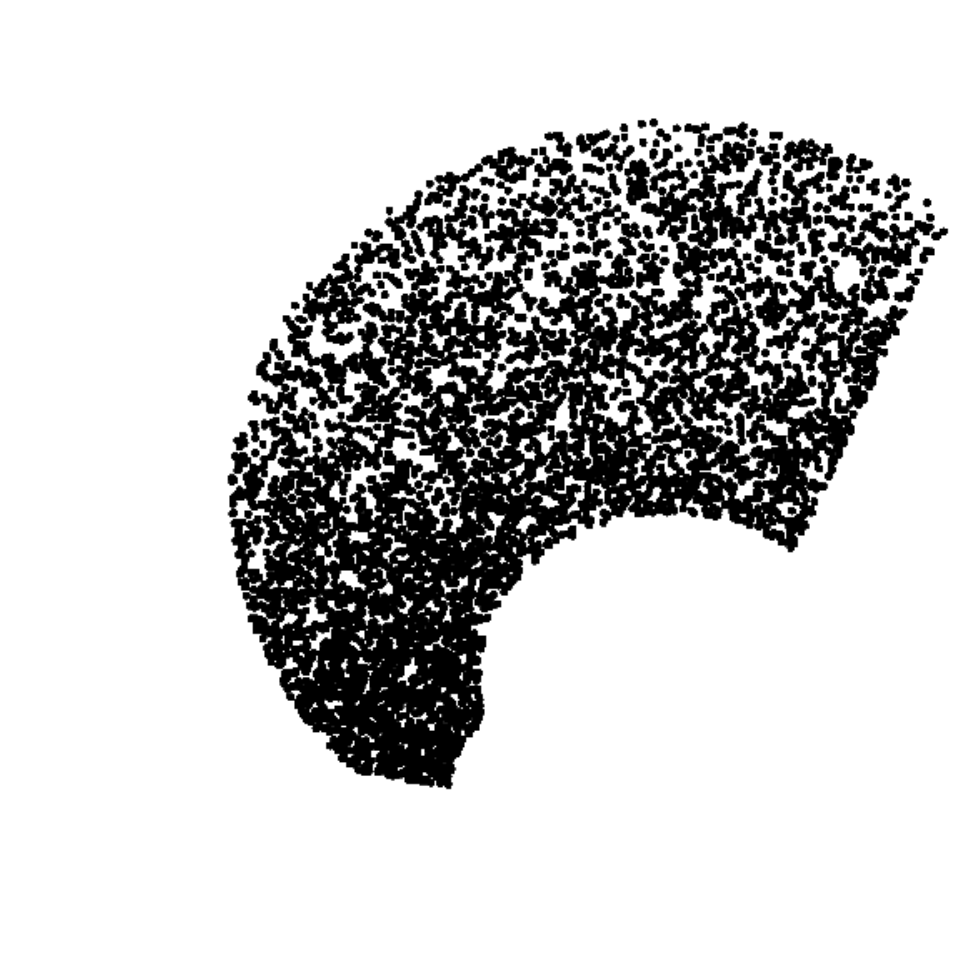}
& 
\includegraphics[scale=0.20, trim={6cm 2.5cm 5cm 2.5cm}, clip]{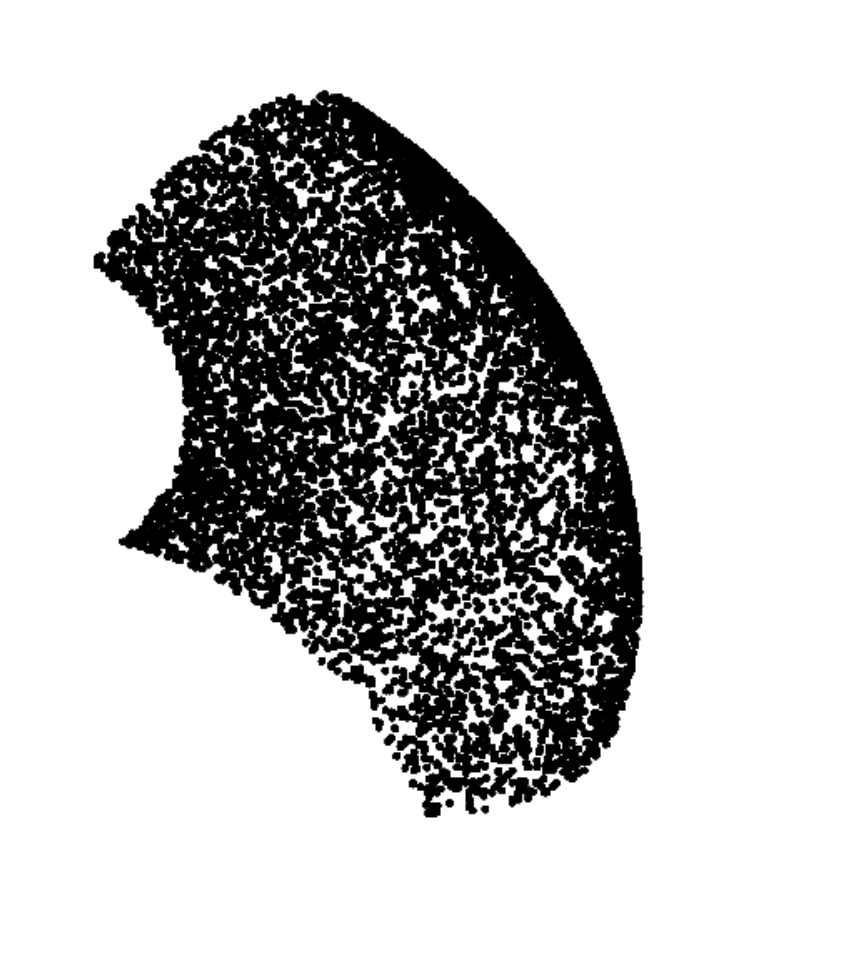}
&
\\
\hhline{------~} 

% UNIFORM NOISE
\multicolumn{1}{|c|}{\cellcolor{BlueViolet!20} \begin{turn}{90}$\text{A1}$\end{turn}} 
& 
\includegraphics[scale=0.155, trim={3.75cm 2.0cm 2.75cm 2cm}, clip]{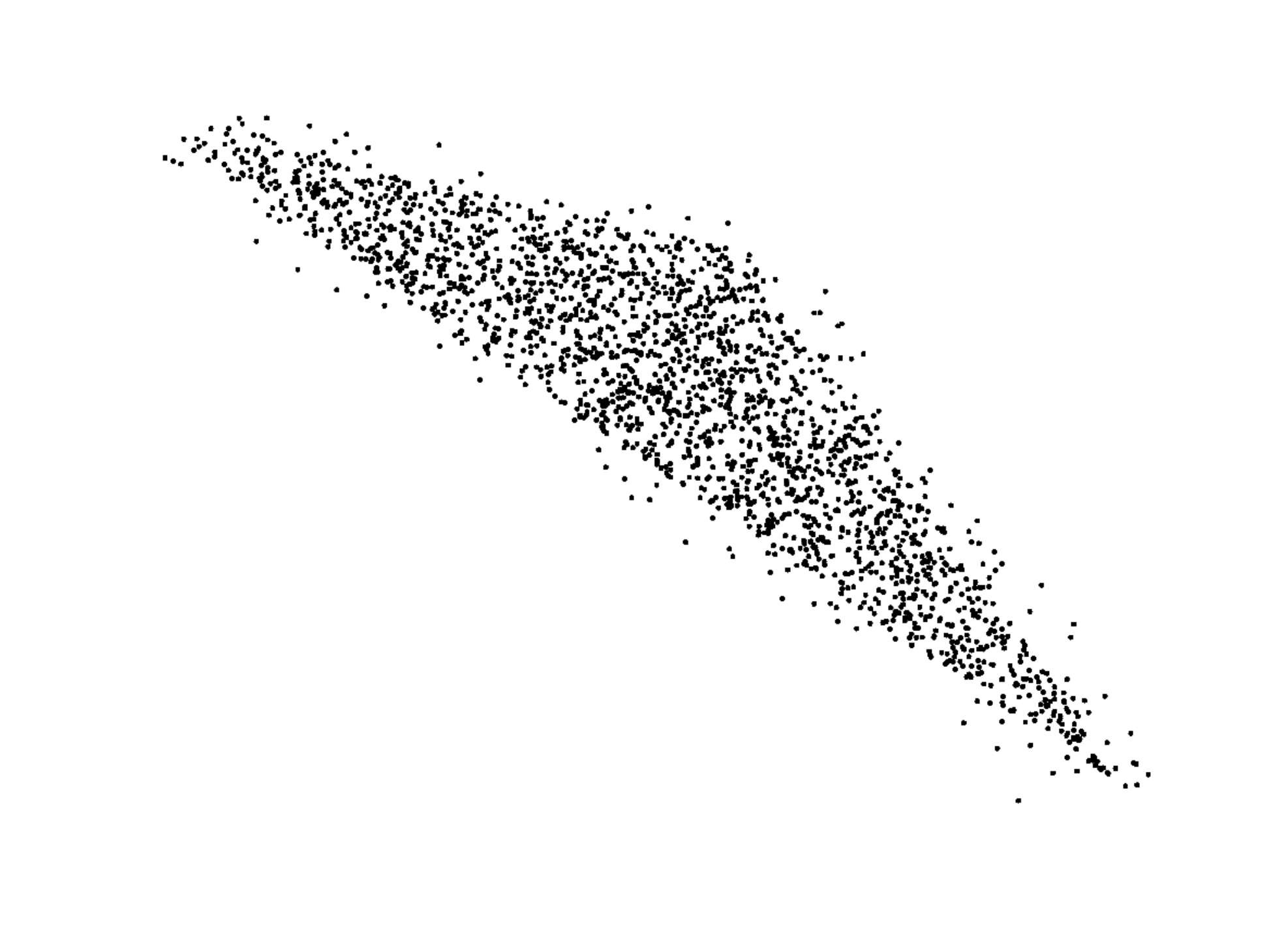}
&
\includegraphics[scale=0.26, trim={5.75cm 2.75cm 5cm 2cm}, clip]{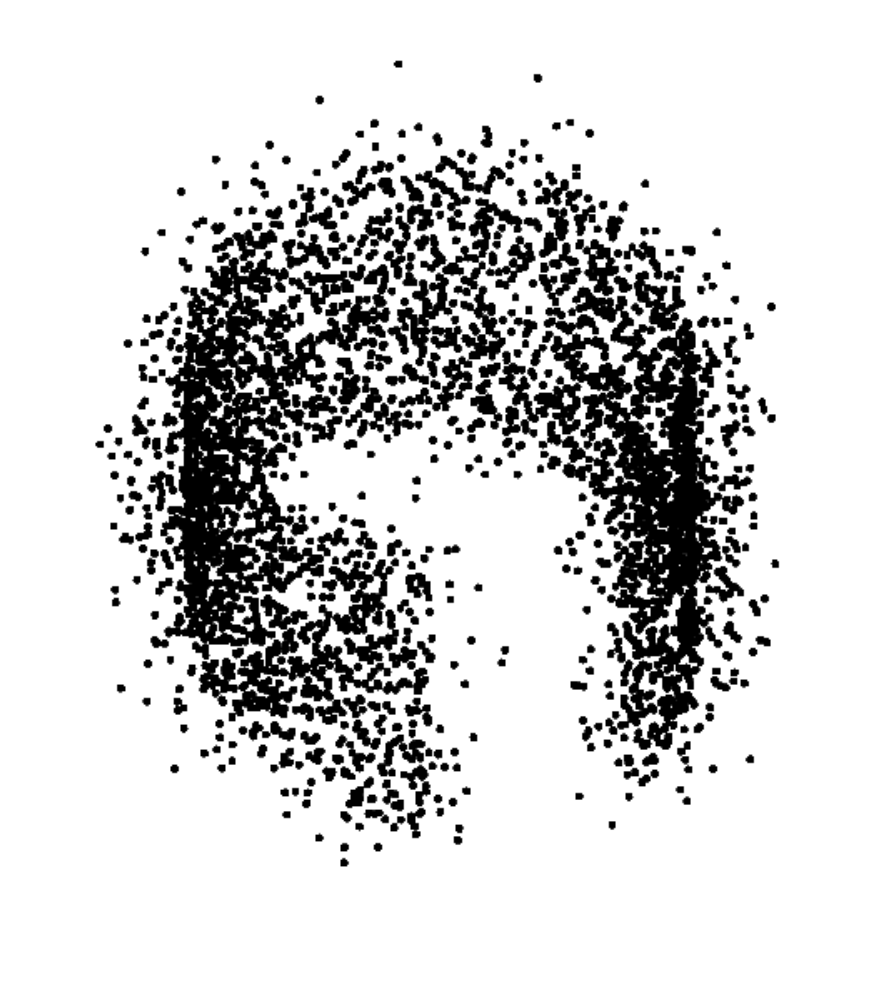}
& 
\includegraphics[scale=0.26, trim={5.75cm 2.75cm 5cm 2cm}, clip]{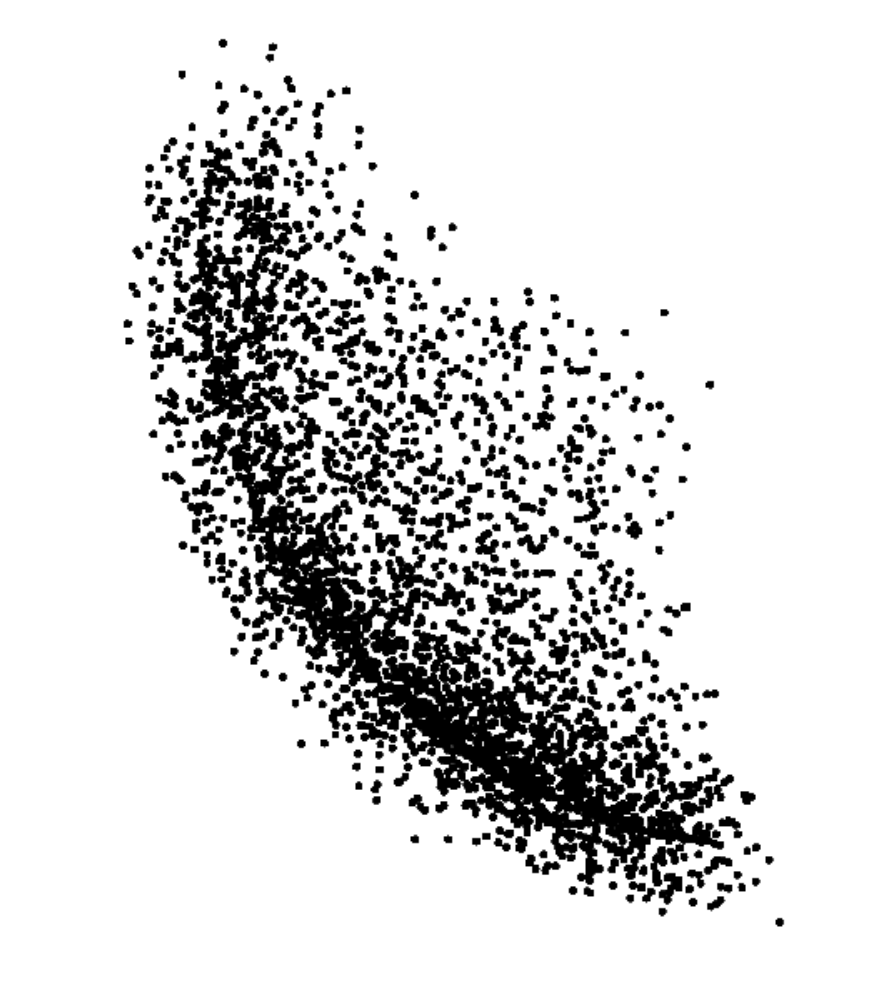}
& 
\includegraphics[scale=0.24, trim={5.75cm 2.75cm 5cm 2.5cm}, clip]{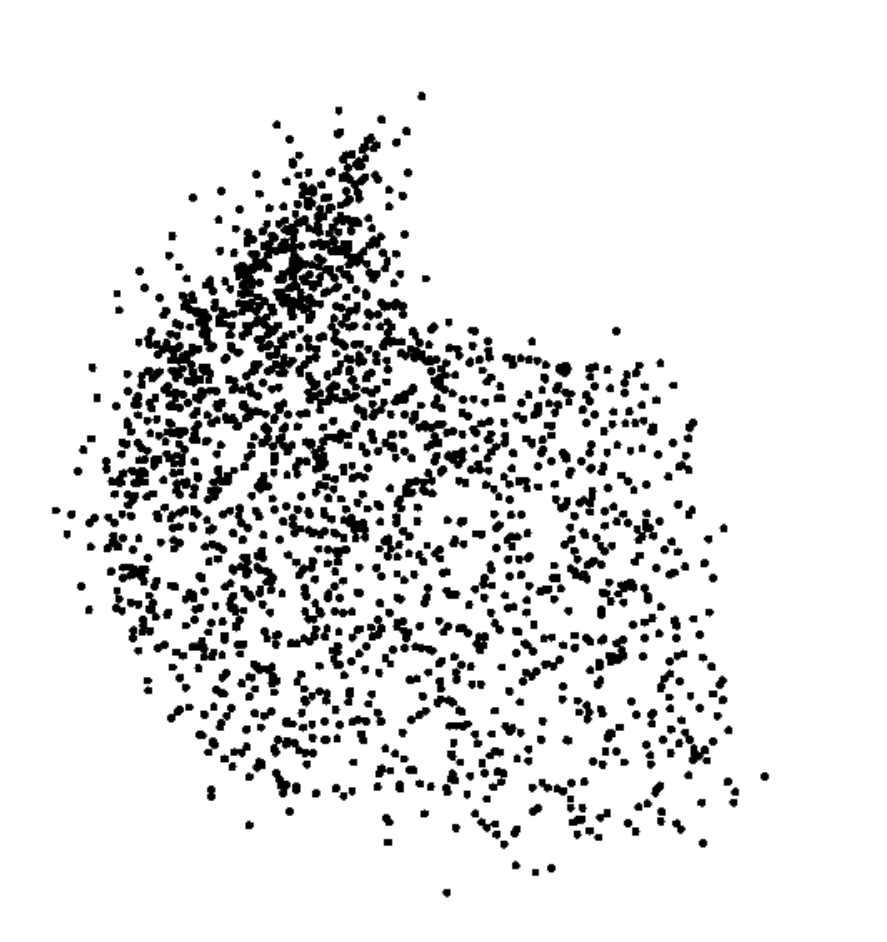}
& 
\includegraphics[scale=0.25, trim={6cm 2.75cm 5cm 2.5cm}, clip]{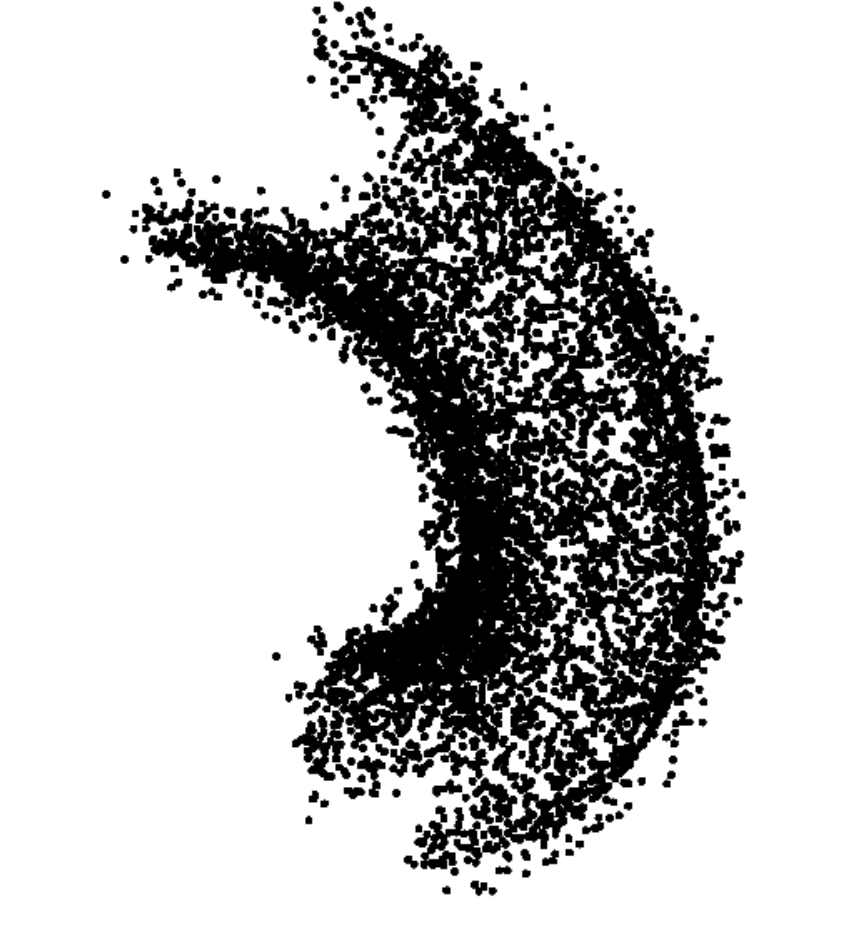}
&
\\
\hhline{------~} 

% GAUSSIAN NOISE
\multicolumn{1}{|c|}{\cellcolor{BlueViolet!20} \begin{turn}{90}$\text{A2}$\end{turn}} 
& 
\includegraphics[scale=0.25, trim={7cm 4cm 6cm 3.0cm}, clip]{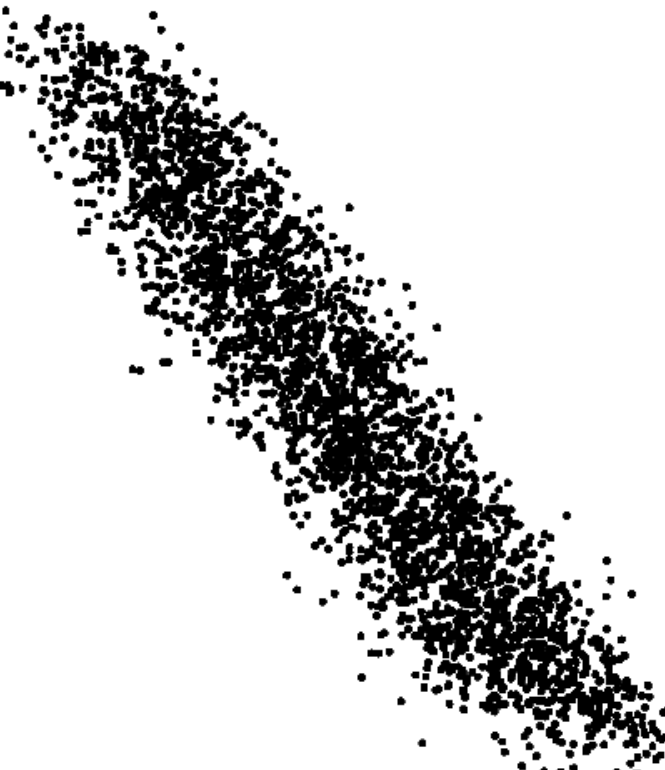}
&
\includegraphics[scale=0.175, trim={5cm 2.5cm 4.5cm 2.5cm}, clip]{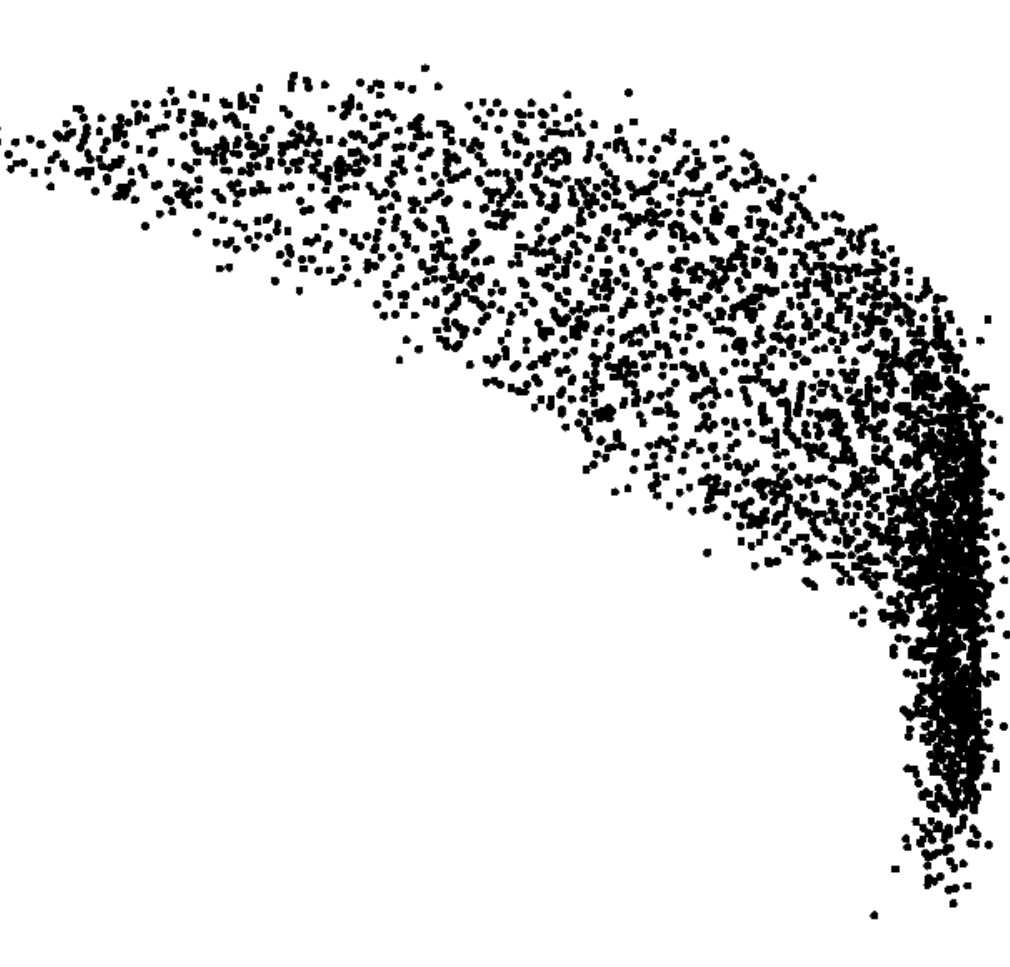}
&
\includegraphics[scale=0.175, trim={5cm 1.5cm 4cm 1cm}, clip]{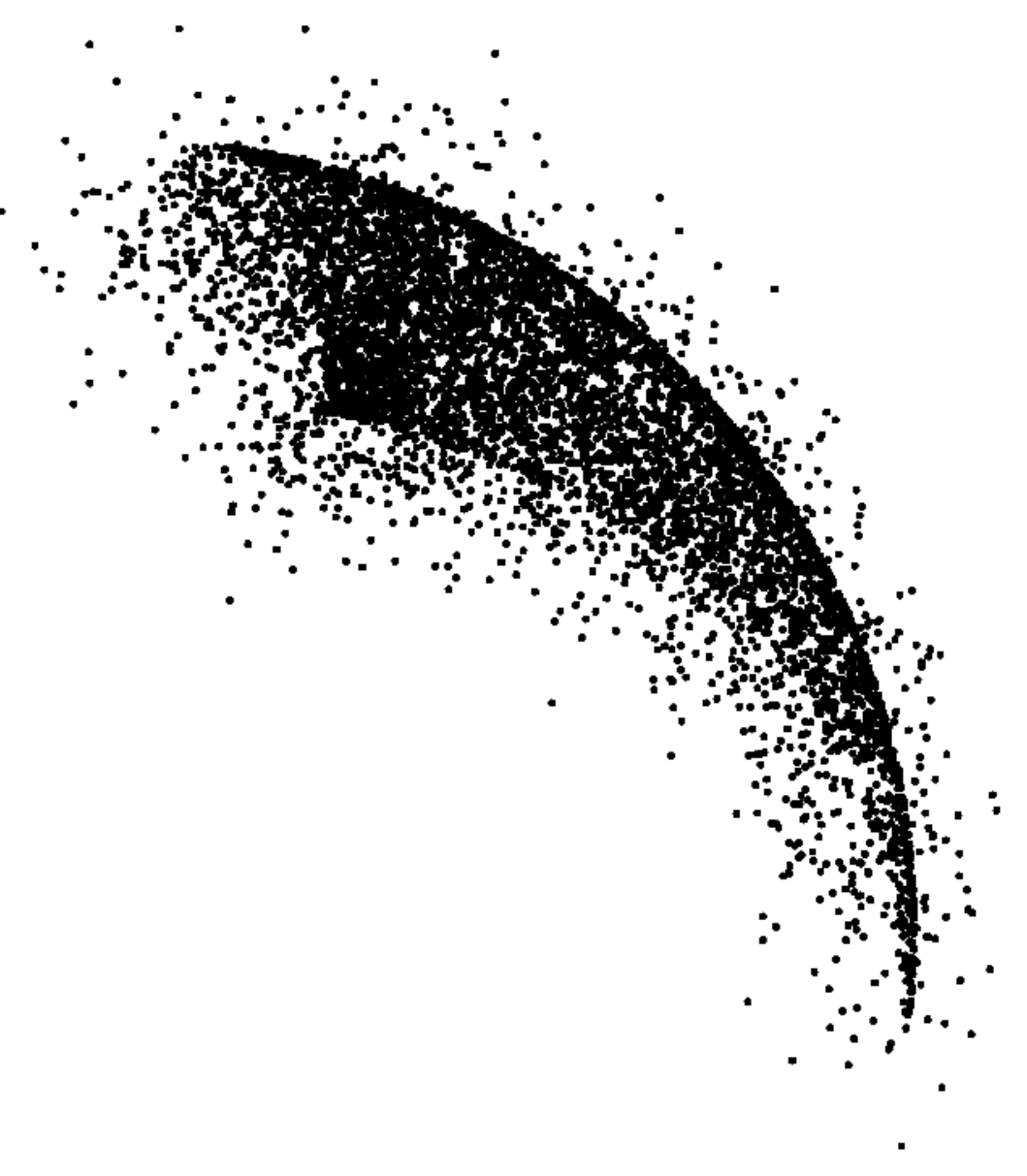}
&
\includegraphics[scale=0.20, trim={5.75cm 4cm 4cm 3cm}, clip]{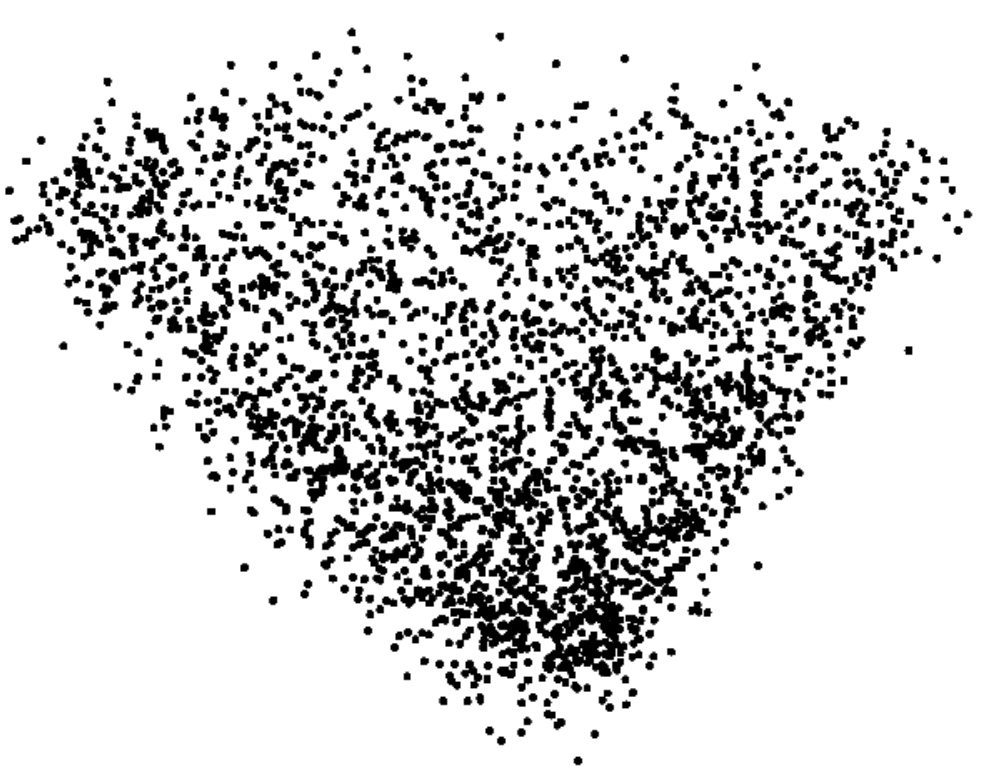}
&
\includegraphics[scale=0.175, trim={5cm 1cm 4cm 1cm}, clip]{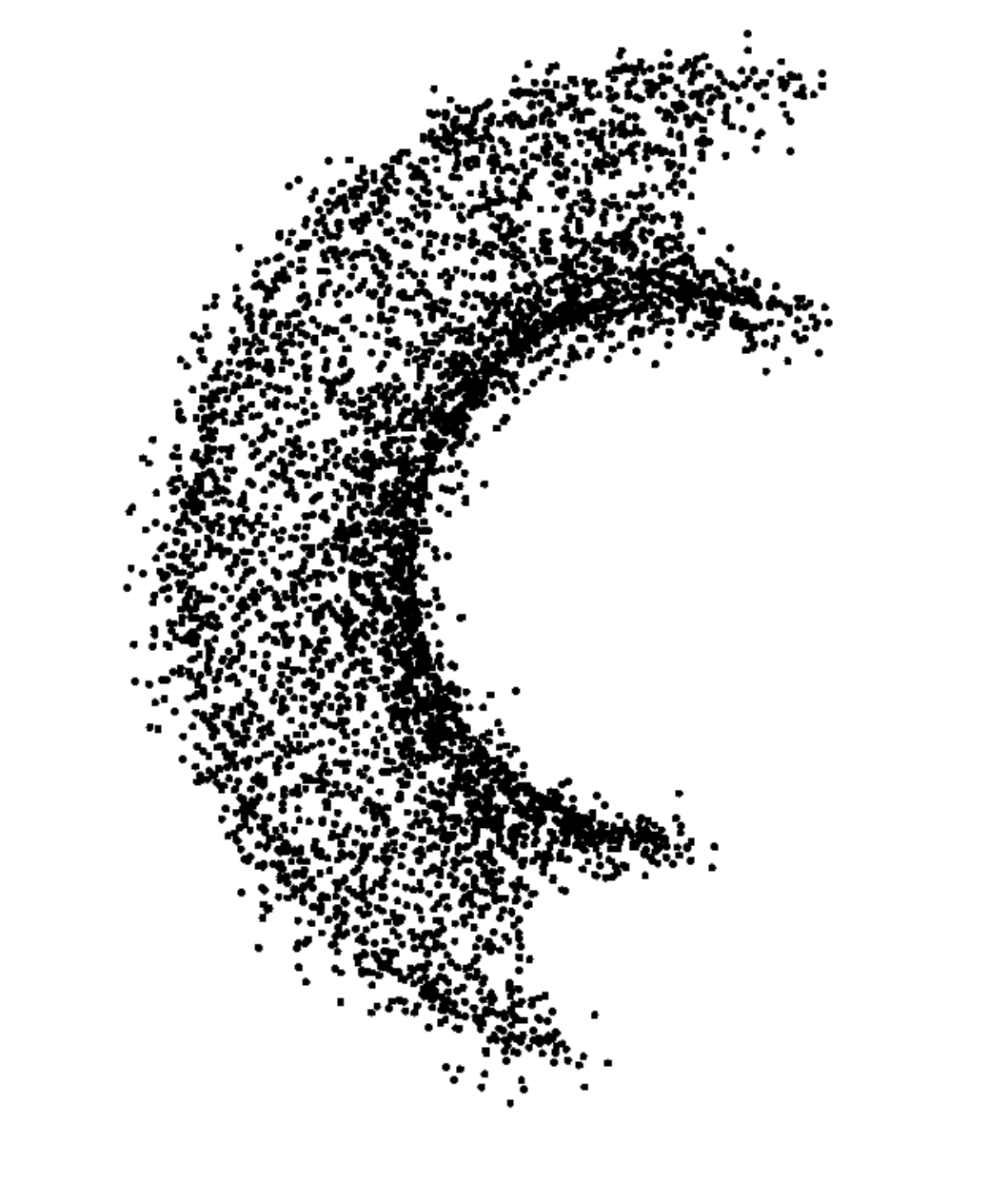}
&
\\
\hhline{------~} 

% UNDERSAMPLING
\multicolumn{1}{|c|}{\cellcolor{BlueViolet!20} \begin{turn}{90}$\text{A3}$\end{turn}} 
& 
\includegraphics[scale=0.175, trim={4.5cm 2cm 4.5cm 2.5cm}, clip]{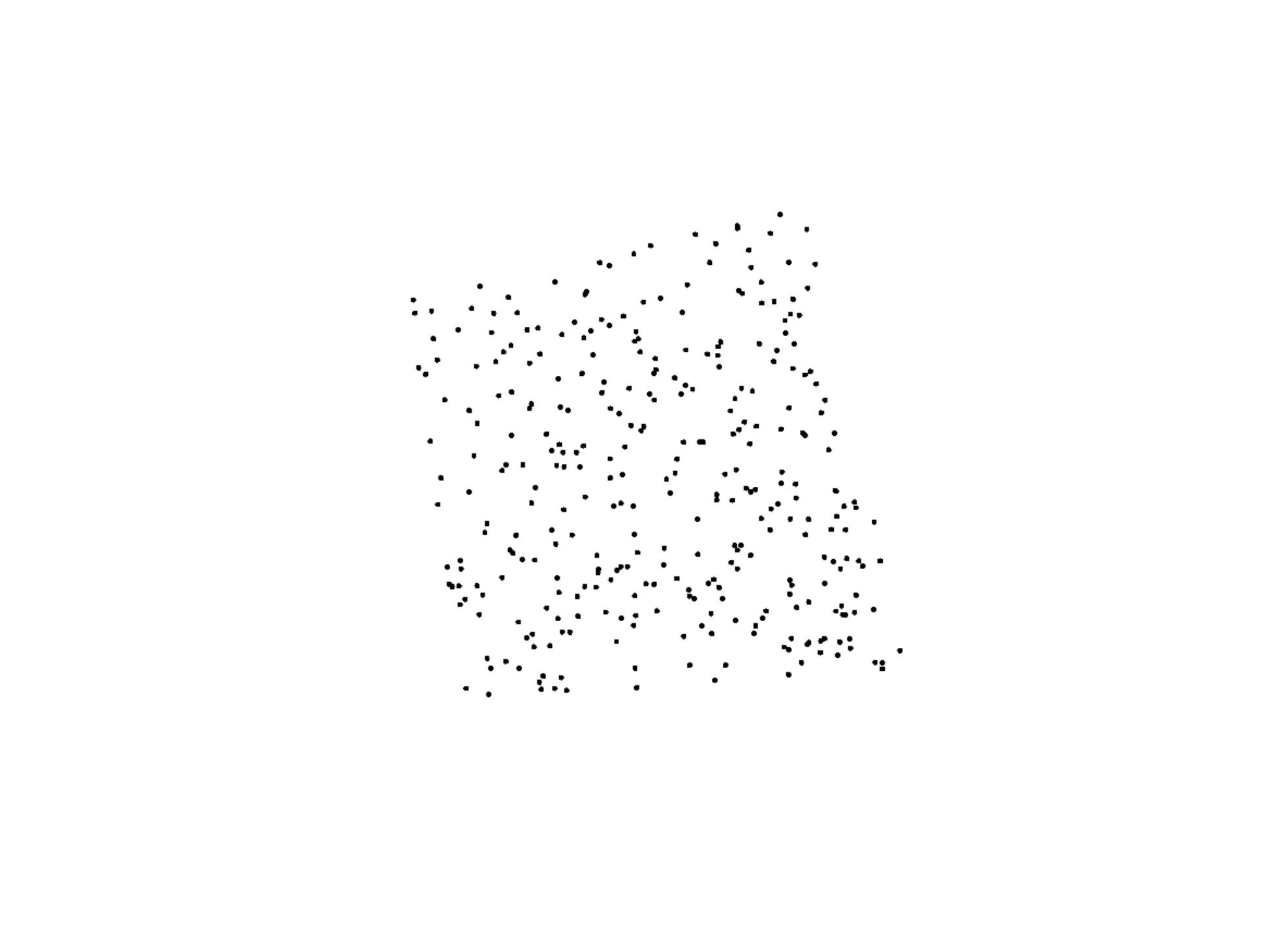}
&
\includegraphics[scale=0.175, trim={4cm 1.5cm 4cm 1.5cm}, clip]{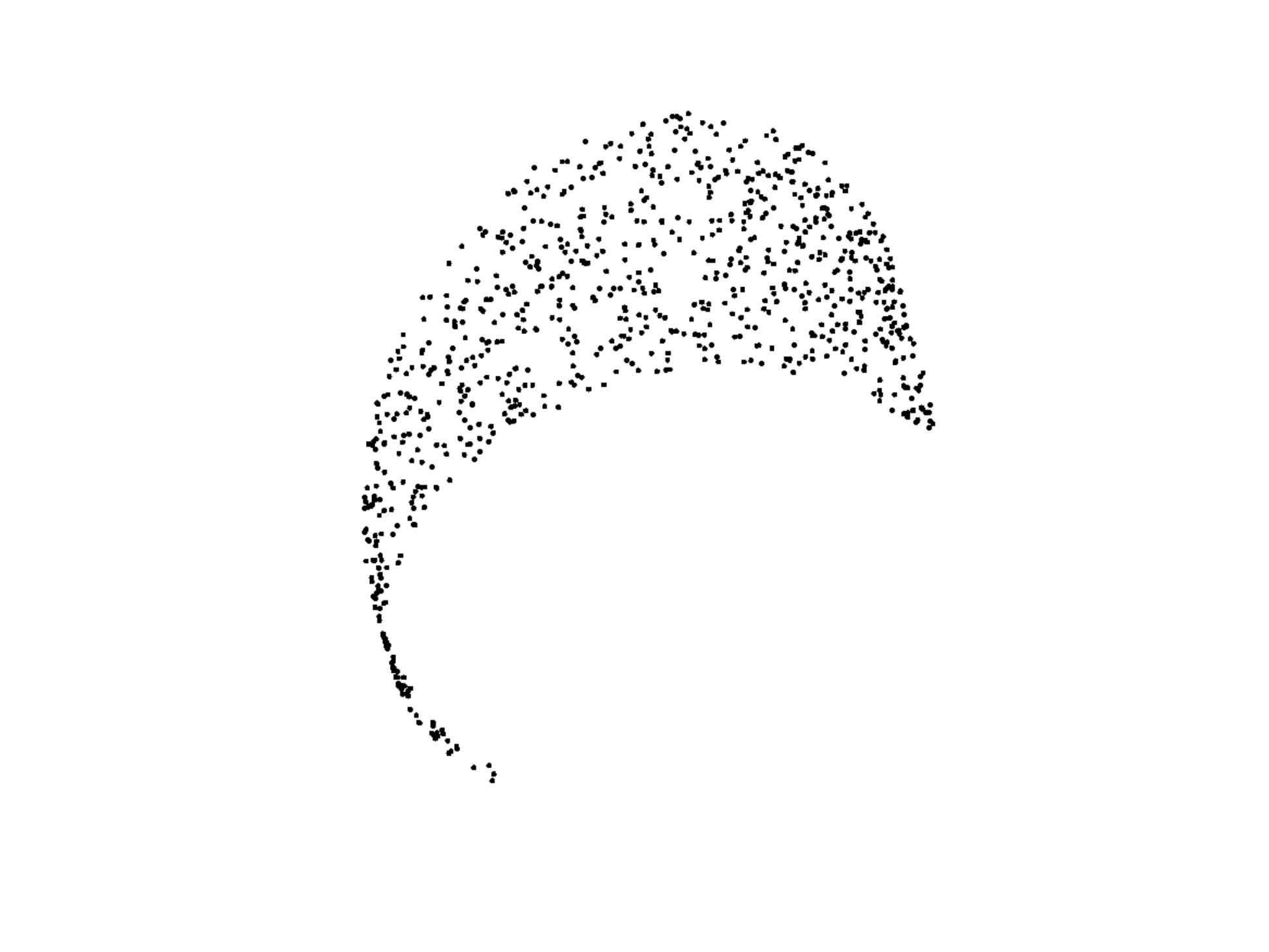}
&
\includegraphics[scale=0.2, trim={6cm 4cm 5cm 3cm}, clip]{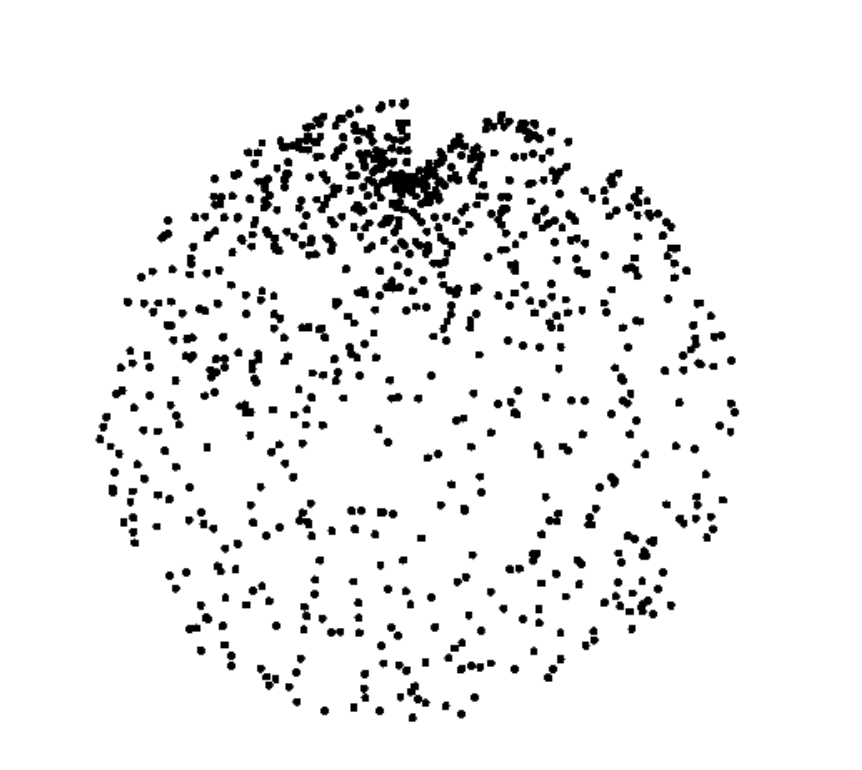}
&
\includegraphics[scale=0.15, trim={3.5cm 2.5cm 2.4cm 1.5cm}, clip]{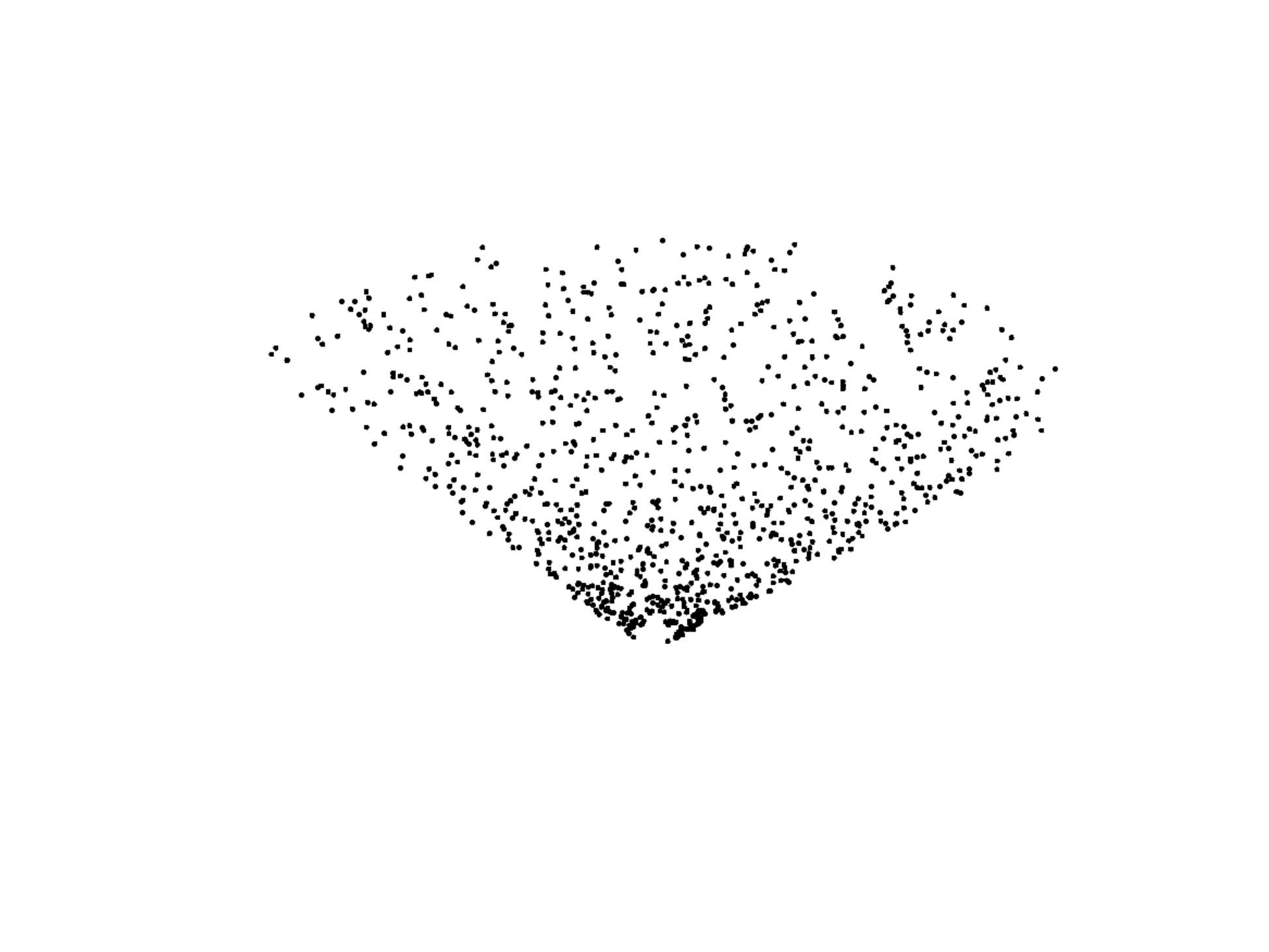}
&
\includegraphics[scale=0.15, trim={3.5cm 1cm 3.4cm 1cm}, clip]{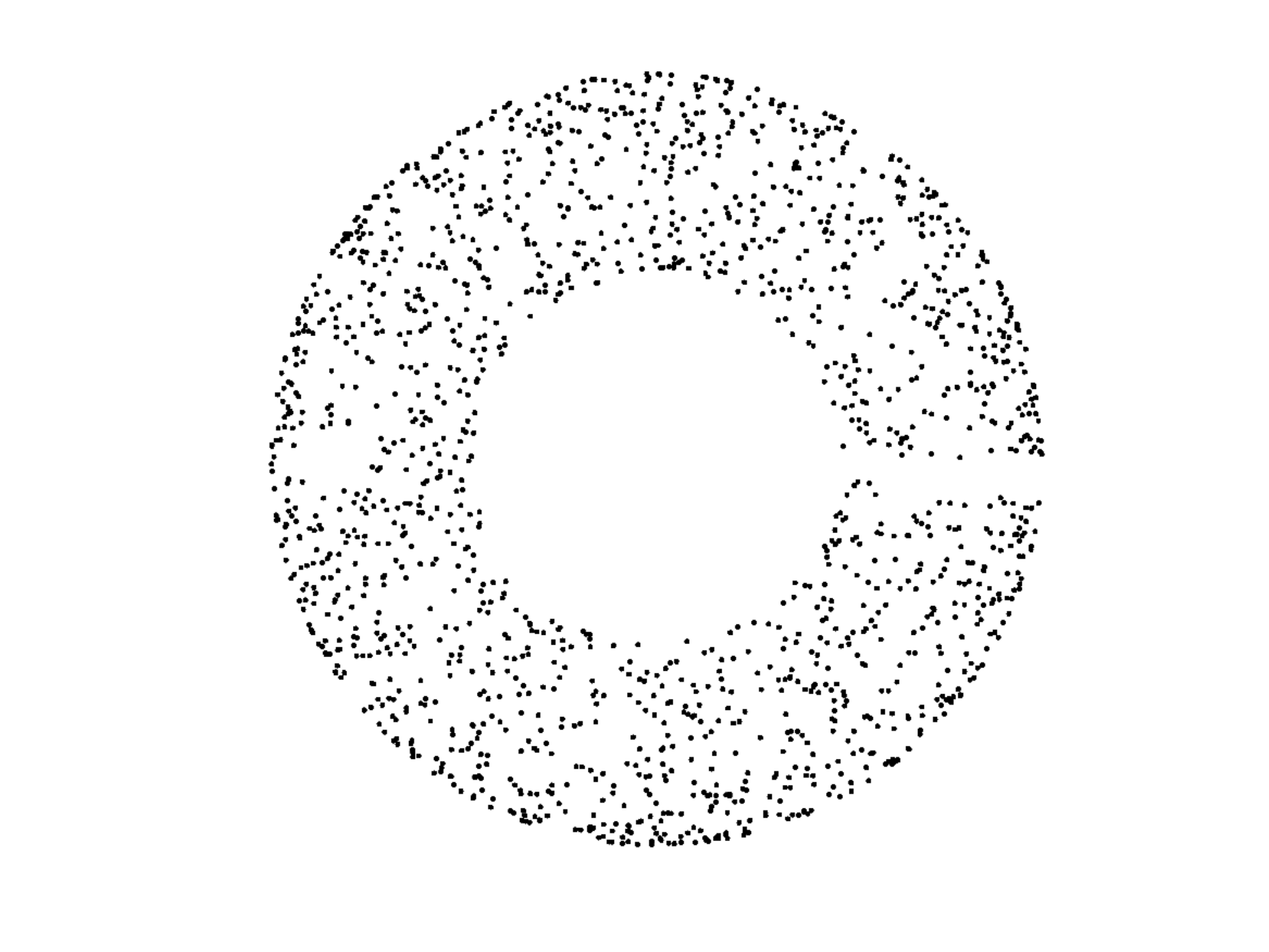}
&
\\
\hhline{------~} 

% MISSING DATA
\multicolumn{1}{|c|}{\cellcolor{BlueViolet!20} \begin{turn}{90}$\text{A4}$\end{turn}} 
& 
\includegraphics[scale=0.20, trim={5cm 2cm 5cm 2.5cm}, clip]{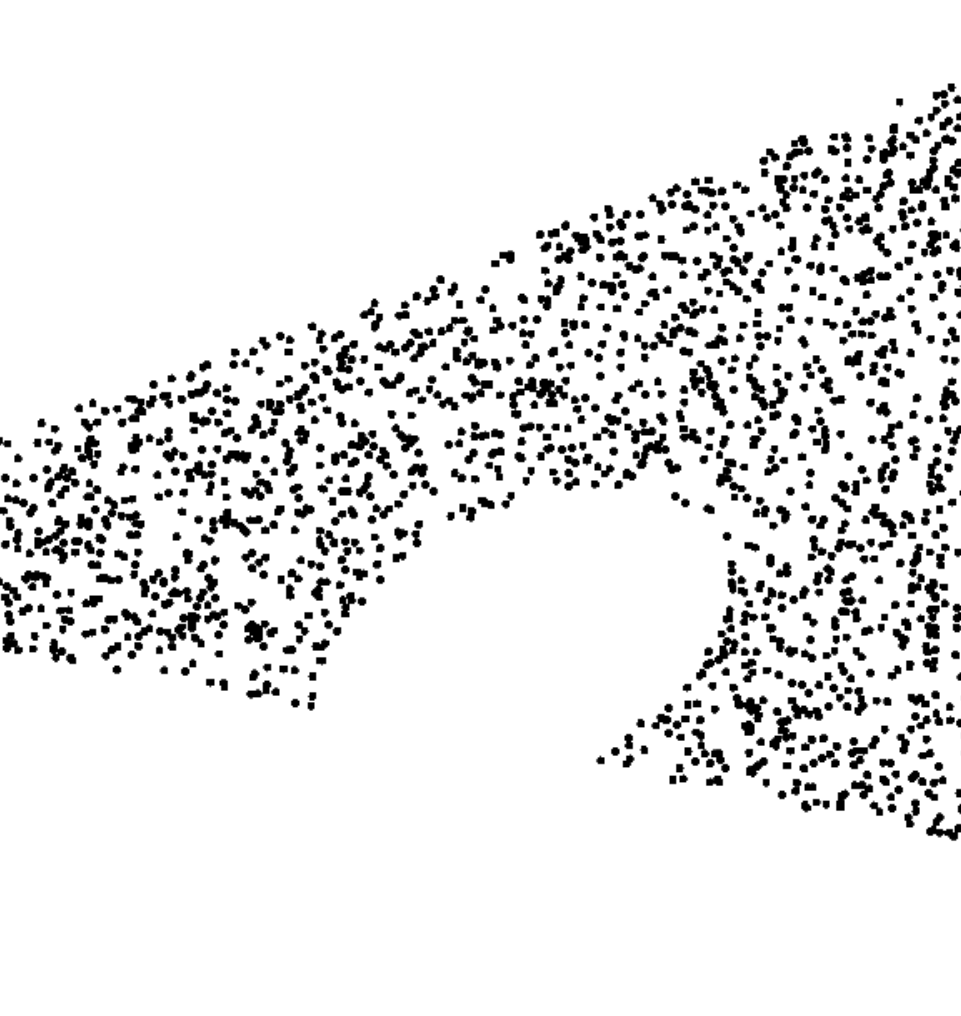}
&
\includegraphics[scale=0.2, trim={5cm 2.5cm 5cm 3cm}, clip]{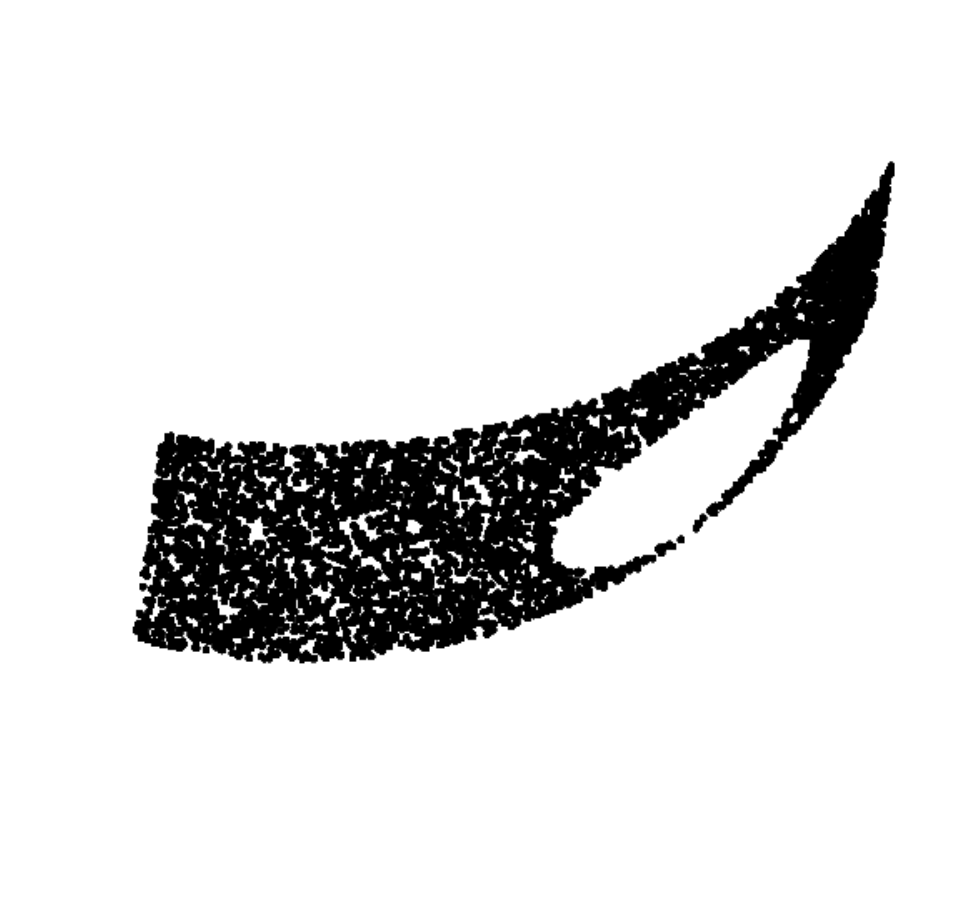}
&
\includegraphics[scale=0.2, trim={5cm 2.5cm 4cm 2cm}, clip]{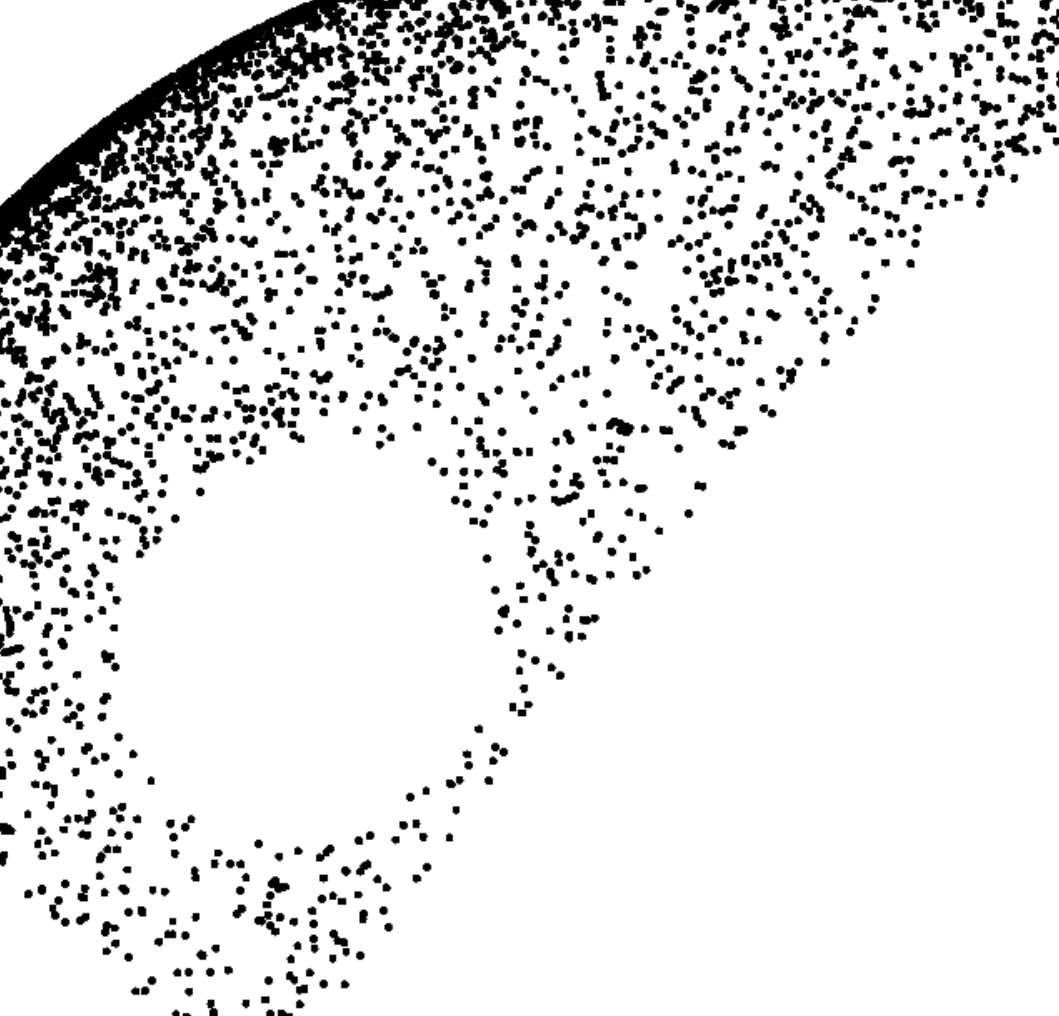}
&
\includegraphics[scale=0.175, trim={5cm 2.5cm 3cm 1.5cm}, clip]{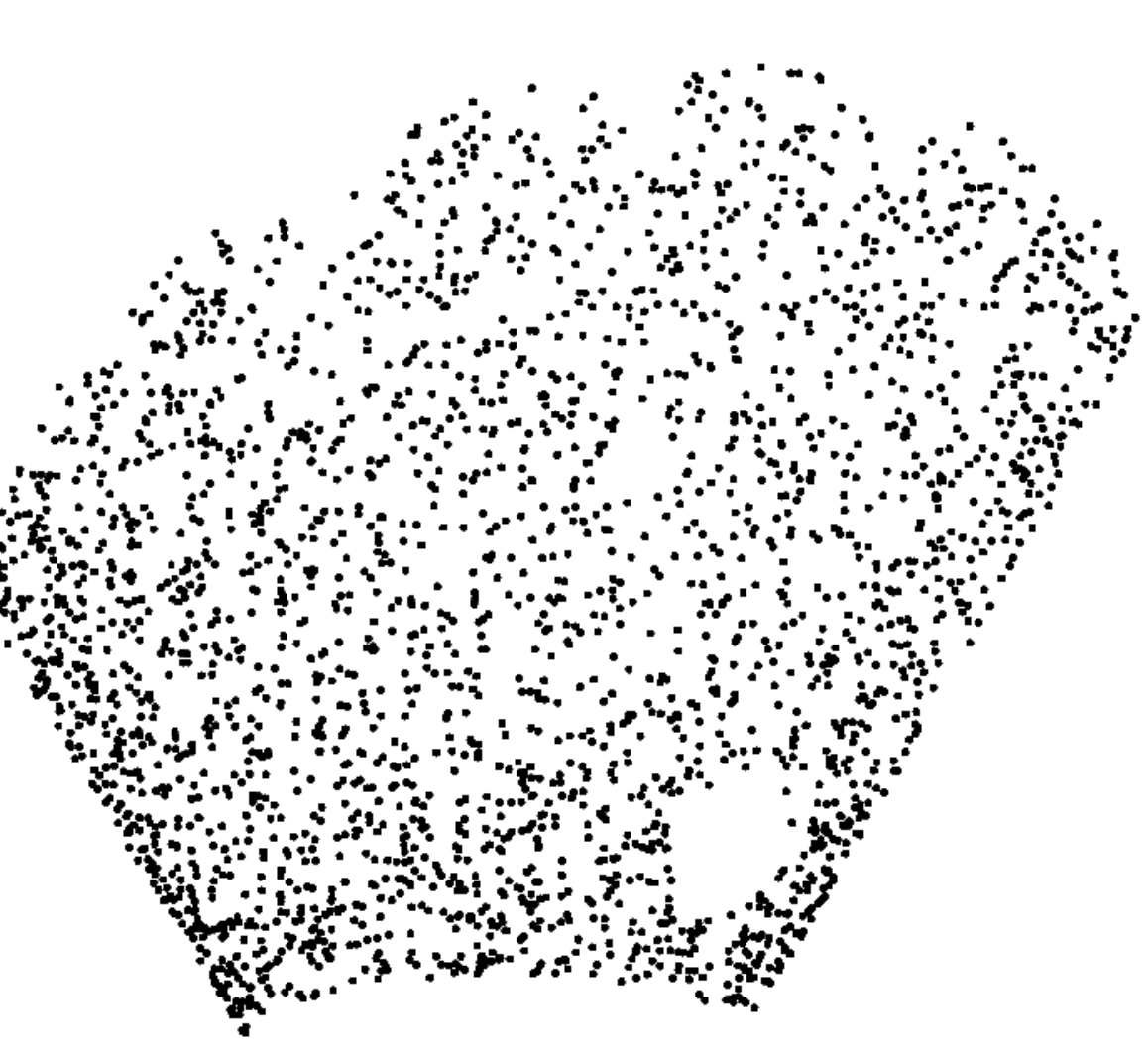}
&
\includegraphics[scale=0.175, trim={5cm 1cm 4cm 1cm}, clip]{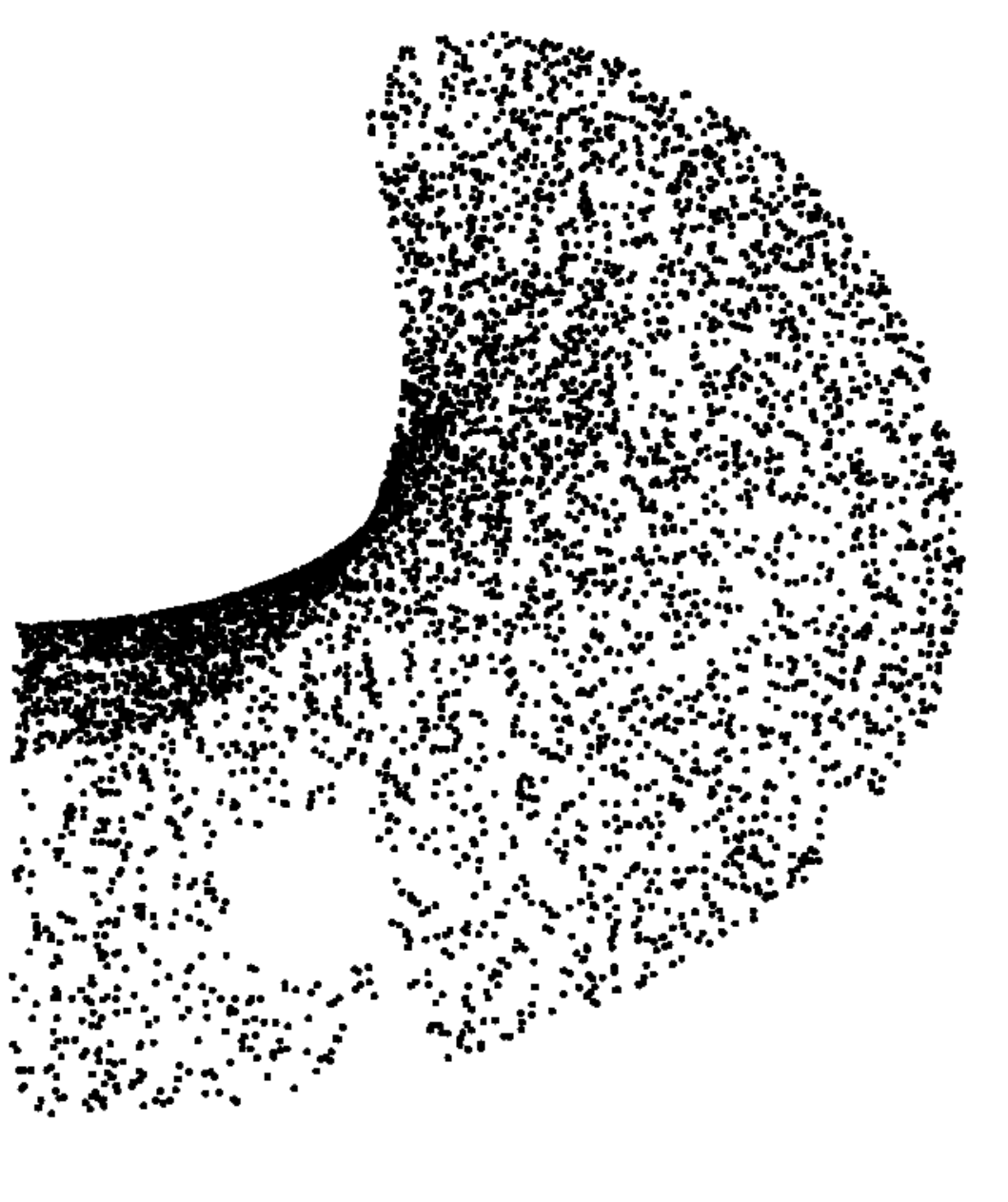}
&
\\
\hhline{------~} 

% UNIFORM NOISE & UNDERSAMPLING
\multicolumn{1}{|c|}{\cellcolor{BlueViolet!20} \begin{turn}{90}$\text{A5}$\end{turn}} 
& 
\includegraphics[scale=0.20, trim={4.5cm 2cm 4.5cm 2.5cm}, clip]{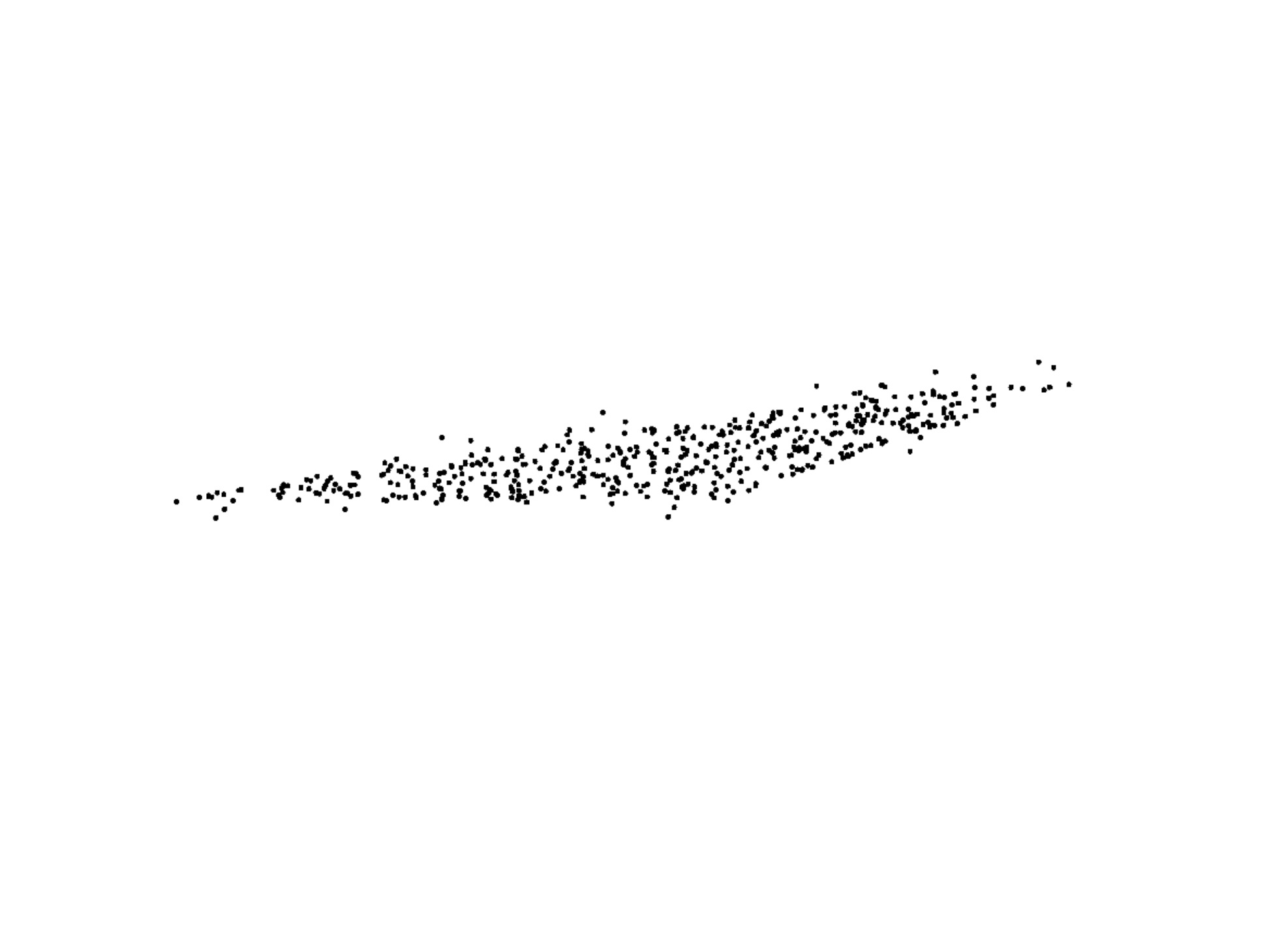}
&
\includegraphics[scale=0.175, trim={5cm 2.5cm 5cm 3cm}, clip]{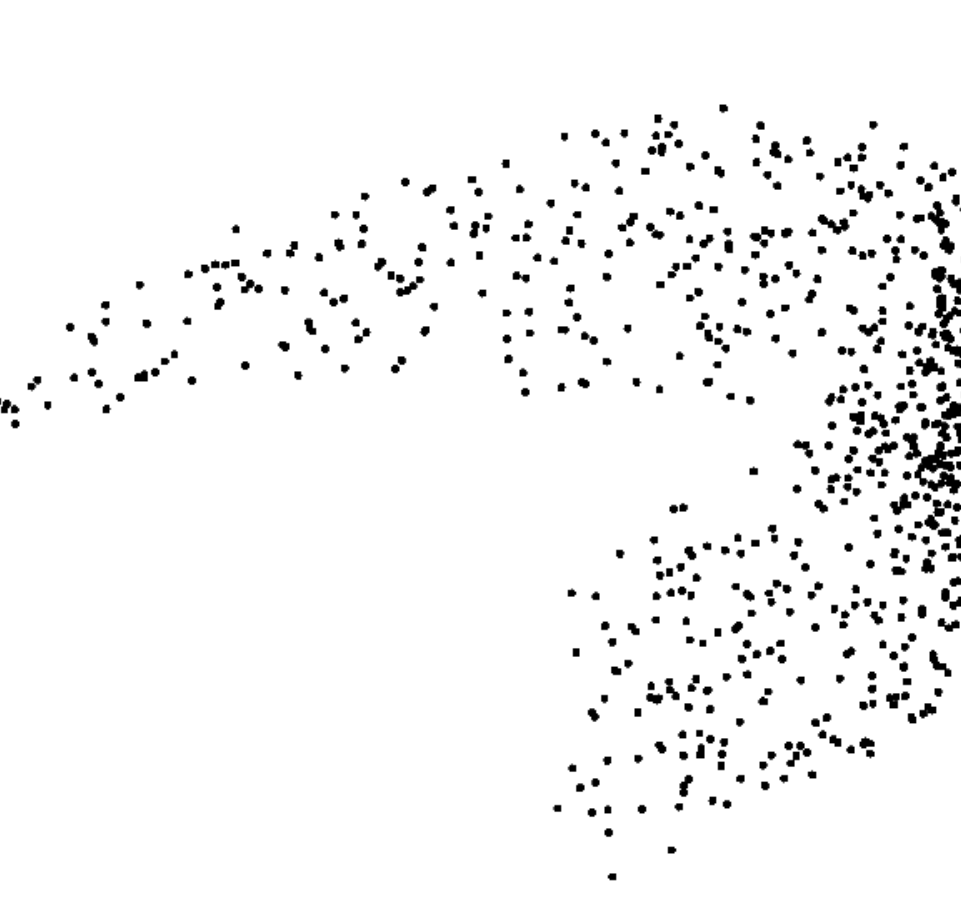}
&
\includegraphics[scale=0.175, trim={5cm 2.5cm 4cm 2cm}, clip]{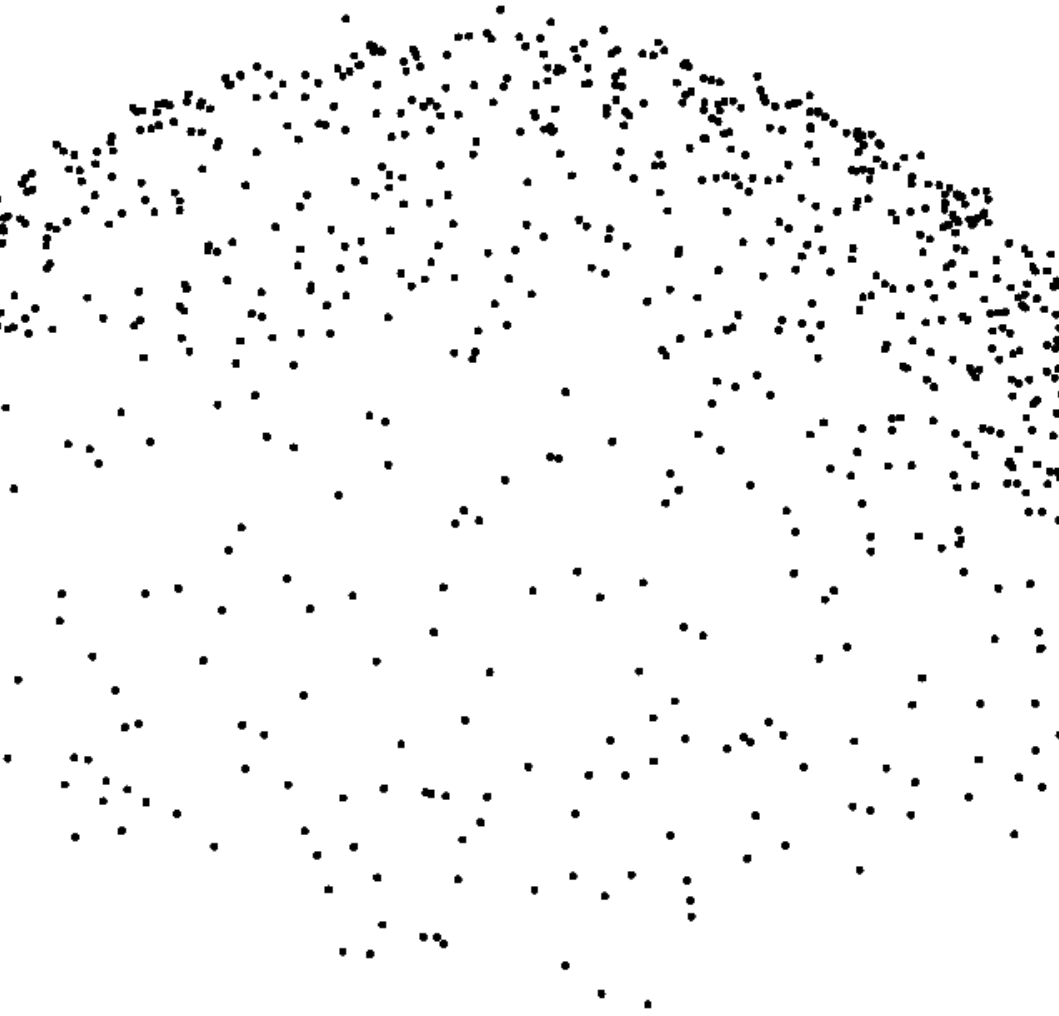}
&
\includegraphics[scale=0.175, trim={4cm 2.5cm 3cm 1.5cm}, clip]{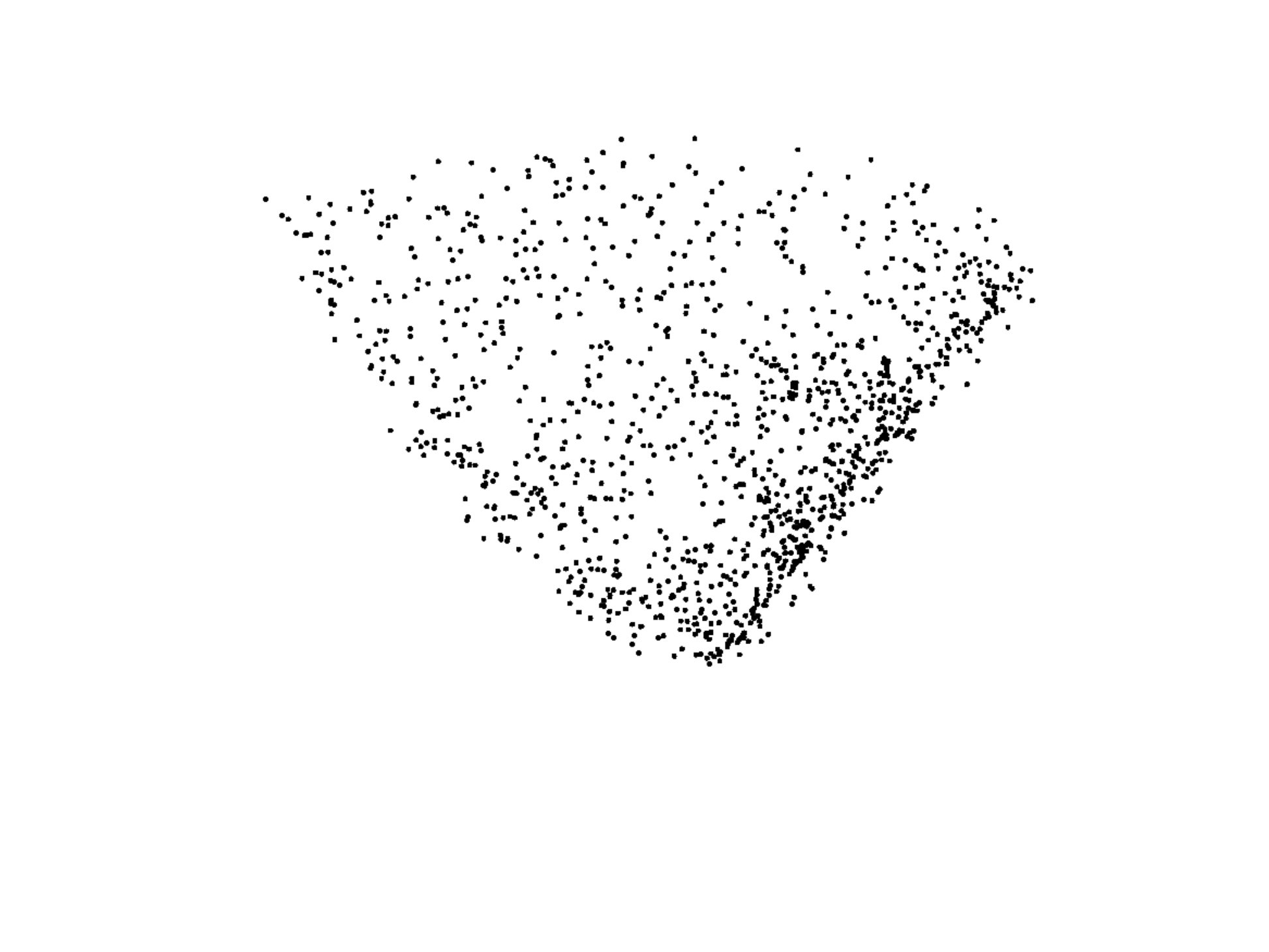}
&
\includegraphics[scale=0.175, trim={5cm 1cm 4cm 1cm}, clip]{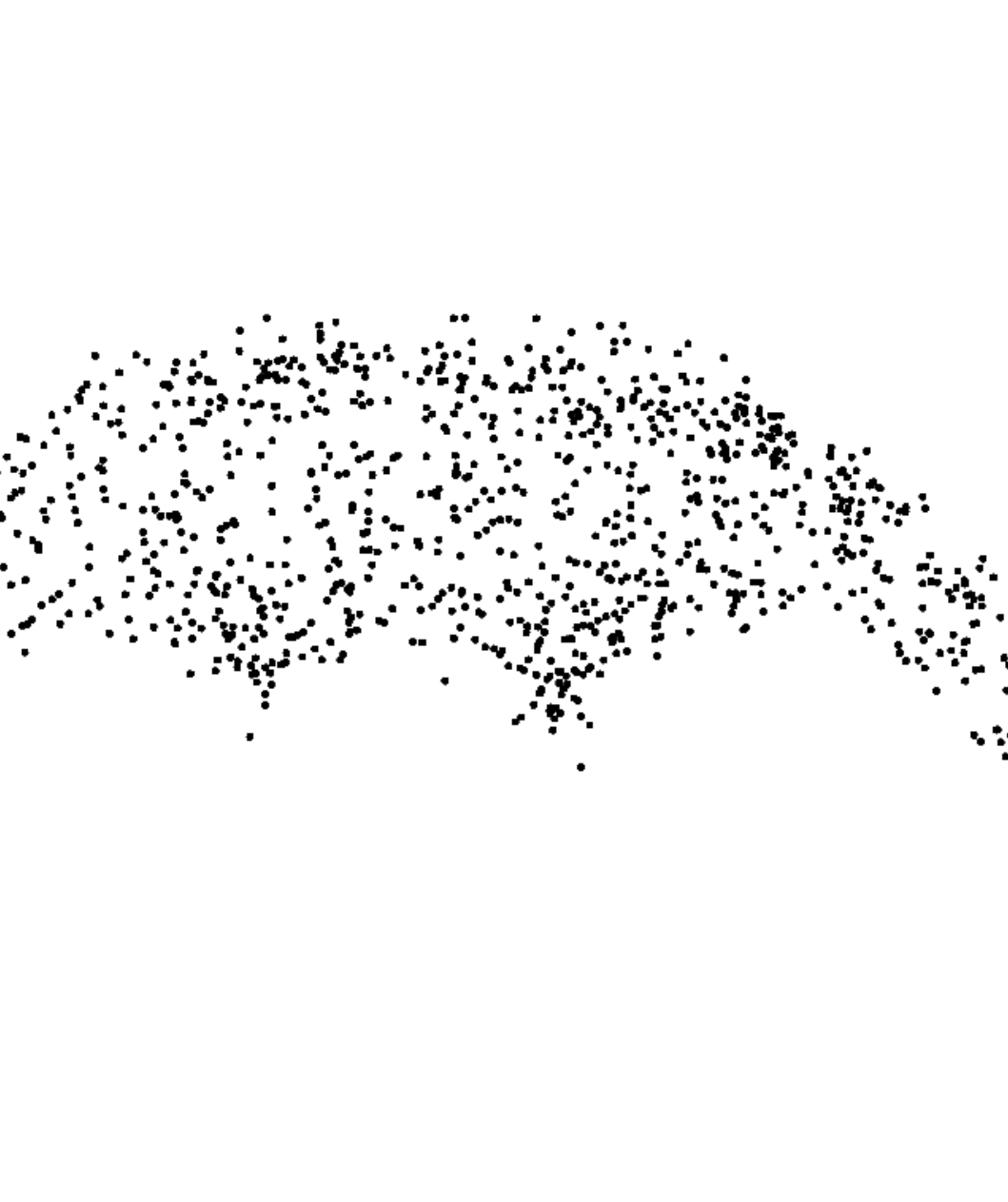}
&
\\
\hhline{------~} 

% GAUSSIAN NOISE & UNDERSAMPLING
\multicolumn{1}{|c|}{\cellcolor{BlueViolet!20} \begin{turn}{90}$\text{A6}$\end{turn}} 
& 
\includegraphics[scale=0.175, trim={4.5cm 2cm 4.5cm 2.5cm}, clip]{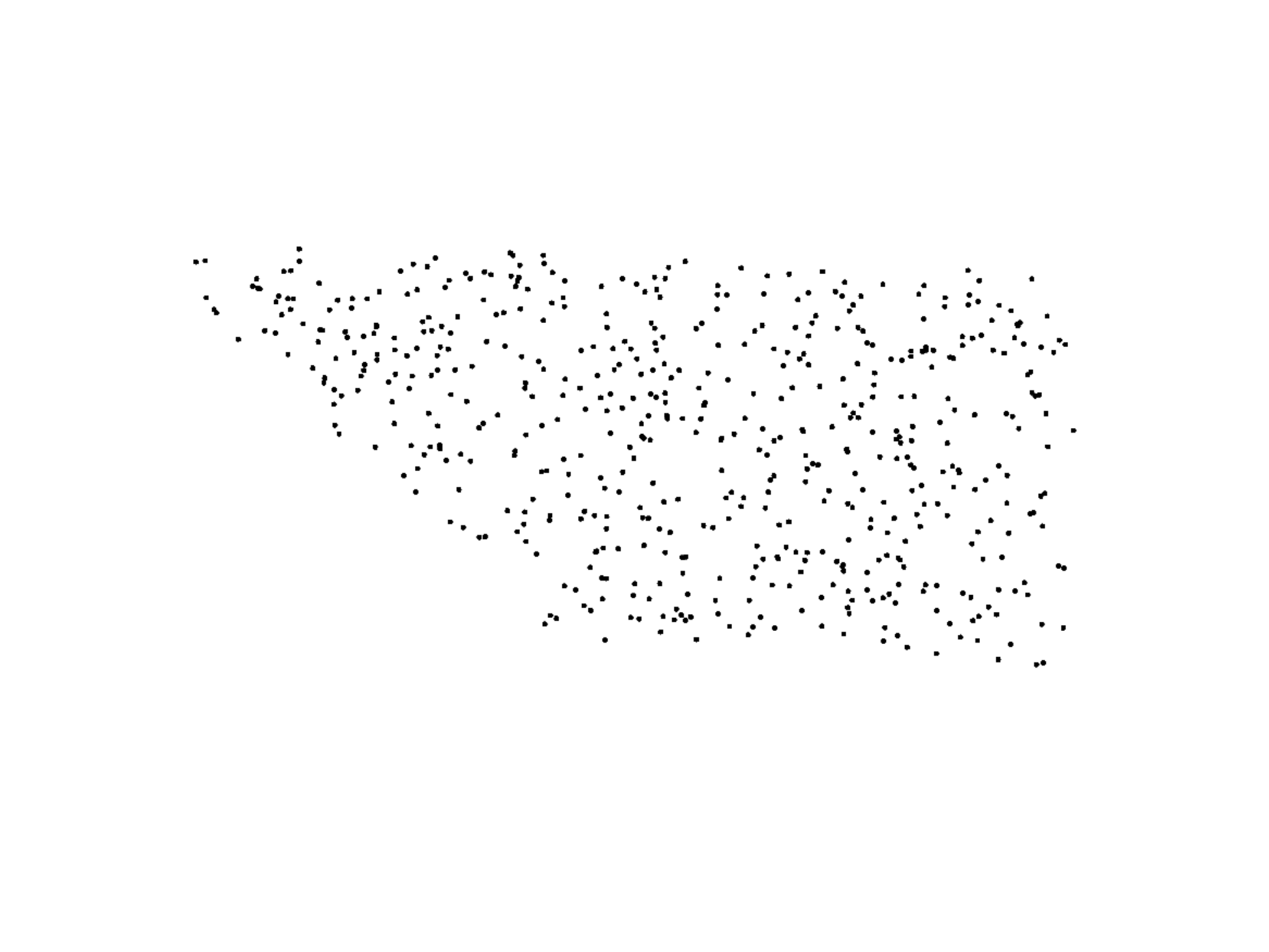}
&
\includegraphics[scale=0.2, trim={5cm 2.5cm 4cm 3cm}, clip]{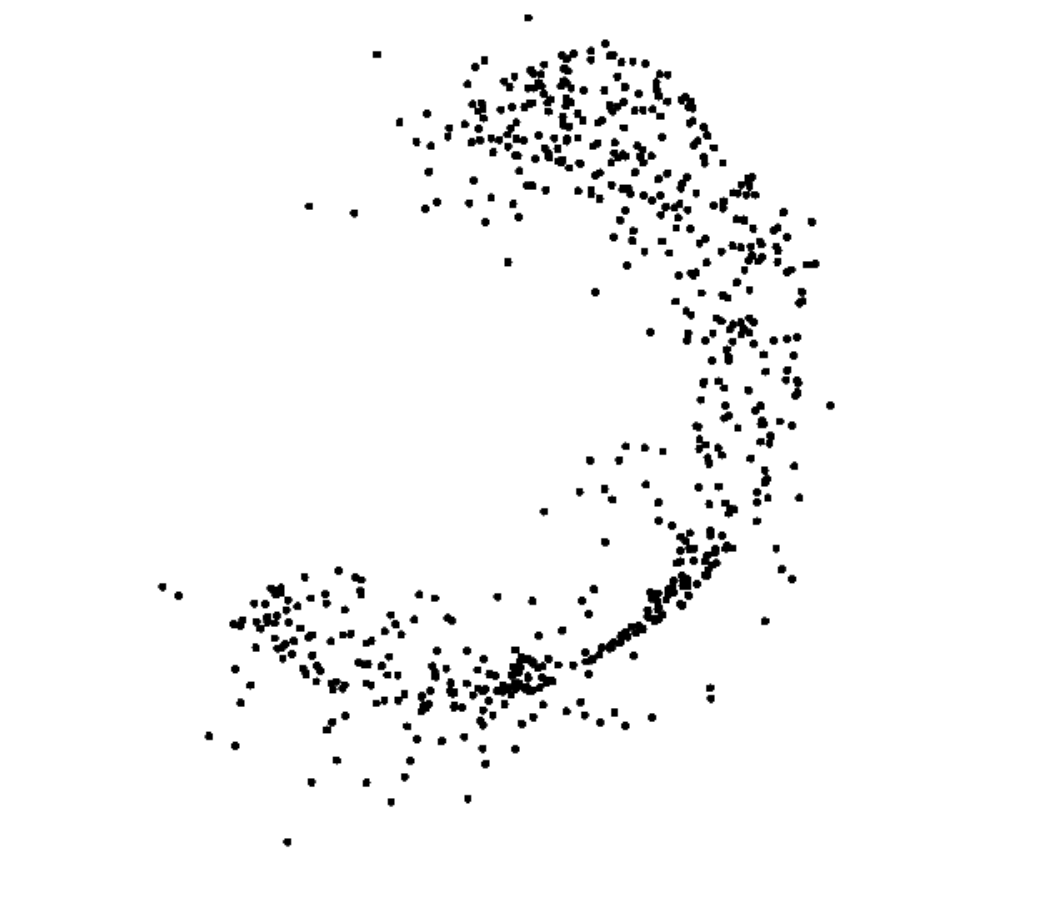}
&
\includegraphics[scale=0.175, trim={5cm 2.5cm 4cm 2cm}, clip]{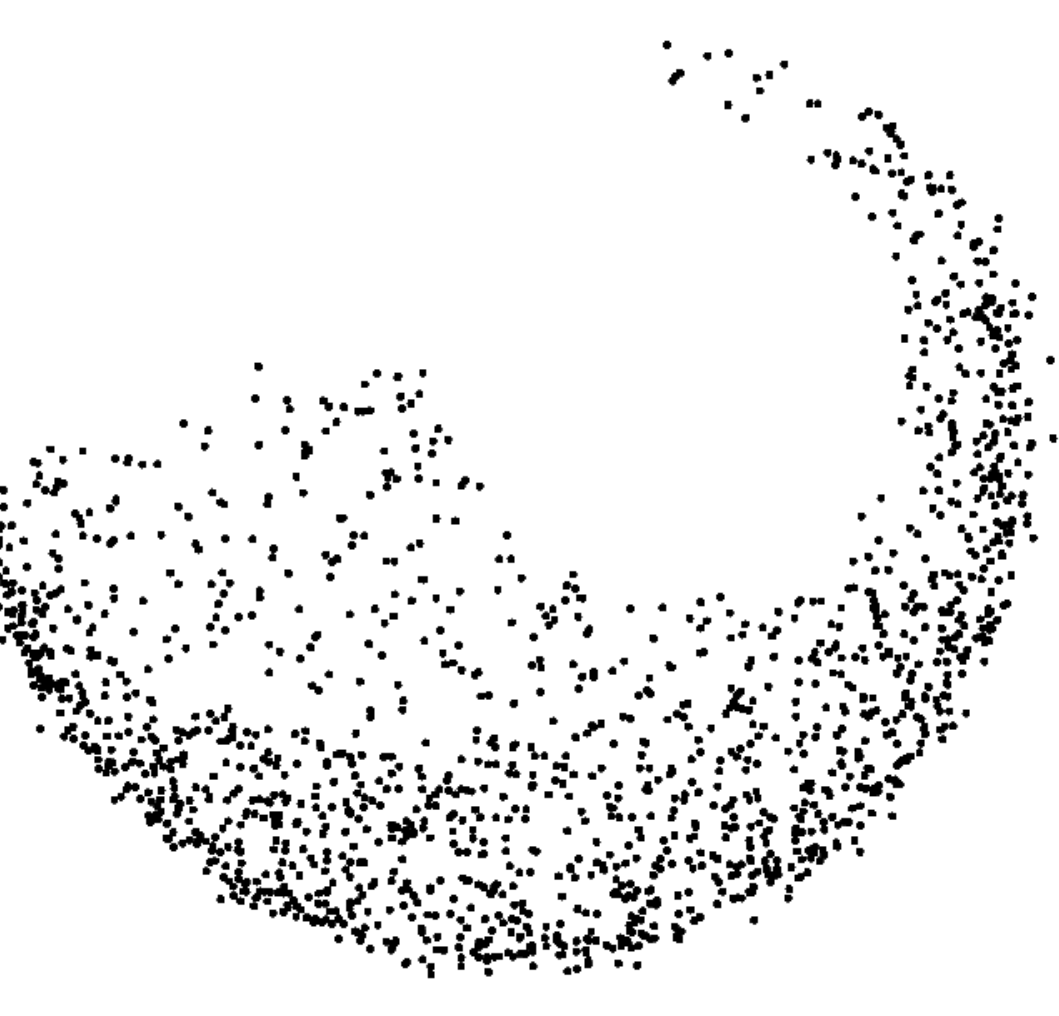}
&
\includegraphics[scale=0.175, trim={5cm 2.5cm 3cm 1.5cm}, clip]{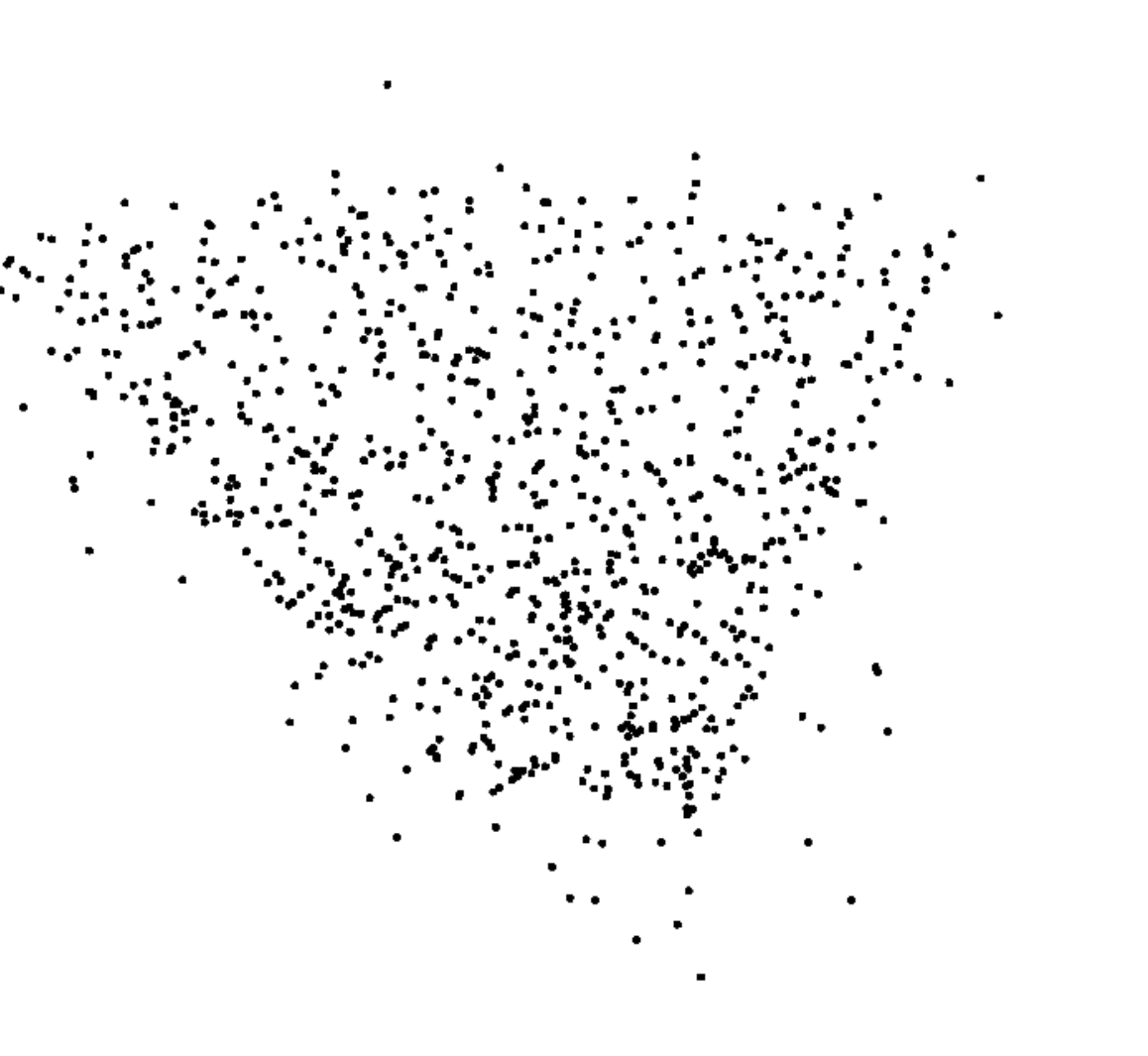}
&
\includegraphics[scale=0.175, trim={5cm 1cm 4cm 1cm}, clip]{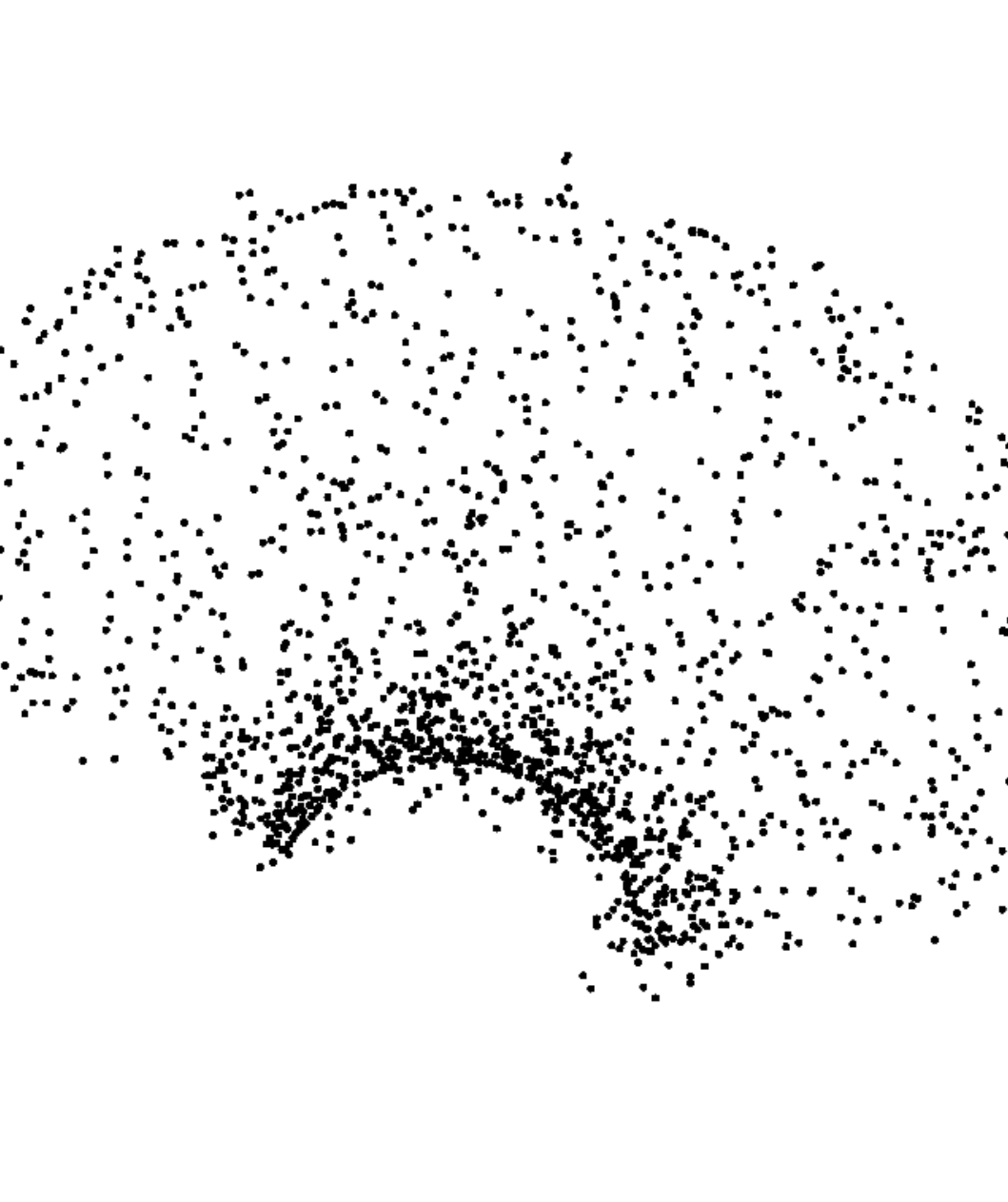}
&
\\
\hhline{------~} 

% UNIFORM NOISE & MISSING DATA
\multicolumn{1}{|c|}{\cellcolor{BlueViolet!20} \begin{turn}{90}$\text{A7}$\end{turn}} 
& 
\includegraphics[scale=0.25, trim={7cm 4cm 6cm 4cm}, clip]{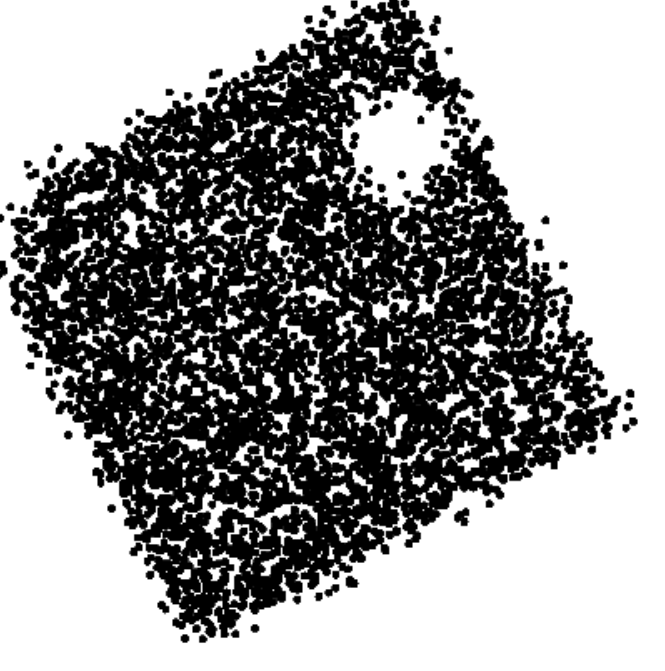}
&
\includegraphics[scale=0.22, trim={6cm 4cm 5cm 3cm}, clip]{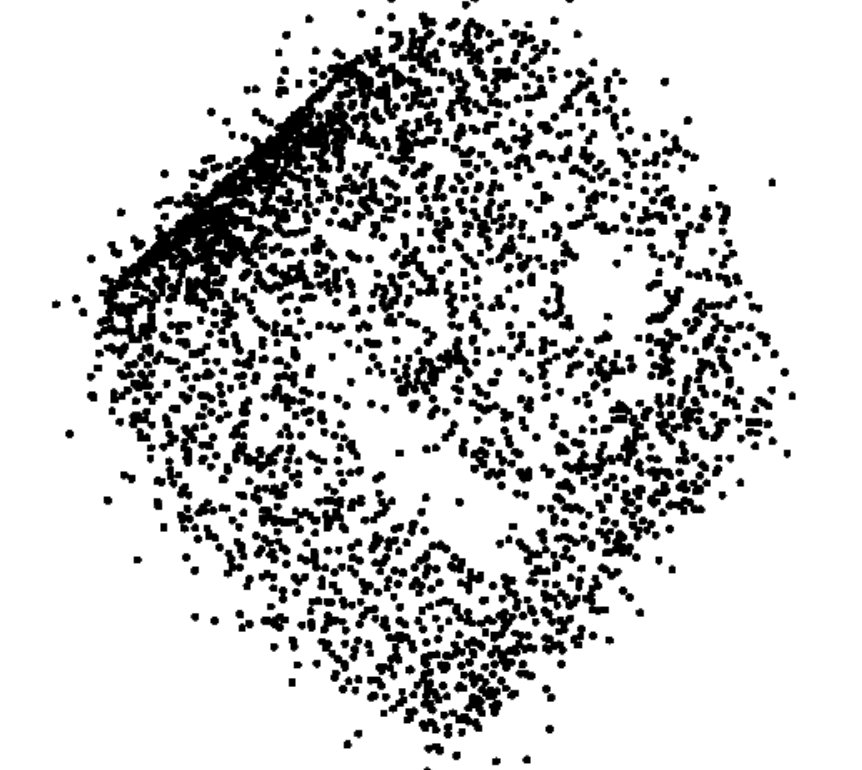}
&
\includegraphics[scale=0.22, trim={6cm 4cm 6cm 3cm}, clip]{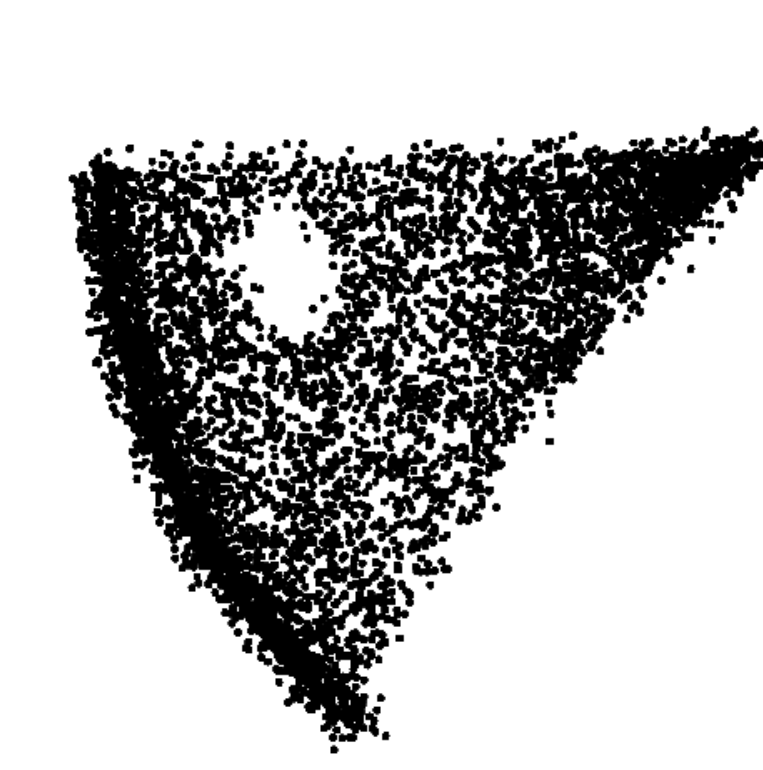}
&
\includegraphics[scale=0.175, trim={5cm 3cm 4cm 3cm}, clip]{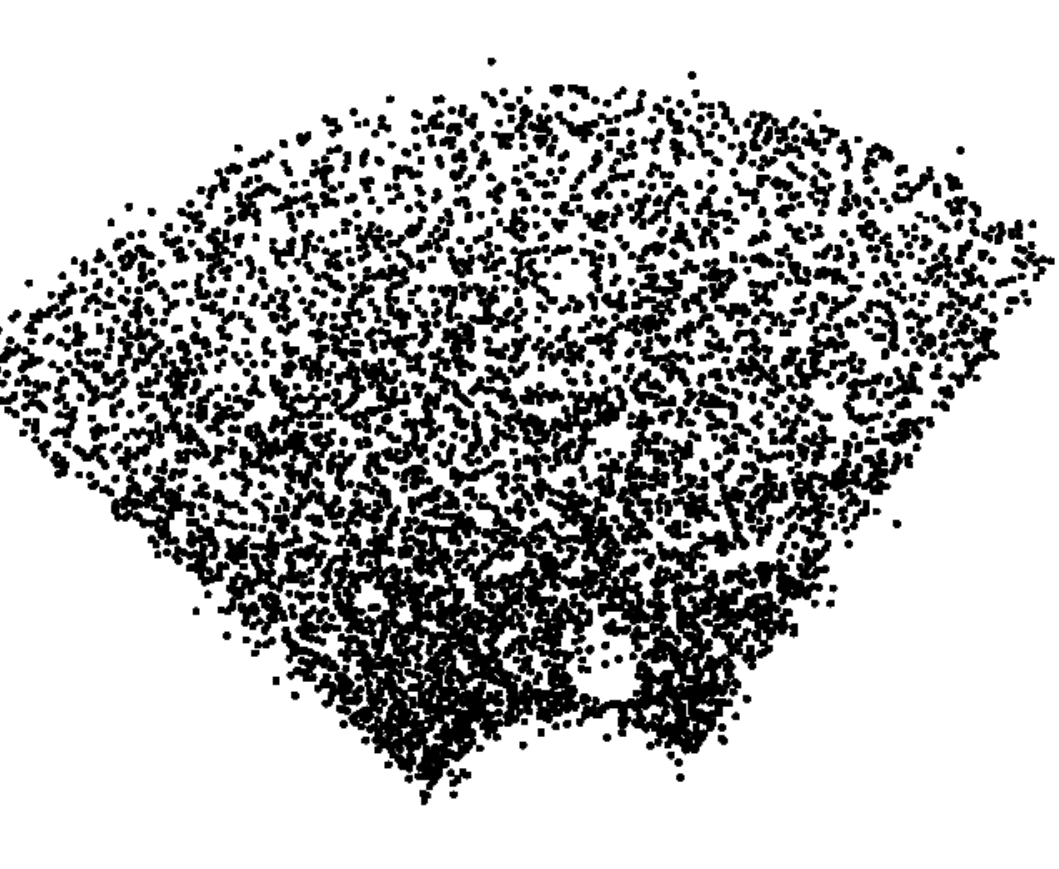}
&
\includegraphics[scale=0.20, trim={6cm 4cm 5.5cm 3cm}, clip]{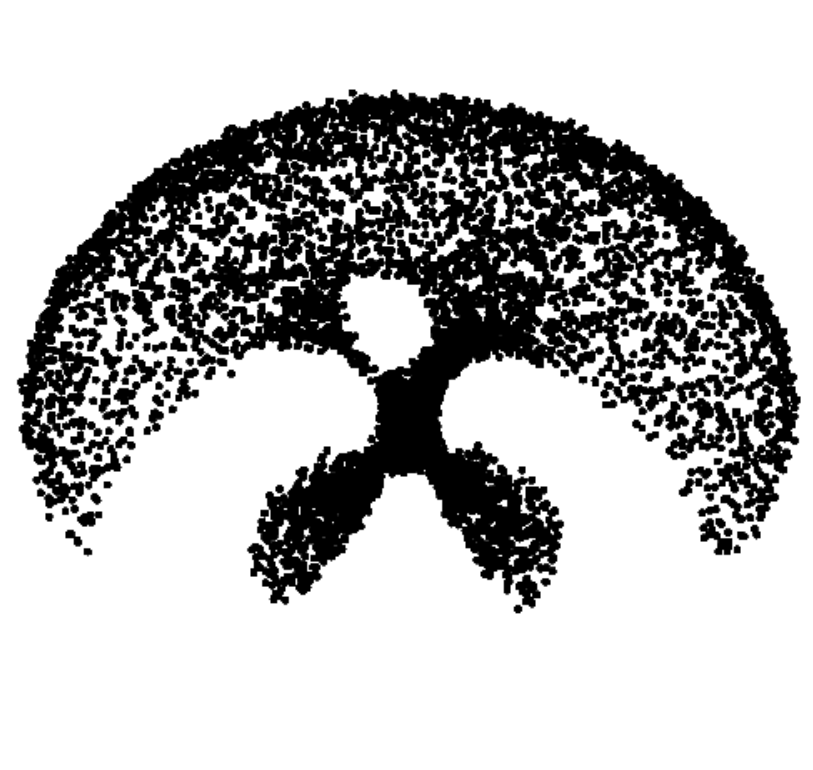}
&
\\
\hhline{------~} 

% GAUSSIAN NOISE & MISSING DATA
\multicolumn{1}{|c|}{\cellcolor{BlueViolet!20} \begin{turn}{90}$\text{A8}$\end{turn}} 
& 
\includegraphics[scale=0.175, trim={5cm 3.5cm 4.5cm 4cm}, clip]{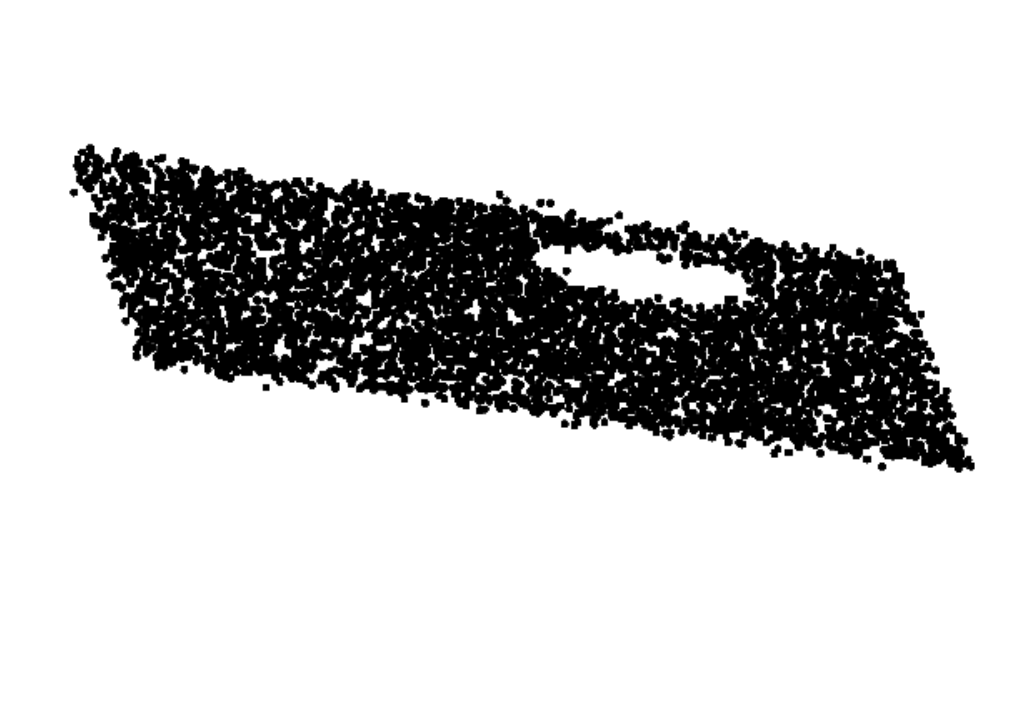}
&
\includegraphics[scale=0.175, trim={5cm 2.5cm 5cm 3cm}, clip]{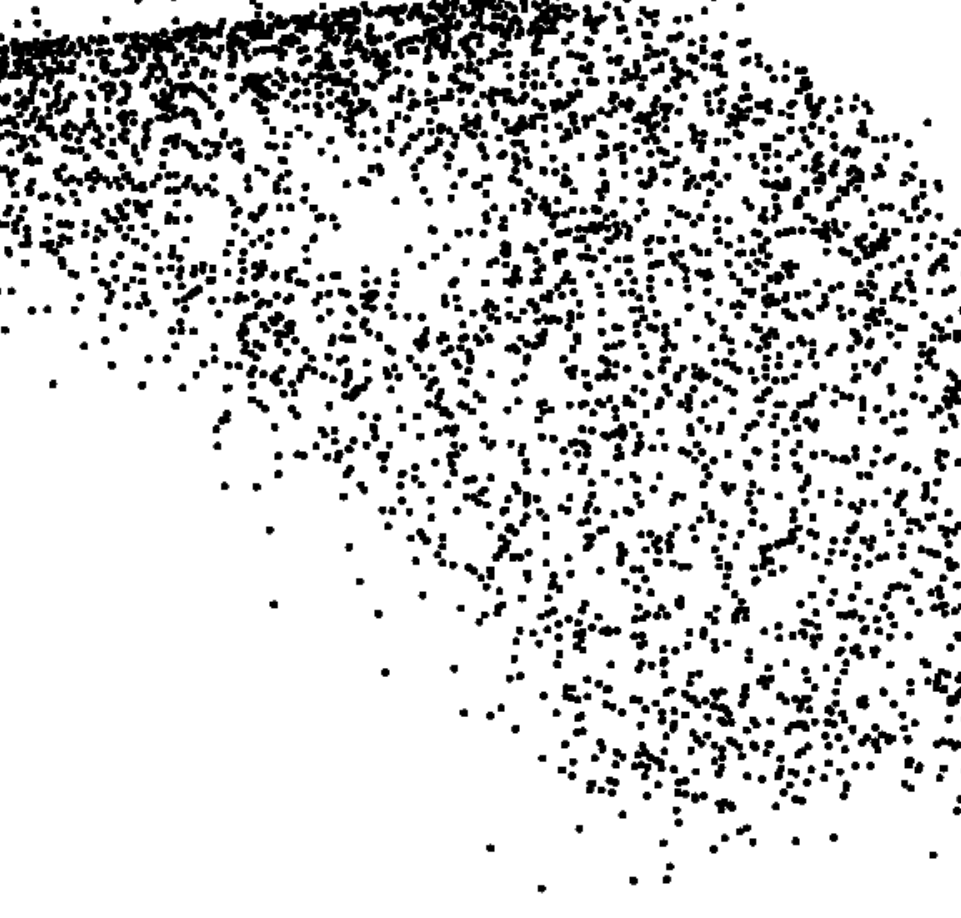}
&
\includegraphics[scale=0.175, trim={5cm 2.5cm 4cm 2cm}, clip]{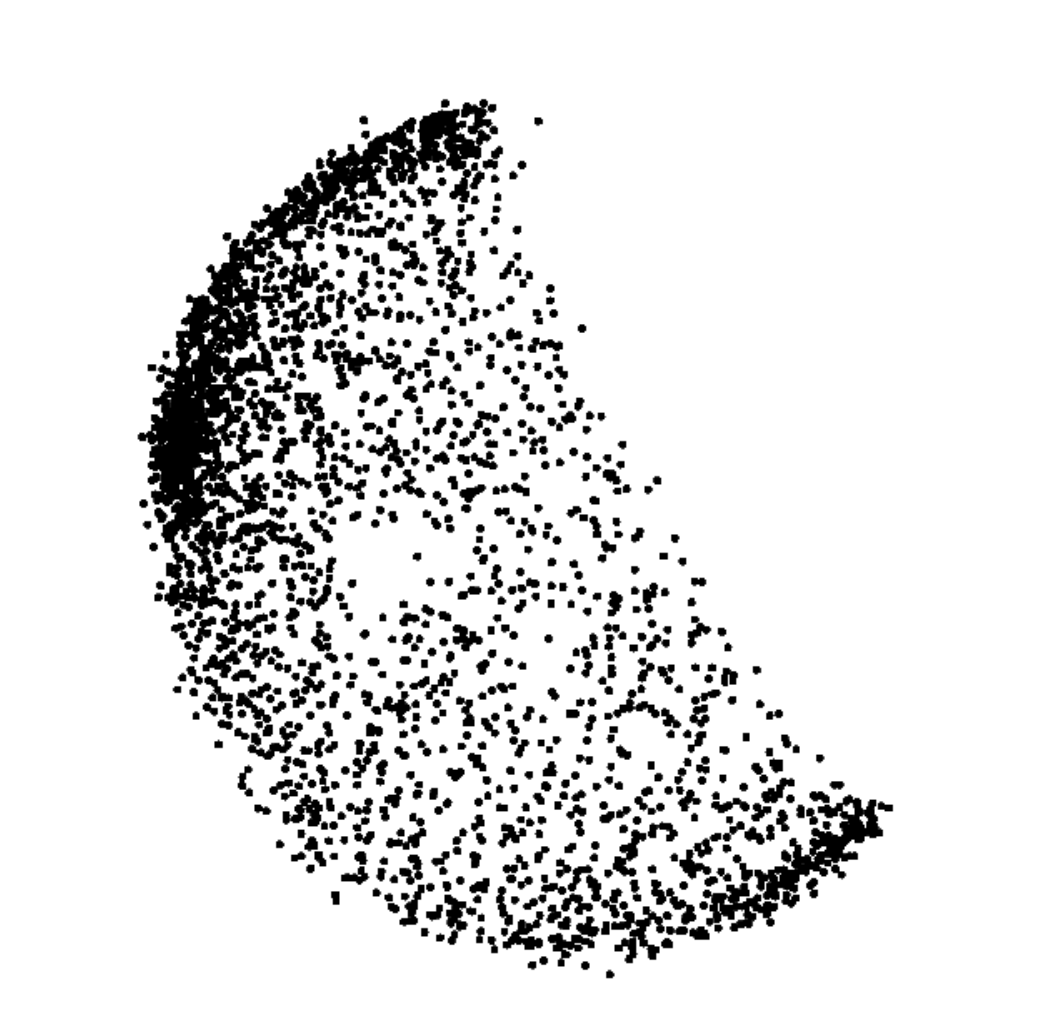}
&
\includegraphics[scale=0.175, trim={5cm 3cm 4cm 2cm}, clip]{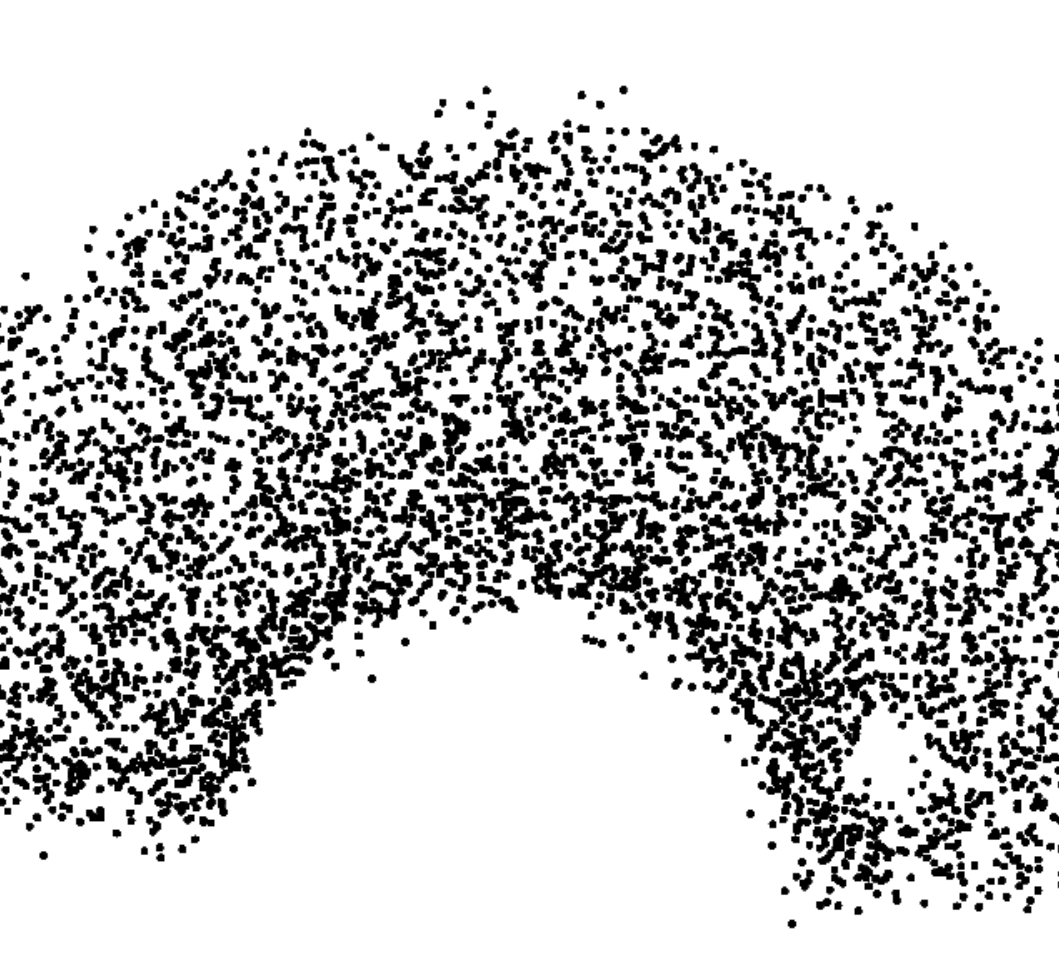}
&
\includegraphics[scale=0.175, trim={5cm 3cm 4cm 3cm}, clip]{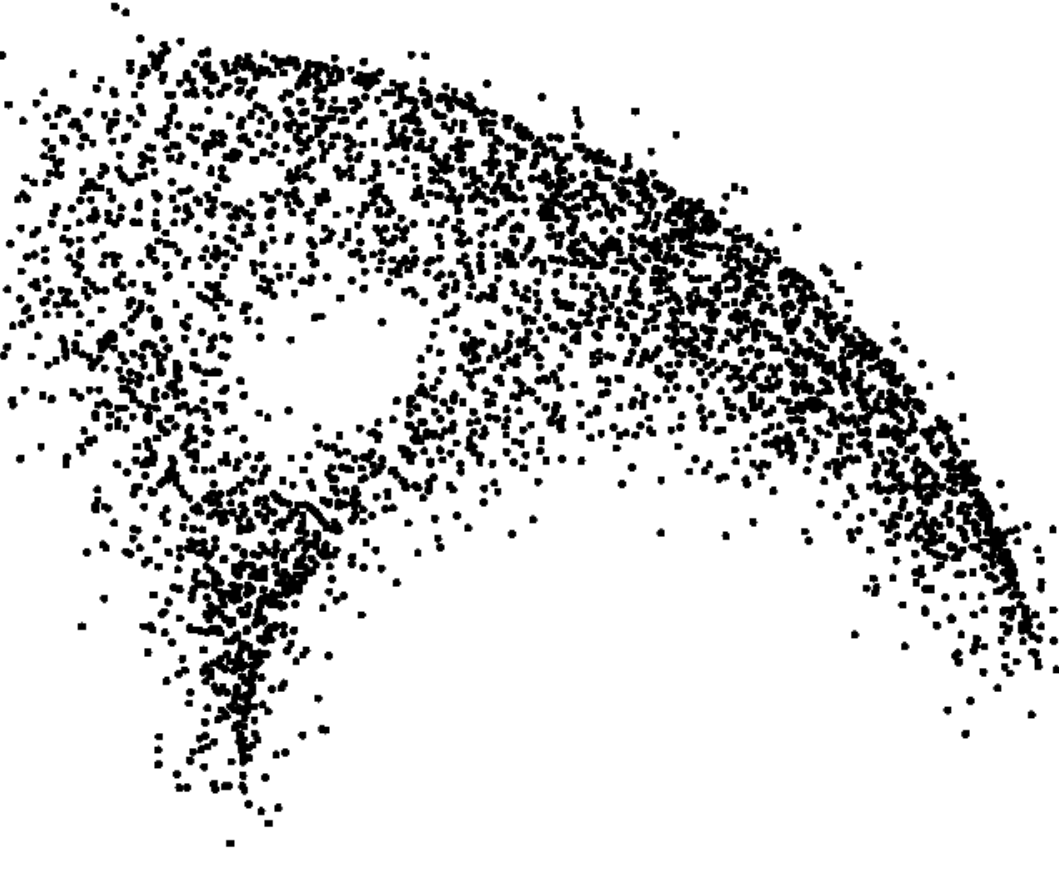}
&
\\
\hhline{------~} 

% SMALL DEFORMATIONS
\multicolumn{1}{|c|}{\cellcolor{BlueViolet!20} \begin{turn}{90}$\text{A9}$\end{turn}} 
& 
\includegraphics[scale=0.2, trim={5cm 2cm 4.5cm 2.5cm}, clip]{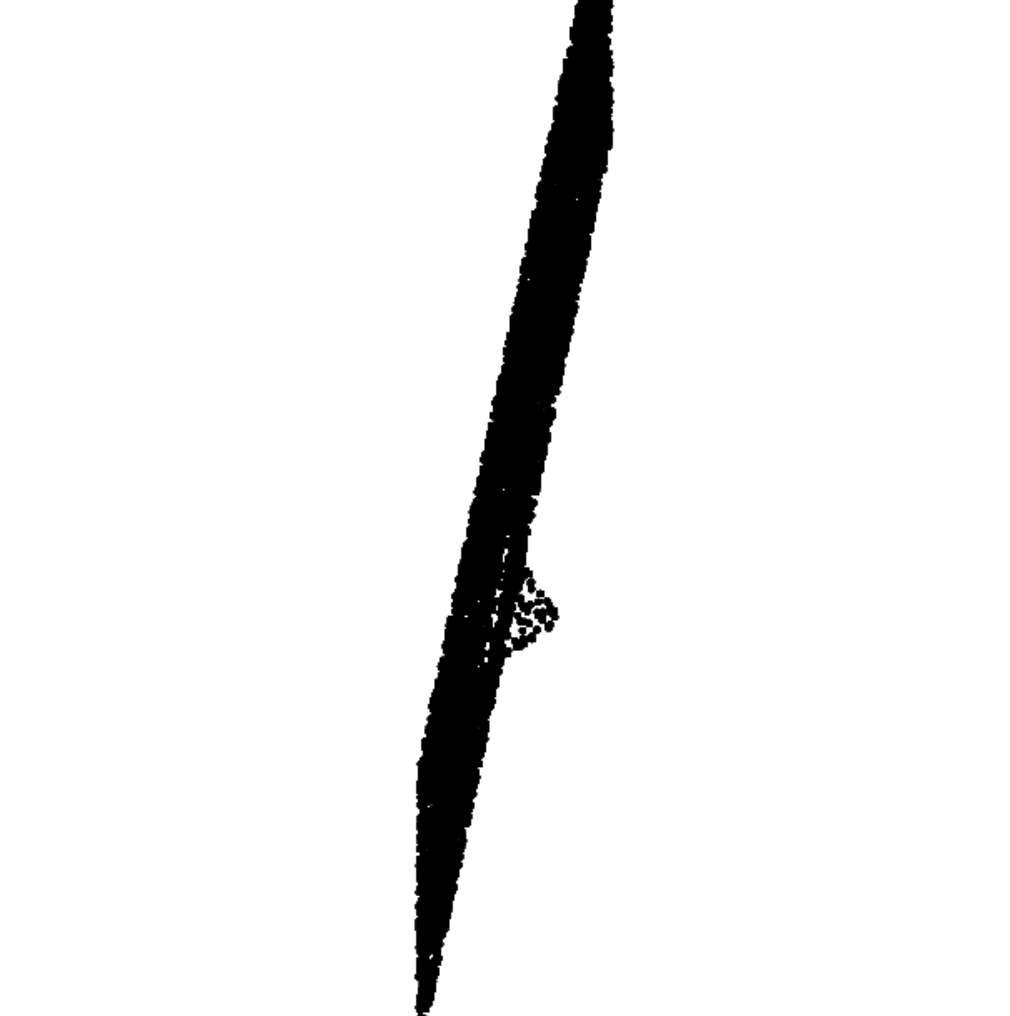}
&
\includegraphics[scale=0.2, trim={5cm 2.5cm 5cm 3cm}, clip]{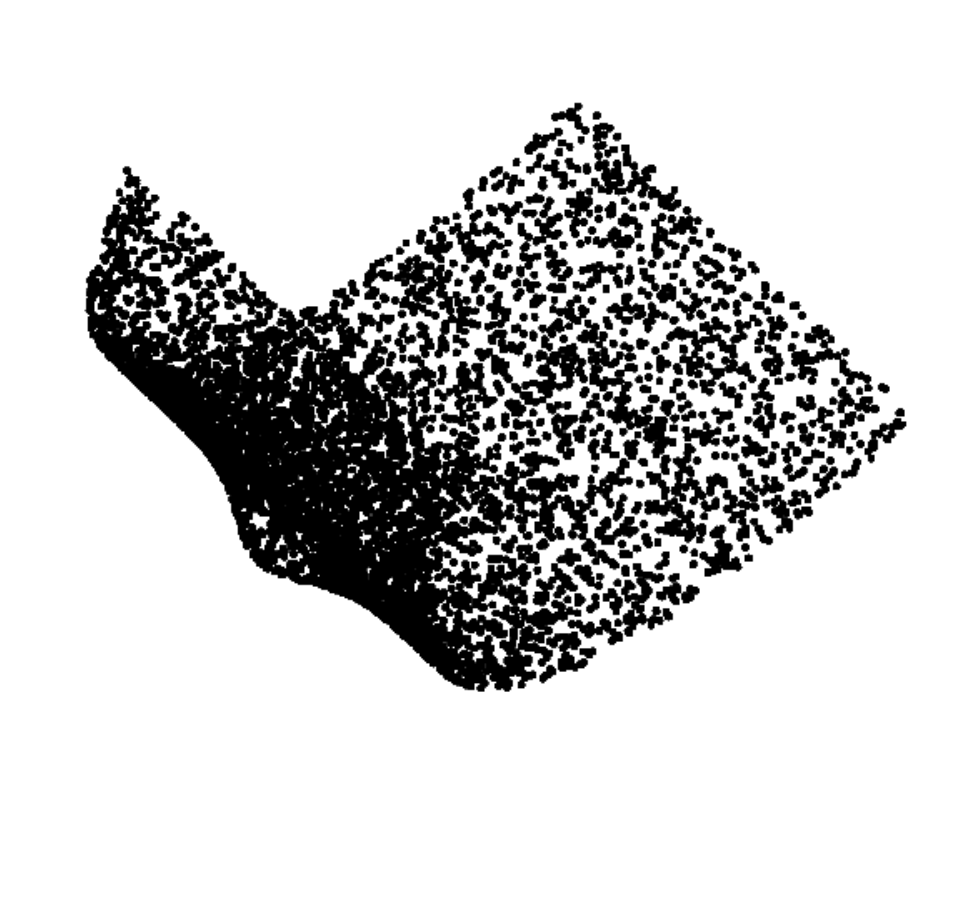}
&
\includegraphics[scale=0.175, trim={5cm 2.5cm 4cm 2cm}, clip]{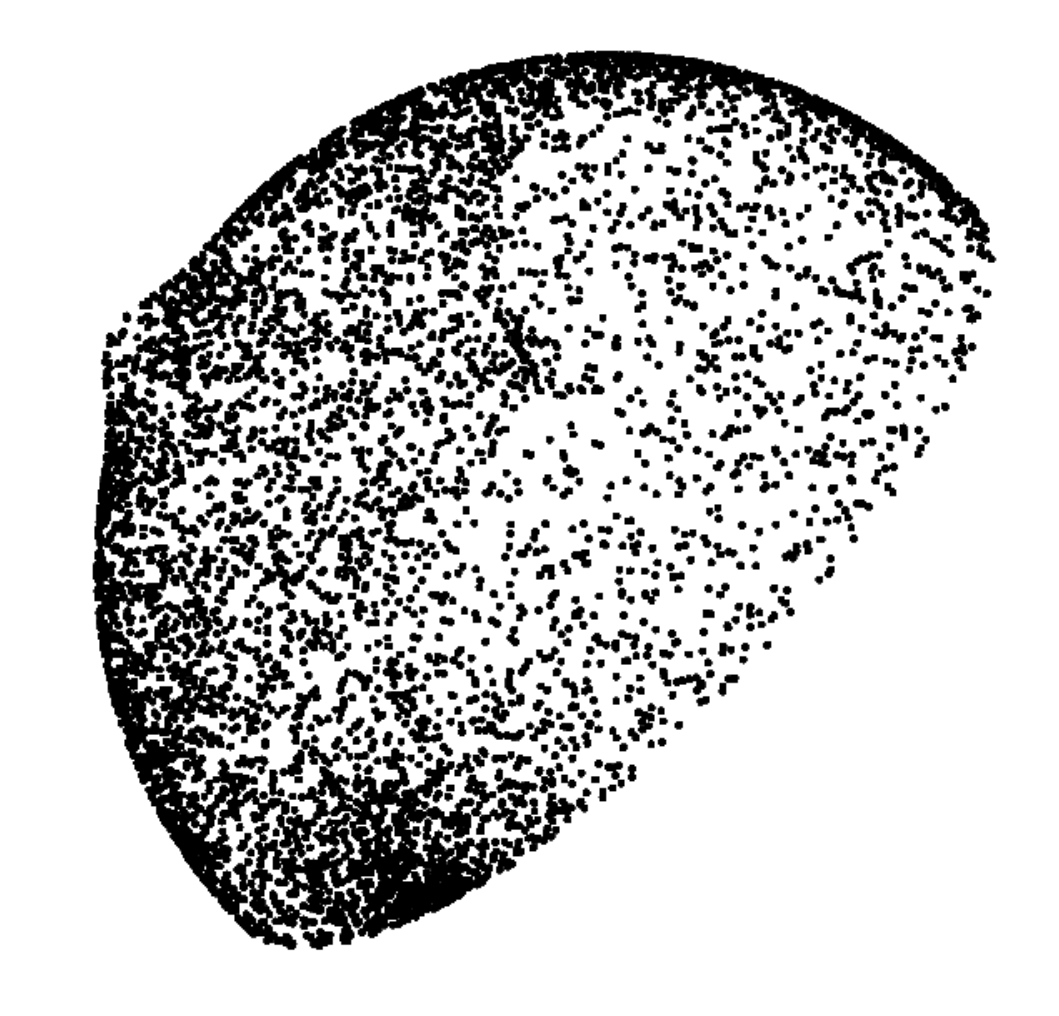}
&
\includegraphics[scale=0.175, trim={5cm 2.5cm 3cm 1.5cm}, clip]{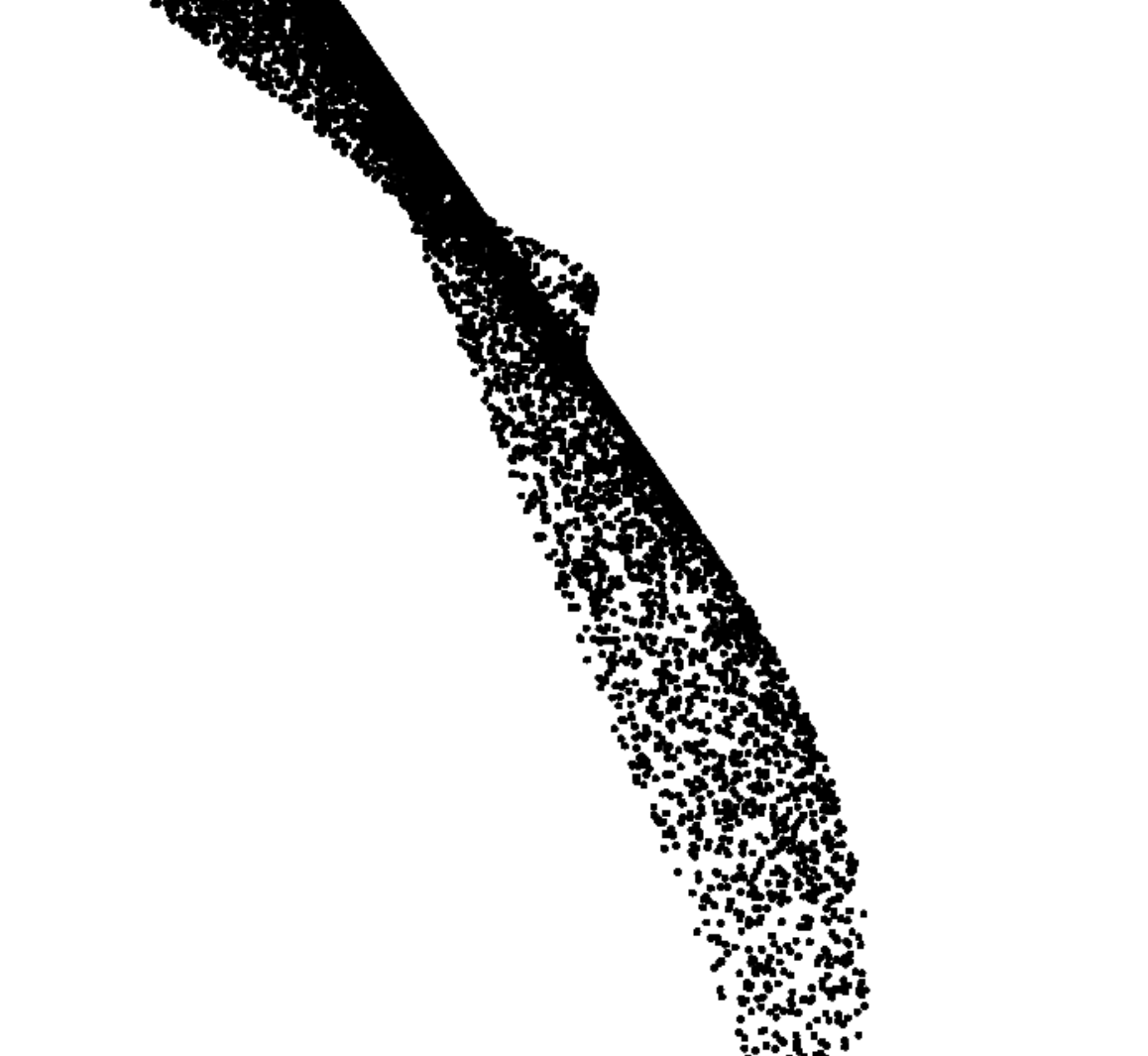}
&
\includegraphics[scale=0.175, trim={5cm 1cm 4cm 1cm}, clip]{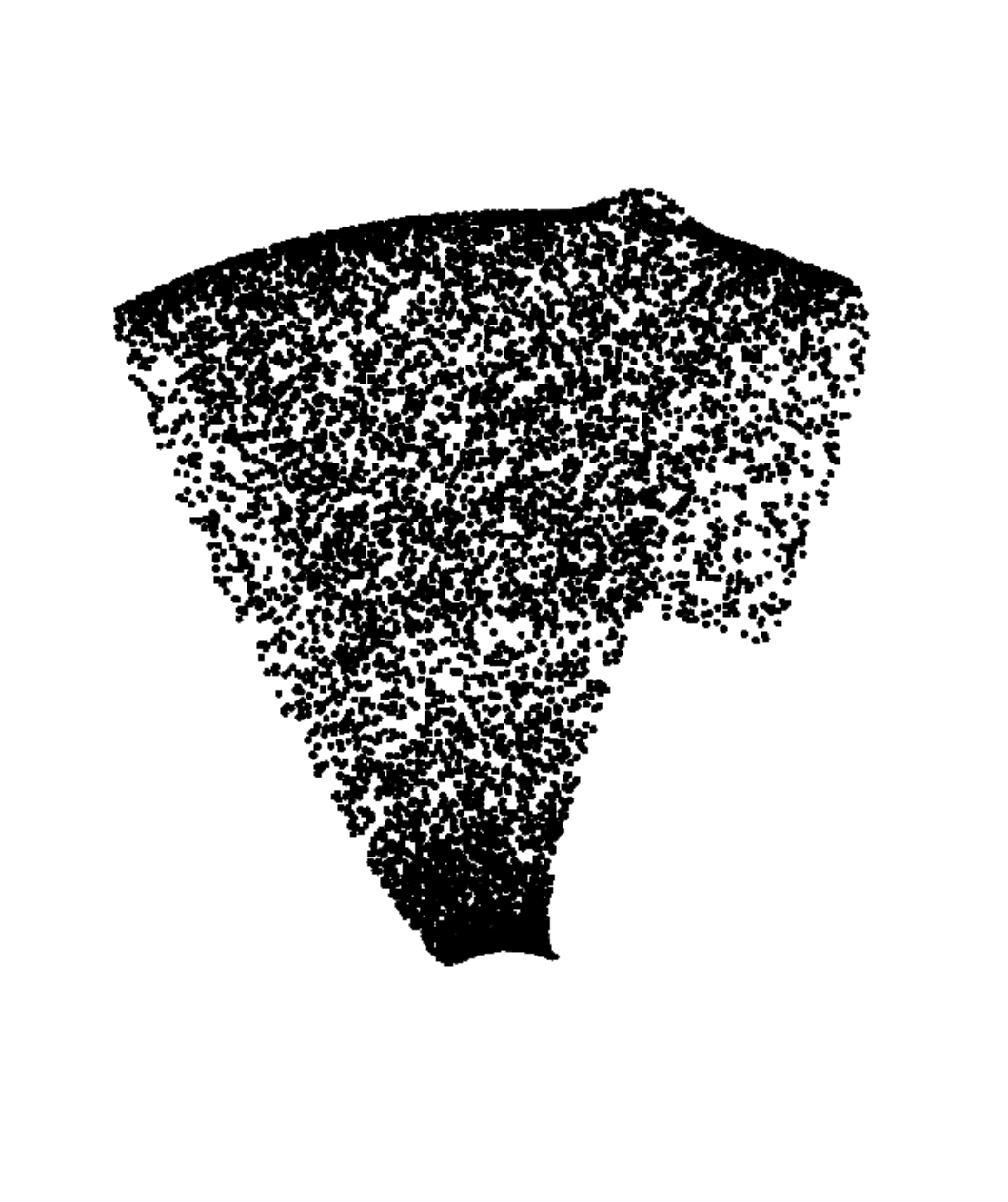}
&
\\
\hhline{------~} 
\end{tabular}
}
\caption{Examples of point clouds from the dataset. Columns identify primitive types: plane (T1), cylinder (T2), sphere (T3), cone (T4) and torus (T5). Rows correspond to point cloud artifacts: none (A0), uniform noise (A1), Gaussian noise (A2), undersampling (A3), missing data (A4),  uniform noise + undersampling (A5), Gaussian noise + undersampling (A6), uniform noise + missing data (A7), Gaussian noise + missing data (A8), small deformations (A9). \label{table:dataset_preview}}
\end{figure}

The point clouds (see Figure \ref{table:dataset_preview}) are generated by sampling classical surface primitives derived from constructive solid geometry (i.e., planes, cylinders, spheres, cones and tori), by means of the following procedure. First, point clouds are sampled by using parametric equations in their canonical form, i.e., centered at the origin of the coordinate axes and with rotational axis aligned with the $z-$axis; the geometric quantities that define each primitive (i.e., amplitudes and radii, if any) are assigned randomly. Second, to obtain segments of different shapes, several cuts are enforced by exploiting random planes. Third, we randomly apply translations and/or rotations to recover primitives in their general position. Lastly, each point cloud is (potentially) processed so that it can be:
\begin{itemize}
    \item \emph{A0 -- Clean}. No perturbation is applied.
    \item \emph{A1 -- Perturbed by uniform noise of different intensities}. The noise is obtained by sampling uniform distributions of the form $\mathcal{U}(-\frac{1}{n},\frac{1}{n})$, being $3\le{}n\le{}20$ random, and adding such perturbations to a random percentage of the points.
    \item \emph{A2 -- Perturbed by Gaussian noise of different intensities}. The noise is obtained by sampling normal distributions among $\mathcal{N}(-\frac{1}{n},\frac{4}{n^2})$, being $10\le{}n\le{}30$ random, and adding such perturbations to a random percentage of the points.
    \item \emph{A3 -- Clean but affected by undersampling}. We randomly select a percentage of points to be removed.
    \item \emph{A4 -- Clean but affected by missing parts}. We randomly choose a point of the point cloud; then, we remove all points contained inside the sphere having the selected point as center and radius assigned randomly.
    \item \emph{A5 -- Perturbed by uniform noise of different intensities and undersampled}.
    \item \emph{A6 -- Perturbed by Gaussian noise of different intensities and undersampled}.
    \item \emph{A7 -- Perturbed by uniform noise of different intensities and affected by missing parts}.
    \item \emph{A8 -- Perturbed by Gaussian noise of different intensities and affected by missing parts}.
    \item \emph{A9 -- Clean but with local deformations}. We randomly select a point of the point cloud and we apply a bivariate Gaussian centered at the selected point with a random covariance matrix.
\end{itemize}
For the sake of brevity, we will often refer to these point cloud artifacts as \emph{perturbation types}. 
For each perturbation type A0, \dots, A8, training set and test set contain, respectively, $5,000$ and $100$ point clouds; for perturbation type A9, training set and test set count, respectively, $1,000$ and $25$ point clouds.

For each point cloud of the training set, the track participants were provided with a ground truth 
TXT file that includes the primitive type and the geometric parameters required to identify each segment. More precisely, each file contains a column vector $\mathbf{v}$ defined as follows:
\begin{itemize}
    \item \emph{Plane}. We list the primitive type (for planes, codified as $1$), the unit normal vector\footnote{To be precise, the definite article ``the" is here used improperly. Indeed, there are two unique unit normal vectors that can be used to defined the same plane, or rotational axis. In our case, we recover uniqueness by enforcing the first nonzero component to be positive.} $\mathbf{a}$ and a point sampled on the plane $\mathbf{P}$. Then, the file provides the vector $\mathbf{v}=[1,\,\mathbf{a}^\mathsf{t},\, \mathbf{P}^\mathsf{t}]^\mathsf{t}$.

    \item \emph{Cylinder}. We list the primitive type (for cylinders, codified as $2$), the radius $r$, the unit vector determining the rotational axis $\mathbf{a}$, and a point $\mathbf{P}$ sampled on the axis. In this case, the file contains the vector $\mathbf{v}=[2,\,r,\,\mathbf{a}^\mathsf{t},\, \mathbf{P}^\mathsf{t}]^\mathsf{t}$.

    \item \emph{Sphere}. We list the primitive type (for spheres, codified as $3$), the radius $r$ and the center $\mathbf{C}$. The ground truth corresponds to the vector $\mathbf{v}=[3,\,r,\,\mathbf{C}^\mathsf{t}]^\mathsf{t}$.
 
    \item \emph{Cone}. We list the primitive type (for cones, codified as $4$), half the aperture\footnote{The aperture of a right circular cone, denoted by $2\alpha$, is the maximum angle between two generatrix lines; we here report half the aperture, i.e., $\alpha$.} $\alpha$, the unit vector determining the rotational axis $\mathbf{a}$ and the vertex $\mathbf{C}$. Then, the file provides the vector $\mathbf{v}=[4,\,\alpha,\,\mathbf{a}^\mathsf{t},\, \mathbf{C}^\mathsf{t}]^\mathsf{t}$. 

     \item \emph{Torus}. We list the primitive type (for tori, codified as $5$), the major and minor radii, respectively $r$ and $R$, the unit vector determining the rotational axis $\mathbf{a}$ and the center $\mathbf{C}$. In this case, the file contains the vector $\mathbf{v}=[5,\,r,\,R,\,\mathbf{a}^\mathsf{t},\, \mathbf{C}^\mathsf{t}]^\mathsf{t}$.

\end{itemize}

For each point cloud, the track participants were required to return a TXT file per point cloud (of the test set), with the same syntax employed for the ground truth files of the training set.

\subsection{Evaluation measures\label{sec:evaluation_measures}}
Three tasks are considered in the evaluation phase:
\begin{itemize}
    \item \emph{Classification task}. Methods are evaluated on the basis of their capability for identifying, for each point cloud of the test set, its primitive type.
    \item \emph{Recognition task}. Methods are studied in terms of their ability to recover the geometric descriptors (e.g, radii of a torus, vertex of a cone) behind each segment. 
     \item \emph{Fitting task}. Regardless of the primitive type, methods' performance is studied with respect to the approximation of each point cloud.
\end{itemize}

To quantify the performance of automatic algorithms in classifying, recognizing and fitting simple geometric primitives on point clouds under different conditions, we therefore consider a variety of classification and approximation measures. This analysis will be further specified according to the type of geometric primitive and  point cloud artifacts. 

\subsubsection{Classification measures}

 The \emph{confusion matrix}~\cite{Kuhn:2018} is a popular way to visualize the classification performance of a method. A confusion matrix $\text{CM}$ is a square matrix whose order equals the number of classes in the dataset under study. Diagonal elements count true positives, i.e., all those items which have been correctly labeled as members of their ground truth classes: precisely, $\text{CM}(i,i)$ is the number of items that have been correctly predicted as elements of class $i$. Off-diagonal elements give the numbers of items 
mislabeled by the classifier; precisely, $\text{CM}(i,j)$, with $j\neq i$, is the number of elements wrongly labeled as belonging to class $j$ rather than to class $i$. Ideals classifiers have diagonal classification matrices.

\paragraph{True Positive and Negative Rates} True Positive Rate (TPR) refers to the method's ability to correctly identify  positives (e.g., the percentage of cats correctly classified as cats). Likewise, True Negative Rate (TNR) measures 
the  classifier's ability to identify negatives (e.g., the percentage of non-cats correctly classified as non-cats). In statistics, TPR and TNR are also called \emph{sensitivity} and \emph{specificity}. A perfect classifier is $100\%$ sensitive and $100\%$ specific.

\paragraph{Positive and negative predicted values}

Besides TPR and TNR, to quantify the likelihood that a method returns a true positive (or true negative) rather than a false-positive (or a false-negative) we consider the Positive Predictive Value (PPV) and the Negative Predictive Value (NPV). PPV is the ratio of \emph{true positives} to the number of items labeled as positives by the classifier. Similarly, NPV is the ratio of \emph{true negatives} to the number of items classified as negatives.

\paragraph{Accuracy} Accuracy describes how often a classifier is correct in its predictions: precisely, it is defined as the proportion of predictions that the classifier got right.

\paragraph{Macro-average} Macro-average permits to evaluate the overall performance of the classifier treating all classes equally, it is defined as the arithmetic mean of one of the metrics mentioned above, computed independently for each class.

\subsubsection{Fitting and recognition measures}\label{sec:recognition_fitting_measures}
To measure the approximation accuracy of a specific primitive, we exploit the geometric descriptors predicted by each method and the corresponding parametric representation to sample densely the recognised surface primitive, say $\mathcal{S}_P$. Let us consider a point cloud $\mathcal{P}$ to be evaluated; we use the following two measures to evaluate the approximation accuracy:

\begin{itemize}
    \item \emph{Mean Fitting Error} (MFE):  \begin{equation}
\text{MFE}(\mathcal{P},\mathcal{S}_P):=\dfrac{1}{|\mathcal{P}|}\sum_{\mathbf{x}\in\mathcal{P}}d(\mathbf{x},\mathcal{S}_P)/l,
\label{eqn:MFE}  
\end{equation}

where $d$ is the minimum Euclidean distance between $\mathbf{x}$ and every point in $\mathcal{S}_P$, and $l$ is the diagonal of the minimum bounding box containing $\mathcal{P}$.

\item \emph{Directed Hausdorff distance}:
		$$d_{\text{dHaus}}(\mathcal{P},\mathcal{S}_P)=\max_{\mathbf{x}\in \mathcal{P}} \min_{\mathbf{y}\in \mathcal{S}_P}d(\mathbf{x},\mathbf{y}),$$ with $d$ the Euclidean distance. 
\end{itemize}

Finally, the recognition accuracy is evaluated in terms of $L^2$ norm between the column vectors of the ground truth and that obtained by each method. Specifically, following the notation introduced in Section \ref{sec:dataset_description}, we compare the geometric descriptors recognised with those of the ground truth considering the $L^2$ distance:
$$d(\mathbf{v}_G,\mathbf{v}_P)=\|\mathbf{v}_G-\mathbf{v}_P \|_2$$
where $\mathbf{v}_G$ and $\mathbf{v}_P$ here denote the ground truth and the prediction vectors, respectively, where we remove the first entry and, for planes and cylinders, the point (since they are not comparable).

\section{Description of the methods\label{sec:description_of_methods}}
Ten groups initially subscribed to this track, with a final number of six methods submitted. In the remainder of this section we describe the methods, here simply referred to as M1, M2, $\dots$, M6.

The six methods can be grouped as:
\begin{itemize}
    \item Fully direct methods: accelerated Hough Transform (M1), and Histograms of local features + parameter fitting (M2).
    \item Fully neural-based methods: PointNet (M3) and 3D ShapeNets (M4).
    \item Mix of neural and direct methods: PoinNet + parameter fitting (M5); AlexNet + parameter fitting (M6).
\end{itemize}

\subsection{M1: Accelerated Hough transform}
This method is proposed by Chiara Romanengo, Andrea Raffo, Silvia Biasotti and Bianca Falcidieno. The method implementation can be found at \url{https://github.com/chiararomanengo/fitting_geometric_primitives.git} and it is introduced in \cite{CAGD:2022}.
\subsubsection{Overall strategy}
The proposed approach is based on an accelerated version of the Hough transform, a well known technique in the pattern recognition community, which have been recently extended to algebraic objects, see \cite{beltrametti2012algebraic}. The general (algebraic) HT-framework deals with the problem of finding a (hyper)surface, within a family of (hyper)surfaces dependent on a set of unknown parameters, that best approximates an input shape. More precisely, given an input point cloud and a family of parametrized primitives with some unknown parameters, the core idea is to consider the parameter space corresponding to the selected family of primitives, to discretize it into cells, and to vote a cell every time the Hough transform of an input point intersects it; the most voted cell uniquely identifies the optimal solution of the recognition problem. 

Despite the robustness to various point cloud artifacts (e.g., noise, outliers, missing data and uneven sampling), a severe limitation of this paradigm is due to the curse of dimensionality; indeed, the dimension of the parameter space corresponds to the number of parameters that are meant to be estimated through the voting procedure. Only a few families of primitives can be considered in their general position, i.e., planes and spheres, but even for ellipsoids it is required that the cloud is centred in the origin of the Cartesian axes \cite{Beltrametti:2020} to make the problem computationally affordable.

\begin{figure}[t!]
    \centering
    \includegraphics[width=17.0cm]{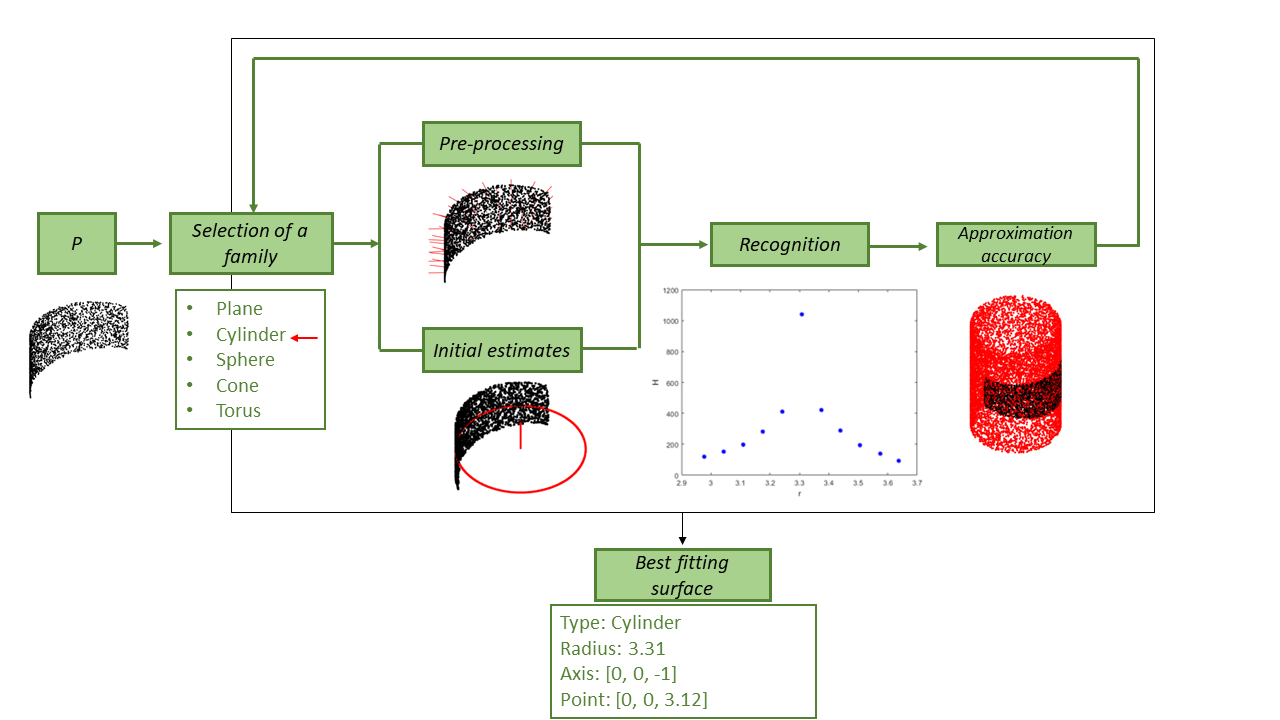}
    \caption{A graphical representation of the strategy adopted in method M1.}
    \label{fig:M1pipeline}
\end{figure}

The proposed method takes in input a point cloud $\mathcal{P}$ and, for each family of primitives $\mathcal{F}$ (among planes, cylinders, spheres, cones and tori), it works in three main steps.

\begin{itemize}
\item \emph{Pre-processing}. Independently from the primitive type, the normal vectors of the points are computed using the Hough transform technique and the family of planes in the Hesse normal form, as described in \cite{LIMBERGER20152043}. The adoption of a voting strategy makes this computation robust to most point cloud artifacts. Once the candidate tangent planes have been estimated, the process is specialized for each type of primitive. Specifically, the rotations and/or translations necessary to bring the point cloud in standard form are based on the geometric properties of the primitive type in question. The result of this process is a new point cloud $\mathcal{P}'$ and an initial estimates of the parameters of the primitive to be recognised. These parameters differ according to the type of family selected; for example in case of cylinders, tori or spheres, they correspond to the radii, while in case of cones they coincide with half the aperture.

\item \emph{HT-based surface recognition}. The new set of points $\mathcal{P}'$ is the input of the classical HT-based recognition algorithm. The estimates obtained in the pre-processing step allow to localize the search in a dimensionally-reduced parameter space. The method returns the parameters of the surface that best fits $\mathcal{P}'$ with respect to the given family of primitives. From these parameters, the geometric descriptors required by the track are obtained.

\item \emph{Evaluation of the approximation accuracy}. To evaluate the recognition accuracy of a specific primitive, the recognised surface is sampled from the parametric equation and the Mean Fitting Error as defined in Equation \ref{eqn:MFE} is calculated between $\mathcal{P}'$ and the sampled surface. Finally, the initial roto-translation is applied backwards in order to obtain the geometric descriptors for the point cloud  $\mathcal{P}$ in its original position.
\end{itemize}

The three steps here described are repeated for each family of surfaces (that is, planes, cylinders, spheres, cones and tori). Then, the primitive type and the geometric descriptors of the surface with the lowest MFE is returned by the method. 

Figure \ref{fig:M1pipeline} provides a graphical abstract of method M1.

\subsubsection{Computational aspects}
The tests were performed on a desktop equipped with an Intel Core i9 processor (at 3.6 GHz), in a Windows system 64 bits. The method was implemented in MATLAB. The execution time for each point cloud of the test set is, on average, $123$ seconds.

\subsection{M2: Histograms of local features and parameter fitting}
This method is proposed by Chi-Bien Chu, Khoi-Nguyen Nguyen-Ngoc, Quang-Thuc Nguyen and Minh-Triet Tran. The method implementation can be found at \url{https://github.com/ccbien/SHREC22-Track2.git}.
\subsubsection{Overall strategy}
The proposed method pre-processes the input point cloud normalizing it in a unit cube. Then, the task is addressed with two main modules: a point cloud classification with $k-$nearest neighbours ($k$-NN) using a histogram of local features (Module 1), divided into several steps, and a shape parameter fitting (Module 2).

In the first step of Module 1, all objects are classified into two types: planar and non-planar segments. The latter are then classified into subcategories following two strategies: one considers cone and cylinder as two categories (Strategy 1), while the other on considers them as one category (Strategy 2). Both of them consider sphere and torus as one category. Finally, the primitive types are further specialized (to distinguish: tori from spheres for Strategy 1; cones, cylinders, tori and spheres for Strategy 2).

\begin{figure}[t!]
    \centering
    \includegraphics[scale=0.45]{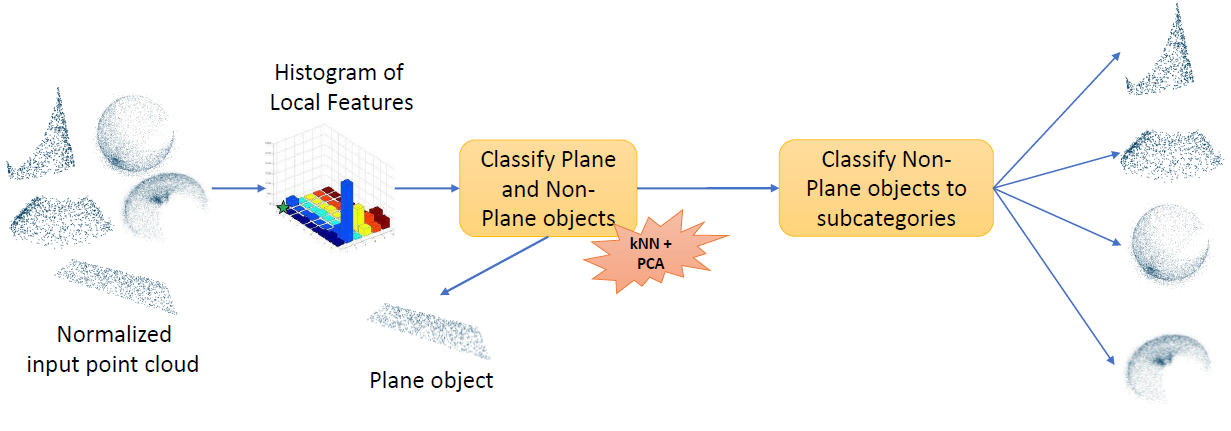}
    \caption{A graphical representation of the strategy adopted in method M2.}
    \label{figure:graph_abstr_M4}
\end{figure}

\begin{itemize}
    \item \emph{Module 1: object shape classification}. In the first step, both
strategies search for all planes. The Histogram of Oriented Normal Vectors (HONV, see \cite{histogram2013}) is used to represent a point cloud, specifically, a 2D histogram (25 x 25). Each cell with a value greater than or equal to $80\%$ of the maximum value in the histogram is shifted to the center of the 2D histogram. In this way, a multiple 2D histogram representation of a single object is created and the $k-$NN, with $k = 3$, is used to classify planar and non-planar objects. If a point cloud is classified as plane, the PCA is applied to further verify its planar property and to remove false-positive cases.

In the second step, the aim is to classify non-planar instances. Once the categories are defined accordingly to the strategy
mentioned before, the HONV feature is used to represent the point cloud and the $k-$NN method is applied with $k = 5$.

To classify sphere and torus, the Surflet-pair-relation histograms \cite{surflet} are used in place of HONVs: in this way, each object can be represented with only one histogram. Finally, the $k-$NN with $k = 15$ is used to distinguish spheres from tori.

In Strategy 2, an additional step is required to classify cylinder and cone. In particular, the HONV feature is used to represent the point cloud and the $k-$NN with $k = 5$ is employed to differentiate between cylinder and cone.

\item \emph{Module 2: finding shape parameters}. For each primitive type, different approaches are considered to estimate and fine tune the shape parameters.
\begin{itemize}
  \item \emph{Plane}. The average value of all the points considered to be on the surface is calculated. Then, PCA is used to find the eigenvector corresponding to the smallest eigenvalue, which is taken as approximation of the normal vector.     
  \item \emph{Sphere}. A least squares sphere fit algorithm\footnote{jekel.me/2015/Least-Squares-Sphere-Fit} is employed to find the center and radius of a sphere. In this case, only $30\%$ of points are used and the process loops 100 times to find the parameters. Using the estimated parameter values, the number of points belonging to the surface of a sphere is computed, through the difference between the distance from the point to the center and the radius. If the difference is larger than $5\%$ of the radius, that point is considered as an outlier. Finally, the set of parameter values with the most number of inliers is chosen.
  \item \emph{Cylinder}. A cylinder fitting approach\footnote{https://github.com/xingjiepan/cylinder fitting} is used to find the rotational axis and the radius by minimizing the common least squares loss.
 \item \emph{Cone}. The RANSAC method proposed in \cite{ransac} is followed to estimate the parameters for conic segments.
 \item \emph{Torus}. The parameter values are initialised: the center in the origin, the rotation axis equivalent to the $z-$axis, the main radius $1$ and the minor radius $0.1$. First, the points are fitted like a 3D circle (find center, radius, axis of rotation) by optimizing the $L^2$ loss function. Then, the minor radius is calculated by averaging the distance of that circle to the set of points.
  \end{itemize}
\end{itemize}
Figure \ref{figure:graph_abstr_M4} provides a graphical abstract of method M2.

\subsubsection{Computational aspects}
The classification method needs at most $20$ seconds on AMD Ryzen 5 3550H, when applying $k-$NN approach to classify all objects in each step of Strategy 1 and Strategy 2; this corresponds to an average of 0.02 seconds per model.  Each point cloud might need multiple steps, and it takes at most 1 second to get the final primitive type prediction for an object. The shape parameter fitting methods take approximately 60 minutes on an Intel i5 10510U for all the 925 objects, which means about 3.89 seconds per model, on average.

\subsection{M3: PointNet}
This method is proposed by Ivan Sipiran. The method implementation can be found at \url{https://github.com/ivansipiran/shrec2022-geometric-primitives.git}.

\subsubsection{Overall strategy}
As for the previous method, M3 is based on a point cloud pre-processing, consisting of a centering and a scaling, and two consecutive steps: shape classification and parameter regression. 

This method considers the PointNet~\cite{Pointnet} architecture as backbone for all networks.
A first neural network is used to return the class probabilities for the five classes: plane, cylinder, sphere, cone, and torus. 

Depending on the predicted class, each point cloud is fed into one of five specific networks to obtain the primitive parameters. Networks differ in the final layer, whose outputs must depend on the primitive type; the last layer of the backend consists of 256 linear neurons, which means that the input for the last layer of the regression is a 256-dimensional vector. Each primitive parameter is thus generated by a different MLP head which consists of a linear layer. For example, the plane network has two heads in the final layer:  a $256\times 3$ linear layer for the normal and a $256\times 3$ linear layer for the point in plane parameter.

When it comes to the classification task, the loss function used to train the network is the cross-entropy loss between the one-hot activation functions from ground-truth and the network output. For the regression step, each primitive type has its own loss, defined as the sum of simpler functions: 
\begin{itemize}
    \item The \textit{scalar loss} $\mathcal{L}_{x, \hat{x}}$ compares a ground truth scalar parameters $x$ with its prediction $\hat{x}$, and is defined as the mean squared error. It can be used for radii and angles.
    \item The \textit{point loss} $\mathcal{L}_{\mathbf{p}, \hat{\mathbf{p}}}$ is defined as the Euclidean distance between ground-truth point $\mathbf{p}$ and output point $\hat{\mathbf{p}}$. It is used for points, e.g., for the sphere centers and the cone vertices.
    \item The \textit{vector loss} controls the direction between a ground-truth vector $\mathbf{v}_G$ and the output vector $\mathbf{v}_O$, and  is defined as follow:
    \begin{equation*}
        \mathcal{L}_{\mathbf{v}_G, \mathbf{v}_O}=1-\cos^8(\mathbf{v}_G, \mathbf{v}_O)  + \|\mathbf{v}_G - \mathbf{v}_O\|_2^2.
    \end{equation*}
    It is used for normal vectors and rotational axes. The $L^2$ regularization between $\mathbf{v}_G$ and $\mathbf{v}_O$ enforces the componentwise similarity between vectors. The exponent of the cosine function is meant to improve the learning of normal directions.
\end{itemize}
The total loss of each primitive type is defined by summing the above-defined losses:
\begin{align*}
\quad
    &\mathcal{L}_{\text{plane}}  &= \quad 
    &\mathcal{L}_{\mathbf{p}, \hat{\mathbf{p}}} + \mathcal{L}_{\mathbf{n}, \hat{\mathbf{n}}}\\
    &\mathcal{L}_{\text{cylinder}} &= \quad &\mathcal{L}_{\mathbf{p}, \hat{\mathbf{p}}} + \mathcal{L}_{\mathbf{a}, \hat{\mathbf{a}}} + \mathcal{L}_{r, \hat{r}}\\
    &\mathcal{L}_{\text{sphere}} &= \quad 
    &\mathcal{L}_{\mathbf{c}, \hat{\mathbf{c}}} + \mathcal{L}_{r, \hat{r}}\\
    &\mathcal{L}_{\text{cone}} &= \quad 
    &\mathcal{L}_{\mathbf{v}, \hat{\mathbf{v}}} + \mathcal{L}_{r, \hat{r}}+ \mathcal{L}_{\alpha, \hat{\alpha}}\\
    &\mathcal{L}_{\text{torus}} &= \quad
    &\mathcal{L}_{\mathbf{c}, \hat{\mathbf{c}}} + \mathcal{L}_{\mathbf{a}, \hat{\mathbf{a}}} + \mathcal{L}_{r_{\min}, \hat{r}_{\min}} + \mathcal{L}_{r_{\max}, \hat{r}_{\max}}\\
\end{align*}
Finally, the parameters are transformed with respect to the inverse translations and scaling applied in the preprocessing. 

The training is performed after splitting the original training set is split into two parts: a (sub-)training set (80\%) and a validation set (20\%).

\begin{figure*}[t!]
    \centering
    \includegraphics[scale=0.75]{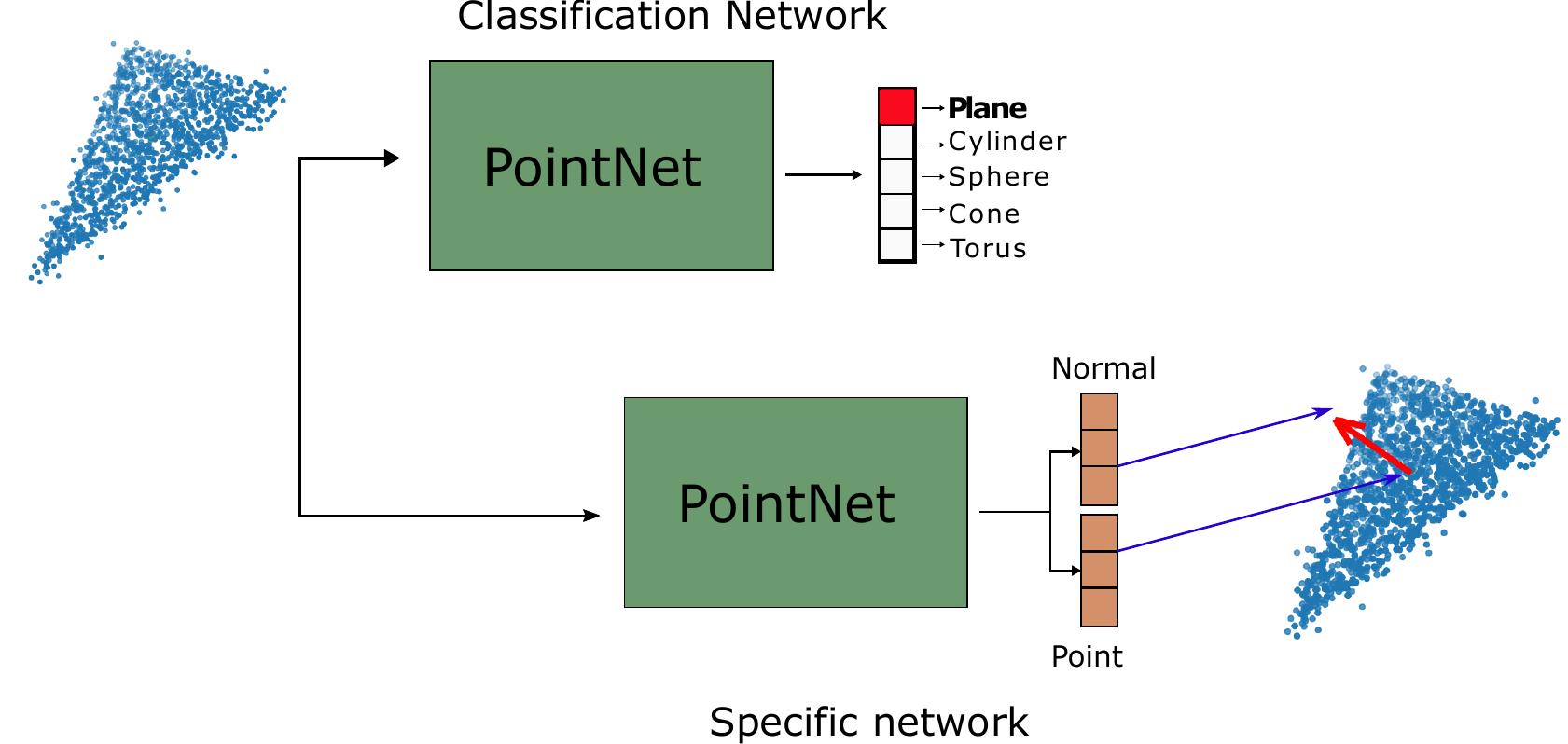}
    \caption{A graphical representation of the strategy adopted in method M3.}
    \label{figure:graph_abstr_M3}
\end{figure*}

Figure \ref{figure:graph_abstr_M3} provides a graphical abstract of method M3.

\subsubsection{Computational aspects}
This method was tested by using a 64-bit Linux OS, with 32GB RAM and a GPU NVIDIA RTX 2080 of 8GB. The method was tested with Python 3.6, by implementing NN architectures via Pytorch. All the training tasks were performed considering 200 epochs. The training time was about 4 hours for the classification step and approximately 6 hours for the regression part, which means an average of 72 and 108 seconds per epoch. The average prediction time, per point cloud, is 280 ms.

\subsection{M4: 3D ShapeNets}
This method is proposed by Vlassis Fotis, Ioannis Romanelis, Eleftheria Psatha and Konstantinos Moustakas. The method implementation can be found at \url{https://github.com/JohnRomanelis/SHREC2022_PrimitiveRecognition.git}.

\subsubsection{Overall strategy}
An initial preprocessing is applied to the input point clouds. First, they are translated to the origin, by subtracting the respective centroids. Following that, a normalization is applied, so that all points lie within the unit sphere. Random rotations about the $x$-, $y$- and $z$-axes are applied, for data augmentation purposes. The point clouds are finally converted into volumetric grids.\\

The backbone neural network for all models was originally proposed in the Minkowski Engine \cite{Choy:2019} as a candidate for the ModelNet classification challenge \cite{Wu:2015}. It is based on sparse voxel convolution, which minimizes the memory requirements associated with voxel based methods.\\

The network comprises of a voxel-wise input MLP and an input convolution. This is followed by three strided convolutions that progressively reduce the resolution of the voxel grid. After that, the output of the convolutional layers are projected on the initial grid, concatenated and finally passed through three strided convolutional layers. To extract the feature vector, Max and Mean pooling is applied to the resulting grid and the results are concatenated. All layers are followed by Batch Normalization and LeakyReLU activation function. For this track a 3-layer MLP head was used, with the number of output parameters changing according to the task. 
\begin{figure}[t!]
    \centering
    \includegraphics[scale=0.40, trim={1cm 0 0 0}, clip]{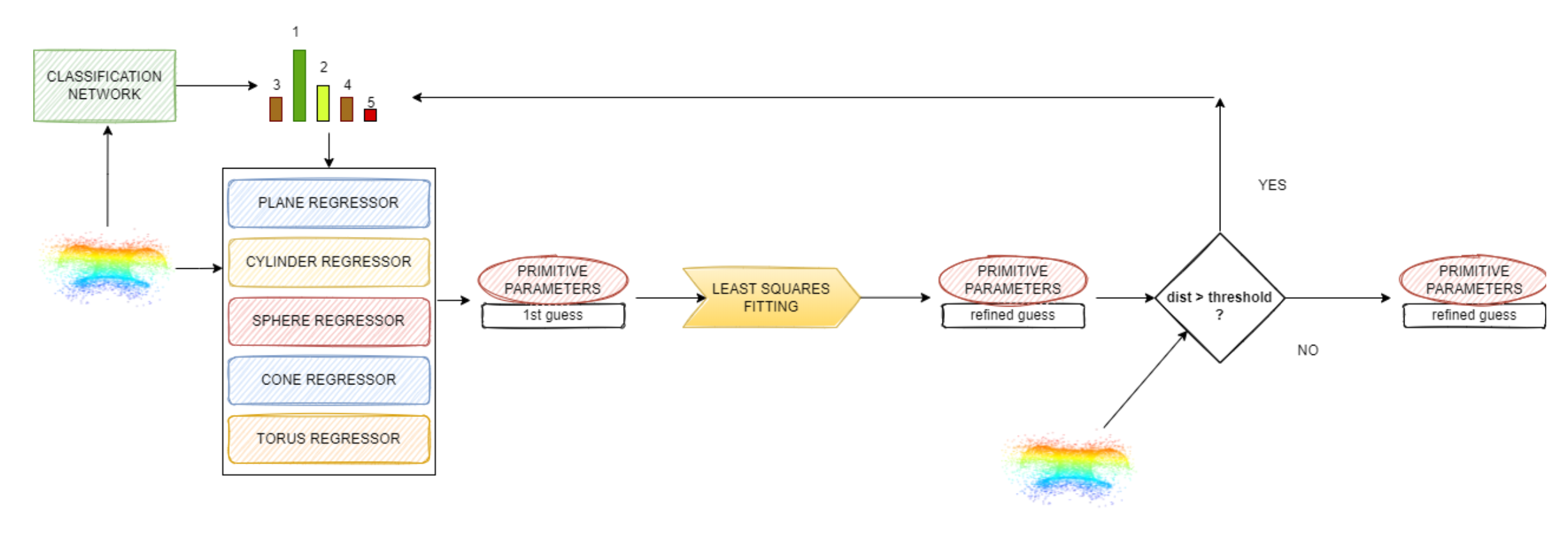}
    \caption{A graphical representation of the strategy adopted in method M4.}
    \label{figure:graph_abstr_M2}
\end{figure}

M2's strategy involves two consequent steps:
\begin{itemize}
    \item \emph{Classification step}. An instance of the model described above, trained for classification, is tasked to find which primitive the input points where sampled from. 
    \item \emph{Fitting step}. Five  more copies of the same model, each one trained on a specific type of primitive, are considered for the fitting task. During this step, the primitive type is known, as it was predicted in the previous step. The model corresponding to that type of primitive is activated to regress the primitive's parameters. Although the network’s performance is generally quite good, it can be increased even further by performing least-squares fitting. This is accomplished by using the network’s prediction as an excellent initial guess, thereby making the fitting process very quick. For the least squares fitting, an existing python library \texttt{geomfitty}\footnote{https://github.com/mark-boer/geomfitty} is used and extended to include cones.
\end{itemize}

For the classification model, M2 minimizes the cross entropy loss. For the regression models, M4 uses a weighted combination of four basic losses, depending on the type of parameters to be predicted for a given primitive type:
\begin{itemize}
    \item \textit{Axis to axis loss.} This loss is applied to predict direction vectors. Then, given two such vectors $\mathbf{v}_1$ and $\mathbf{v}_2$ the loss is defined as:
    $$\mathcal{L}_1=1-\frac{(\mathbf{v}_1\cdot{} \mathbf{v}_2)^2}{\| \mathbf{v}_1\|\,\| \mathbf{v}_2\|}.$$
    Vectors are forced to have the same direction by minimizing the angle between them.
    \item \textit{Point to axis loss.} This loss is used for anchor points that help to define the axes of various shapes. It is defined as the distance between the query point and the line that passes through the shape’s axis.
    \item \textit{Point to point loss.} This loss is used for the centers of spheres and tori. For a given point and a ground truth center it is defined as the Euclidean distance between them.
    \item \textit{Scalar to scalar loss.} The remaining parameters such as radii and angles are simply scalars, so this loss is calculated via the mean squared error.
\end{itemize}

It is important to note that neural networks perform very poorly when tasked to predict numbers that are distributed over a wide range of values. The unit sphere normalization applied to the input helps with this problem, as it restricts said range to roughly $(0,1)$ for most types of parameters. In order to map these values back to the correct range we apply the inverse transformations that we applied to the input points.

The network was trained and evaluated on a $85\%-15\%$ training-validation split of the given dataset and was found to perform very well, reaching over $95\%$ validation accuracy. To further improve the performance, a  \emph{backtracking strategy} is implemented to deal with heavily perturbed primitives. After the first shape’s parameters are regressed, the distance between the shape and the input points is measured. If the distance exceeds a pre-defined threshold, then it is assumed that our initial classification guess was wrong, so the network’s guess with the second highest probability is taken and the process is repeated. For peculiar cases where the least-squares fitting diverges, the regressor network’s predicted parameters is used as the final estimation.

Figure \ref{figure:graph_abstr_M2} provides a graphical abstract of method M4.

\subsubsection{Computational aspects}
After experimenting on the training set, the voxel size was set to 0.05. For the minimization problem, M4 consider the optimization algorithm Adam \cite{Adam}. The training of each model was performed on a machine equipped with a single Nvidia RTX 3090 and a Ryzen 7 3800x processor with 32GB of RAM. The classification network required 80 epochs to converge, while the regressor networks required 100 epochs. The average training times, per epoch, are: 123 seconds for the classification network, 14 seconds for the plane and cylinder regressors and 15, 16, 22 seconds for the sphere, cone and torus regressors respectively. The entire pipeline takes 39 seconds to complete processing the test set. No backtracking was required on the test set. The average execution times, per point cloud, for the classification network, regression network, and least squares fitting are $0.010$, $0.009$ and $0.021$ seconds, respectively.

\subsection{M5: PointNet bundle and parameter estimation with least square fitting}
This method is proposed by Dinh-Khoi Vo, Tuan-An To, Nham-Tan Nguyen, Nhat-Quynh Le-Pham, Hai-Dang Nguyen and Minh-Triet Tran. The method implementation can be found at \url{https://github.com/ToTuanAn/SHREC2022.git}.
\begin{figure}[t!]
    \centering
    \includegraphics[scale=0.5]{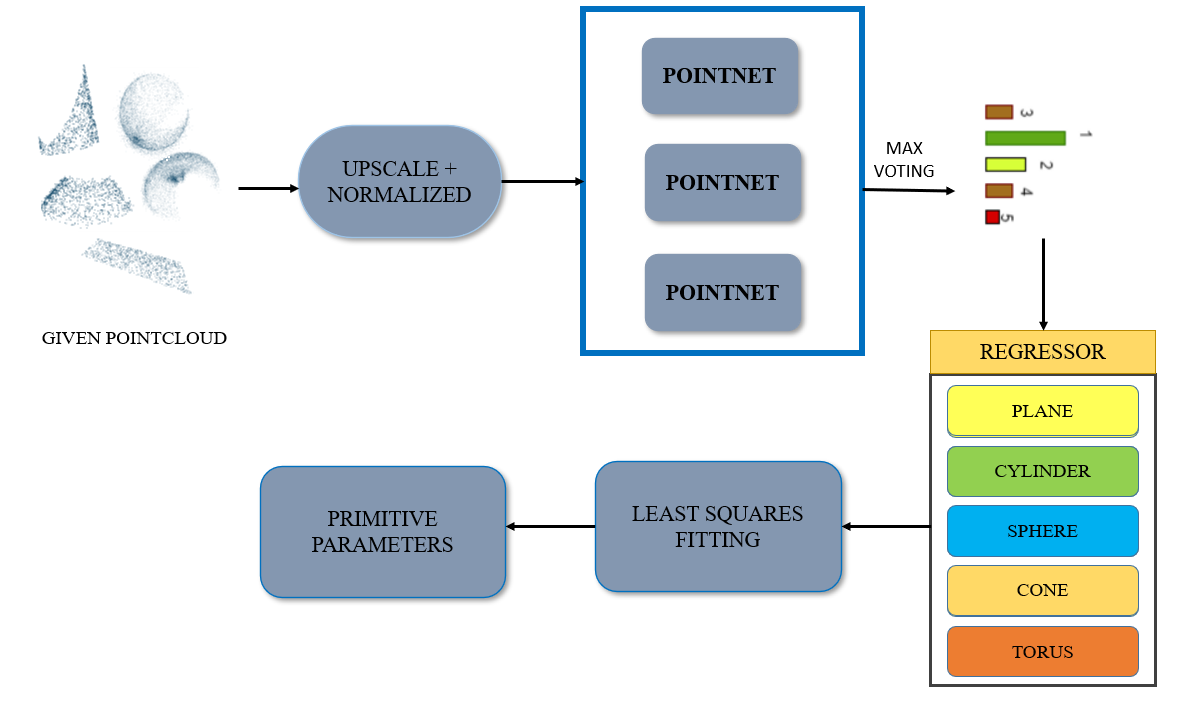}
    \caption{A graphical representation of the strategy adopted in method M5.}
    \label{figure:graph_abstr_M5}
\end{figure}
\subsubsection{Overall strategy}
The input point clouds are pre-processed; specifically, each segment is re-sampled in order to obtain the same number of points (8100 points), as the number of points in each object is not constant. This strategy augments the data by holding two of the three $(x, y, z)$ position and adds a small shift in the remainder coordinate to create new data points. Finally, normalization and random rotations about the $z-$axis are applied to increase the effectiveness of the training process. 

Similarly to M3, the core component for point cloud classification is the PointNet~\cite{Pointnet} architecture. This architecture has three main modules: the max pooling layer to aggregate information symmetrically from all the points symmetrically, a global
and local feature combination structure and two joint alignment networks. The symmetrical max pooling layer helps to make model invariant to the orders of the input data. Input points are fed in a collection of multi-layer perceptron networks to extract global features. After computing the global point cloud feature vector, data are fed back to concatenate the global feature with each of the point features. M5's strategy works in two main steps:
\begin{itemize}
    \item \emph{Classification step}. The data is trained by the above model with different parameters and the top three models are obtained with the highest accuracy in the validation test. The final label is returned after considering the majority votes from the models.
    \item \emph{Shape parameter fitting}. For each primitive type, different ways are considered to estimate the shape parameters of each point cloud.
    \begin{itemize}
        \item \emph{Plane}. The regression with least mean square loss is used to fit a 3D plane on the input point cloud.
        \item \emph{Sphere}. The least mean square method is used to calculate the most appropriate sphere to the point cloud. 
        \item \emph{Cylinder}. The least square fitting method is used to fit the point clouds to an infinitely long cylinder, parameterized by an anchor point, the direction and the radius of the cylinder.
        \item \emph{Cone}. Tangent planes are computed to approximate the vertex of the cone; then the circle that fits the set of points having the same distance from the vertex is used to find the rotational axis and half the aperture.
        \item \emph{Torus}. The \texttt{SciPy} library is used to determine the torus that best fits the point cloud.
    \end{itemize}
   
    For the least squares fitting, the python library \texttt{geomfitty} is used (see method M4).
 \end{itemize}

The networks are trained and evaluated by splitting the initial dataset in two parts, training set ($80\%$) and validation set ($20\%$); the validation accuracy is above $80\%$.

Figure \ref{figure:graph_abstr_M5} provides a graphical abstract of method M5.

\subsubsection{Computational aspects}
In the experiments, a learning rate of $2.5\times10^{-4}$ is used, and Adam is used for optimization. The training is run on Google Colab Pro - GPU P100, for a total of 10 epochs. The average training times, per epoch, are: 50 minutes for the PointNet network and 8 min for all 3D geometry regressors. The execution time is, on average, $36$ seconds for the classification of the 925 test segments, while it is $10$ minutes to estimate their shape parameters.

\subsection{M6: AlexNet and ISO reconstruction standards}
This method is proposed by Yifan Qie and Nabil Anwer. The method implementation can be found at \url{https://github.com/YifanQie/SHREC22_fitting_LURPA.git}.

\subsubsection{Overall strategy}
Input point clouds are first preprocessed by centering them to the origin and scaling them into a unit sphere. Then, they are rotated so that the first principal axis of each point cloud lies along $z-$axis. 

The primitive type identification of this method is based on AlexNet \cite{AlexNet} which is a convolutional neural network architecture with 60 million parameters and 650,000 neurons. Segments are turned into images as input for training the neural network; images are collected from the principal directions for feature extraction and recognition. All the layers, except for the last three layers, are  extracted form AlexNet. The updated last three layers consists of a fully connected layer, a softmax layer, and a classification output layer, where the fully connected layer is set to return as many numbers as the number of primitive types defined in contest (i.e., 5). Regarding the specificity of the input point clouds, transfer learning is used in the approach since the first layers in the pretrained networks have proved to be able to address simple pattern recognition in the images. 

\begin{figure}[t!]
    \centering
    \includegraphics[scale=0.375]{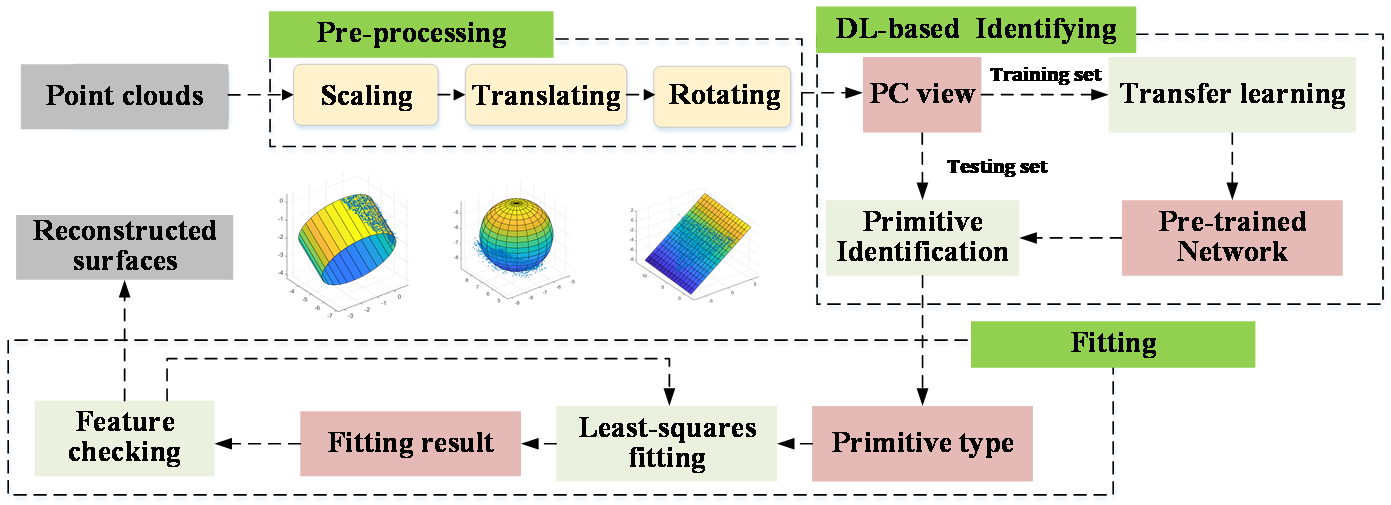}
    \caption{A graphical representation of the strategy adopted in method M6.}
    \label{figure:graph_abstr_M6}
\end{figure}

This strategy involves two main steps:
\begin{itemize}
\item \textit{Classification step.} The pre-trained neural network is used for the test set to provide the primitive type that identify the input point cloud.
\item  \textit{Fitting step.} The surface reconstruction is tested using the methods considered in the ISO standards on geometrical product specifications and verification (GPS), detailed in \cite{QIE2020152}. Once the primitive type of a point cloud, that is the invariance class in ISO GPS terminology, is identified by the fine-tuned AlexNet, an ideal feature defined by a parametrized equation can be guaranteed using least square fitting to obtain a reconstructed surface, see \cite{LSFitting}. 

\end{itemize}

The network was trained and evaluated on a $95\%-5\%$ training-validation split of the given dataset, reaching over $87\%$ validation accuracy.

Figure \ref{figure:graph_abstr_M6} provides a graphical abstract of method M6.

\subsubsection{Computational aspects}
All tests are performed on a workstation equipped with a 1.70 GHz Intel(R) Xeon(R) CPU, an 16 GB RAM, and the Windows 10 operating system.  The classification network required 6 epochs to converge. 
The training process takes about 50 minutes per epoch.
The execution time for each model of the test set is about 0.5 second on average.

\section{Comparative analysis}\label{sec:analysis}
The performance of each method presented in Section \ref{sec:description_of_methods} is here quantitatively evaluated on the basis of the measures described in Section \ref{sec:evaluation_measures}. 

For the sake of conciseness (and of the readers), we here report only tables referring to the whole test set; additional tables obtained by focusing  on specific point cloud artifacts are provided as additional material in \ref{sec:appB} and \ref{sec:appC}.
\begin{figure*}[h!]
    \centering
    \begin{tabular}{ccc}
        M1 & M2 & M3 \\
        \includegraphics[scale=0.485, trim={1cm 0cm 1cm 0.5cm}, clip]{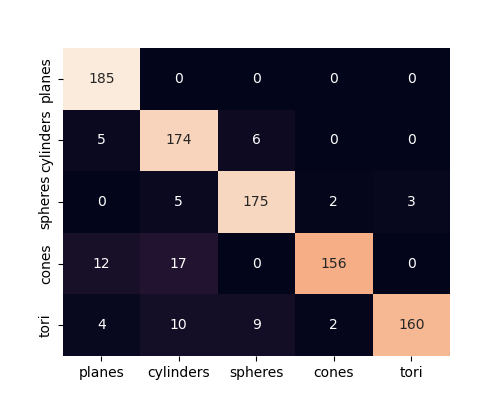}
        &
        \includegraphics[scale=0.485, trim={1cm 0cm 1cm 0.5cm}, clip]{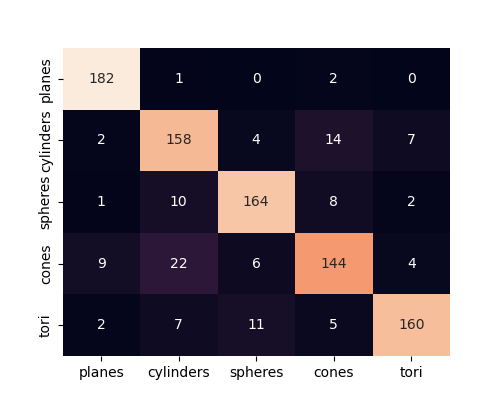}
        &
        \includegraphics[scale=0.485, trim={1cm 0cm 1cm 0.5cm}, clip]{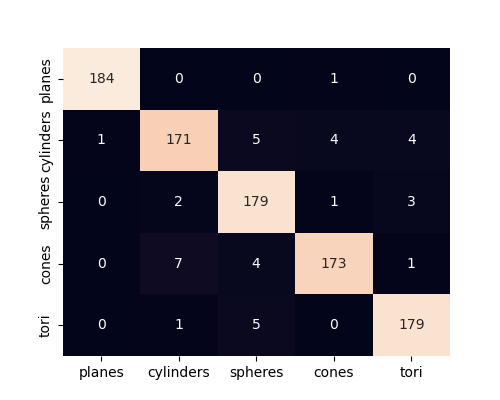}
        \\
        M4 & M5 & M6 
        \\
        \includegraphics[scale=0.485, trim={1cm 0cm 1cm 0.5cm}, clip]{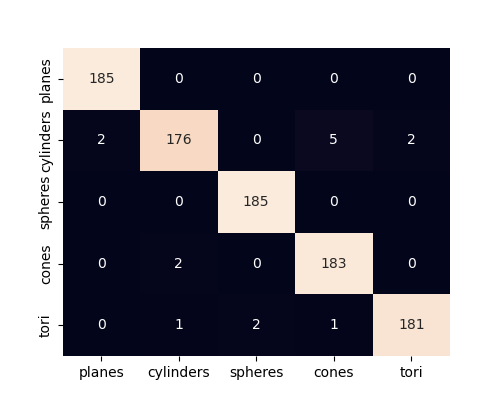}
        &
        \includegraphics[scale=0.485, trim={1cm 0cm 1cm 0.5cm}, clip]{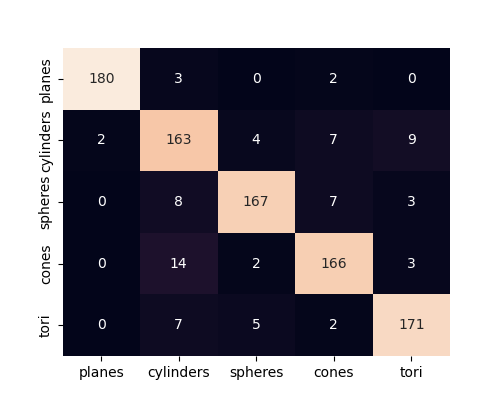}
        &
        \includegraphics[scale=0.485, trim={1cm 0cm 1cm 0.5cm}, clip]{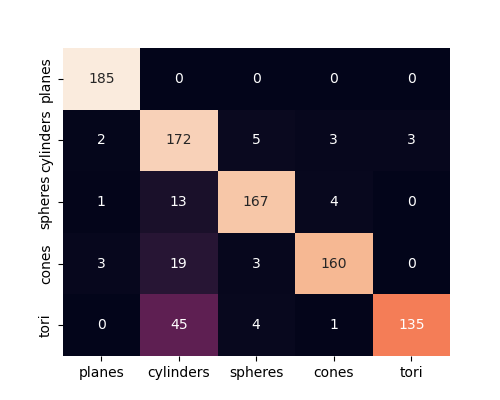}
    \end{tabular}
    \caption{Confusion matrices $\text{CM}$ for the whole test set, with respect to each primitive type. Here, the entry $\text{CM}(i,j)$ indicates the number of samples with true label being the $i$-th class and predicted label being the $j$-th class.}
    \label{fig:globalCM}
\end{figure*}

\subsection{Classification task}
Figure \ref{fig:globalCM} shows the confusion matrices over the whole test set. All the six methods have a high number of true positives, being cones and tori the most difficult primitive types to be recognized. A more thorough analysis is provided by Table \ref{table:classif_measures_WHOLE}, which contains the following information: Positive Predicted Value (PPV), Negative Predicted Value (NPV), True Positive Rate (TPR), True Negative Rate (TNR), Accuracy (ACC) and their macro-averages. We can notice that:
\begin{itemize}
    \item Methods M2, M3, M4 and M5 have NPV (slightly) higher than PPV: it is more likely for these methods to be right when reporting a negative rather than a positive. Methods M1 and M6 have mixed outcomes.
    \item Methods M3 and M5 have TNR higher than TPR, making them more reliable in correctly finding true negatives rather than true positives. Methods M1, M2, M4 and M6 presents, for some primitive types, a TPR higher than the TNR: this means that, in some cases, the classifier is more likely to identify correctly true positives rather than true negatives.
    \item All methods have the accuracy over $90\%$, with the slightly lower scores for cylinders, cones and tori. Macro-averaged values of accuracy are always above $95\%$, with two of the three best scores reached by neural methods.
\end{itemize}

\begin{table}[h!]
\centering
\footnotesize	
\begin{tabular}{l|c|ccccc|c|}
%First row
\hhline{~-------}
& \cellcolor{ForestGreen!15}Method & \cellcolor{ForestGreen!15}T1
& \cellcolor{ForestGreen!15}T2 & \cellcolor{ForestGreen!15}T3 & \cellcolor{ForestGreen!15}T4 & 
\cellcolor{ForestGreen!15}T5 & 
\cellcolor{ForestGreen!15}Avg \\ \hline

%PPV
\multicolumn{1}{|l|}{\cellcolor{ForestGreen!15}} & M1 & 0.8981 & \textcolor{bronze}{0.8447} & 0.9211 & \textcolor{gold}{0.9750} & \textcolor{silver}{0.9816} & \textcolor{bronze}{0.9241} \\
\multicolumn{1}{|l|}{\cellcolor{ForestGreen!15}} & M2 & 0.9286 & 0.7980 & 0.8865 & 0.8324 & 0.9249 & 0.8741 \\
\multicolumn{1}{|l|}{\cellcolor{ForestGreen!15}} & M3 & \textcolor{gold}{0.9946} & \textcolor{silver}{0.9448} & 0.9275 & \textcolor{bronze}{0.9665} & 0.9572 & \textcolor{silver}{0.9581} \\
\multicolumn{1}{|l|}{\cellcolor{ForestGreen!15}} & M4 & \textcolor{silver}{0.9893} & \textcolor{gold}{0.9832} & \textcolor{gold}{0.9893} & \textcolor{silver}{0.9683} & \textcolor{gold}{0.9891} & \textcolor{gold}{0.9838} \\
\multicolumn{1}{|l|}{\cellcolor{ForestGreen!15}} & M5 & \textcolor{bronze}{0.9890} & 0.8359 & \textcolor{silver}{0.9382} & 0.9022 & 0.9194 & 0.9169 \\
\multicolumn{1}{|l|}{\multirow{-6}{*}{\cellcolor{ForestGreen!15} \begin{turn}{90}PPV\end{turn}}} & M6 & 0.9686 & 0.6908 & \textcolor{bronze}{0.9330} & 0.9524 & \textcolor{bronze}{0.9783} & 0.9046 \\
\hline

%NPV
\multicolumn{1}{|l|}{\cellcolor{ForestGreen!15}} & M1 & \textcolor{gold}{1.0000} & \textcolor{silver}{0.9847} & \textcolor{bronze}{0.9864} & 0.9621 & 0.9672 & \textcolor{bronze}{0.9801} \\
\multicolumn{1}{|l|}{\cellcolor{ForestGreen!15}} & M2 & \textcolor{bronze}{0.9959} & 0.9629 & 0.9716 & 0.9455 & 0.9668 & 0.9685 \\
\multicolumn{1}{|l|}{\cellcolor{ForestGreen!15}} & M3 & \textcolor{silver}{0.9986} & \textcolor{bronze}{0.9812} & \textcolor{silver}{0.9918} & \textcolor{silver}{0.9839} & \textcolor{silver}{0.9919} & \textcolor{silver}{0.9895} \\
\multicolumn{1}{|l|}{\cellcolor{ForestGreen!15}} & M4 & \textcolor{gold}{1.0000} & \textcolor{gold}{0.9879} & \textcolor{gold}{1.0000} & \textcolor{gold}{0.9973} & \textcolor{gold}{0.9946} & \textcolor{gold}{0.9960} \\
\multicolumn{1}{|l|}{\cellcolor{ForestGreen!15}} & M5 & 0.9933 & 0.9699 & 0.9759 & \textcolor{bronze}{0.9744} & \textcolor{bronze}{0.9811} & 0.9789 \\
\multicolumn{1}{|l|}{\multirow{-6}{*}{\cellcolor{ForestGreen!15} \begin{turn}{90}NPV\end{turn}}} & M6 & \textcolor{gold}{1.0000} & 0.9808 & 0.9759 & 0.9670 & 0.9365 & 0.9720 \\
\hline

%TPR
\multicolumn{1}{|l|}{\cellcolor{ForestGreen!15}} & M1 & \textcolor{gold}{1.0000} & \textcolor{silver}{0.9405} & \textcolor{bronze}{0.9459} & 0.8432 & 0.8649 & \textcolor{bronze}{0.9189} \\
\multicolumn{1}{|l|}{\cellcolor{ForestGreen!15}} & M2 & \textcolor{bronze}{0.9838} & 0.8541 & 0.8865 & 0.7784 & 0.8649 & 0.8735 \\
\multicolumn{1}{|l|}{\cellcolor{ForestGreen!15}} & M3 & \textcolor{silver}{0.9946} & 0.9243 & \textcolor{silver}{0.9676} & \textcolor{silver}{0.9351} & \textcolor{silver}{0.9676} & \textcolor{silver}{0.9578} \\
\multicolumn{1}{|l|}{\cellcolor{ForestGreen!15}} & M4 & \textcolor{gold}{1.0000} & \textcolor{gold}{0.9514} & \textcolor{gold}{1.0000} & \textcolor{gold}{0.9892} & \textcolor{gold}{0.9784} & \textcolor{gold}{0.9838} \\
\multicolumn{1}{|l|}{\cellcolor{ForestGreen!15}} & M5 & 0.9730 & 0.8811 & 0.9027 & \textcolor{bronze}{0.8973} & \textcolor{bronze}{0.9243} & 0.9157 \\
\multicolumn{1}{|l|}{\multirow{-6}{*}{\cellcolor{ForestGreen!15} \begin{turn}{90}TPR\end{turn}}} & M6 & \textcolor{gold}{1.0000} & \textcolor{bronze}{0.9297} & 0.9027 & 0.8649 & 0.7297 & 0.8854 \\
\hline

%TNR
\multicolumn{1}{|l|}{\cellcolor{ForestGreen!15}} & M1 & 0.9716 & 0.9568 & \textcolor{bronze}{0.9797} & \textcolor{gold}{0.9946} & \textcolor{silver}{0.9959} & \textcolor{bronze}{0.9797} \\
\multicolumn{1}{|l|}{\cellcolor{ForestGreen!15}} & M2 & 0.9811 & 0.9459 & 0.9716 & 0.9608 & 0.9824 & 0.9684 \\
\multicolumn{1}{|l|}{\cellcolor{ForestGreen!15}} & M3 & \textcolor{gold}{0.9986} & \textcolor{silver}{0.9865} & 0.9811 & \textcolor{silver}{0.9919} & \textcolor{bronze}{0.9892} & \textcolor{silver}{0.9895} \\
\multicolumn{1}{|l|}{\cellcolor{ForestGreen!15}} & M4 & \textcolor{silver}{0.9973} & \textcolor{gold}{0.9959} & \textcolor{gold}{0.9973} & \textcolor{silver}{0.9919} & \textcolor{gold}{0.9973} & \textcolor{gold}{0.9959} \\
\multicolumn{1}{|l|}{\cellcolor{ForestGreen!15}} & M5 & \textcolor{silver}{0.9973} & \textcolor{bronze}{0.9568} & \textcolor{silver}{0.9851} & 0.9757 & 0.9797 & 0.9789 \\
\multicolumn{1}{|l|}{\multirow{-6}{*}{\cellcolor{ForestGreen!15} \begin{turn}{90}TNR\end{turn}}} & M6 & \textcolor{bronze}{0.9919} & 0.8959 & \textcolor{bronze}{0.9838} & \textcolor{bronze}{0.9892} & \textcolor{silver}{0.9959} & 0.9714 \\
\hline

%ACCURACY
\multicolumn{1}{|l|}{\cellcolor{ForestGreen!15}} & M1 & 0.9773 & \textcolor{bronze}{0.9535} & \textcolor{bronze}{0.9730} & \textcolor{bronze}{0.9643} & \textcolor{bronze}{0.9697} & \textcolor{bronze}{0.9676} \\
\multicolumn{1}{|l|}{\cellcolor{ForestGreen!15}} & M2 & 0.9816 & 0.9276 & 0.9546 & 0.9243 & 0.9589 & 0.9494 \\
\multicolumn{1}{|l|}{\cellcolor{ForestGreen!15}} & M3 & \textcolor{gold}{0.9978} & \textcolor{silver}{0.9741} & \textcolor{silver}{0.9784} & \textcolor{silver}{0.9805} & \textcolor{silver}{0.9849} & \textcolor{silver}{0.9831} \\
\multicolumn{1}{|l|}{\cellcolor{ForestGreen!15}} & M4 & \textcolor{gold}{0.9978} & \textcolor{gold}{0.9870} & \textcolor{gold}{0.9978} & \textcolor{gold}{0.9914} & \textcolor{gold}{0.9935} & \textcolor{gold}{0.9935} \\
\multicolumn{1}{|l|}{\cellcolor{ForestGreen!15}} & M5 & \textcolor{bronze}{0.9924} & 0.9416 & 0.9686 & 0.9600 & 0.9686 & 0.9663 \\
\multicolumn{1}{|l|}{\multirow{-6}{*}{\cellcolor{ForestGreen!15} \begin{turn}{90}ACC\end{turn}}} & M6 & \textcolor{silver}{0.9935} & 0.9027 & 0.9676 & \textcolor{bronze}{0.9643} & 0.9427 & 0.9542 \\
\hline

\end{tabular}
\caption{Classification metrics for the whole test set, with respect to each primitive type: T1=plane, T2=cylinder, T3=sphere, T4=cone, T5=torus. Here, gold, silver and bronze identify the first, second and third best performance. The last column contains the macro averages.\label{table:classif_measures_WHOLE}}
\end{table}

\begin{figure}[h!]
    \centering
    \begin{tabular}{c}
        \includegraphics[scale=.825, trim={0.5cm 0cm 1cm 0cm}, clip]{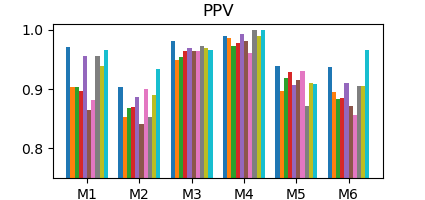}
        \\
        \includegraphics[scale=.825, trim={0.5cm 0cm 1cm 0cm}, clip]{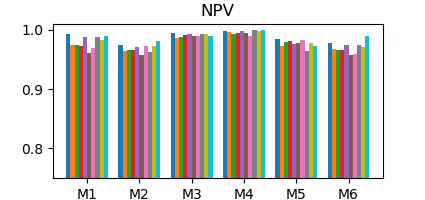}
        \\
        \includegraphics[scale=.825, trim={0.5cm 0cm 1cm 0cm}, clip]{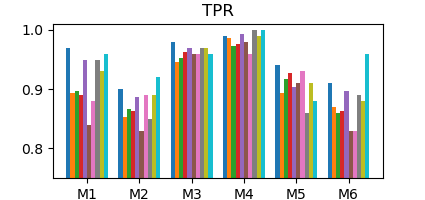}
        \\
        \includegraphics[scale=.825, trim={0.5cm 0cm 1cm 0cm}, clip]{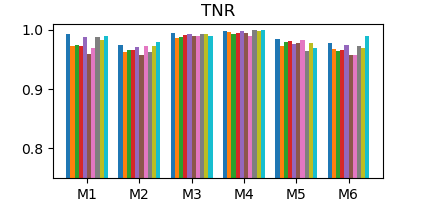}
        \\
        \includegraphics[scale=.825, trim={0.5cm 0cm 1cm 0cm}, clip]{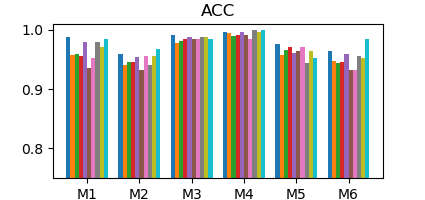}
        \\
        \includegraphics[scale=.70, trim={0.85cm 3cm 0.75cm 3cm}, clip]{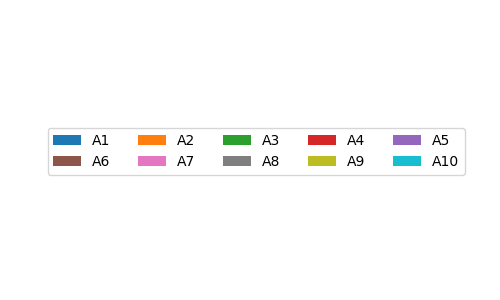}
        \\
    \end{tabular}
    \caption{Bar graphs of the macro averages for the classification measures, grouped by method. The legend reflects the color encoding of the perturbation types.}
    \label{fig:macro_avgs_CLASSIF}
\end{figure}

Figure \ref{fig:macro_avgs_CLASSIF} clarifies the (global) impact of different point cloud artifacts upon the six methods, with respect to the classification measures. We observe that:
\begin{itemize}
    \item M1, M2, M4 and M6 reach their best performance on clean data. Surprisingly enough, M2, M4 and M6 perform best on data with small deformations: these three approaches appear to be able to extract some additional hidden information to improve their prediction capability.
    \item M1 suffers the most from undersampled data. On the other hand, coherently with the Hough paradigm on which the method relies on, it is not much affected by missing data.
    \item M3 and M4 have their lowest performance, respectively, in case of uniform and Gaussian noise. However, the grouped bar charts suggest these two methods are the least prone to misclassification. This result confirms the power of neural methods in handling classification tasks.
    \item In spite of having, in most cases, the lowest classification performance, M2 has still no measure that goes below $80\%$. Compared to the other method using PointNet (M3), here the implementation seems to overfit way more the training data.
    \item M5 has a relatively high performance on undersampled data, when compared to other point cloud artifacts.
    \item The combination of noise and undersampling has proved to be the most challenging perturbation for M6 which, however, exhibits one of the best overall performance on small deformations.
\end{itemize}
By checking which segments are misclassified, it can be additionally observed that the segments which results being more problematic are, in many cases, the small ones.

Tables \ref{table:classif_measures_CLEAN}-\ref{table:classif_measures_DEFORM} in \ref{sec:appB} provide a complete overview of classification measures, specified with respect to the primitive type.

\subsection{Recognition and fitting tasks}
Table \ref{table:fitting_measures_CWHOLE} provides various statistics of the measures introduced in Section \ref{sec:recognition_fitting_measures}, when the whole test set is considered. It contains the following information: first, second and third quartiles, mean value and standard deviation.
More specifically, these quartiles split a set of sorted real numbers into four parts of approximately equal cardinality: the first (Q1) and the third (Q3) quartiles are defined as the values such that, respectively, 25\% and 75\% of the numbers lie below them; the second quartile (Q2) is the median of the set. Quartiles' significance is due to their ability to identify possible outliers. Based on quartiles, we can notice that:
\begin{itemize}
    \item The $L^2$ distance provides low values for all methods with regard to the three quartiles. The lowest Q1 has an order of magnitude of $-14$ and is obtained by method M6; the highest order of magnitude is that of method M3, and corresponds to $-1$. Q2 and Q3 have order of magnitudes between $-3$ and $-1$.
    \item With the sole exception of the third quartile of M5, quartiles of the MFEs lie under the order of magnitude of $-2$. The lowest values are obtained by M2.
    \item When it comes to quartiles, the Hausdorff distance the order of magnitude is generally (non-strictly) lower than $-1$; the only exceptions fare the third quartiles of methods M3 and M5.
\end{itemize}
Based on the arithmetic mean and standard deviation, we can conclude that:     
\begin{itemize}
    \item  Methods M2, M5 and M6 are affected by numerical instability issues, which result in the presence of outliers; it is worth noting, however, this problem is limited to the top $25\%$ of the errors.
    \item M2 presents 11 outliers corresponding to the recognition of  cones in which the estimated half-angle estimate is close to zero and the estimated vertex is far away from the origin of the coordinate axes.  
    \item M5 provides different outliers depending on the type of task. Specifically, it has 61 outliers in the parameter recognition of various cylinders, cones and tori; however, for these clouds, fitting errors are low. We therefore conclude that, in all these cases, the approximation is successful but at the cost of a poor recognition. From the fitting task point of view, MFE and directed Hausdorff distance have high values in correspondence of 18 outliers correctly identified as planes: here, the normal is correctly identified (thus, the low value of $L^2$ distance); on the other hand, the method fails in returning a point passing through each returned plane.
    \item M6 presents 30 outliers, which correspond to point clouds classified as cones; the problem these estimates suffer from are analogue to that of M2. 
    \item M1 and M4 alternate the first and second places in terms of means and standard deviations, suggesting that direct and neural methods can both provide competitive solutions to the problem under study.
 \end{itemize}

\begin{figure}[t!]
    \centering
    \begin{tabular}{c}
        \includegraphics[scale=.75, trim={0.5cm 0cm 1cm 0cm}, clip]{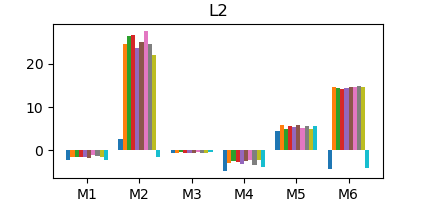}
        \\
        \includegraphics[scale=.75, trim={0.5cm 0cm 1cm 0cm}, clip]{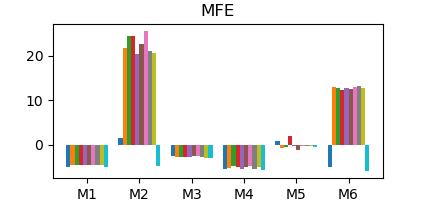}
        \\
        \includegraphics[scale=.75, trim={0.5cm 0cm 1cm 0cm}, clip]{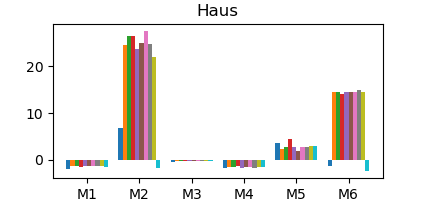}
        \\
        \includegraphics[scale=.75, trim={0.85cm 3cm 0.75cm 3cm}, clip]{legend.png}
        \\
    \end{tabular}
    \caption{Bar graphs of the log macro averages for the recognition and fitting measures, grouped by method. The legend reflects the color encoding of the perturbation types.}
    \label{fig:macro_avgs_RECFIT}
\end{figure}

By checking the bar graphs of the log mean errors with respect to the perturbation type, see Figure \ref{fig:macro_avgs_RECFIT}, it is possible to note a generally low variation of the (average) performance of each method, with the exception of M2 and M6; for these results, the logarithmic scale was preferred because of the presence of outliers. It is indicative that, when the methods (i.e., all but M2 and M6) are right in predicting the primitive type, their estimates are not much influenced by the perturbation type. Moreover:
\begin{itemize}
    \item Methods M2 and M6 have better recognition and fitting performance when dealing with clean data or with point clouds having just small deformations. However, it is worth noticing once more that the poor performance mostly depends on the presence of outliers (see  Table \ref{table:fitting_measures_CWHOLE}).
    \item M5 is on average better in fitting than in recognition: despite being the mean $L^2$ errors rather high, mean MFEs and directed Hausdorff distances are lower.
    \item M1 and M4 behave similarly in both recognition and fitting tasks.
\end{itemize}

\begin{table}[t]
\centering
\footnotesize	
\begin{tabular}{c|>{\centering}m{0.723cm}|>{\centering}m{1.2cm} >{\centering}m{1.2cm} >{\centering}m{1.2cm}  >{\centering}m{1.2cm} c|}
%First row
\hhline{~------}
& \cellcolor{BlueViolet!20}Method & \cellcolor{BlueViolet!20}Q1
& \cellcolor{BlueViolet!20}Q2 & \cellcolor{BlueViolet!20}Q3 & \cellcolor{BlueViolet!20}mean & 
\cellcolor{BlueViolet!20}std \\ \hline

%L2
\multicolumn{1}{|c|}{\cellcolor{BlueViolet!20}} & M1 & 2.16e-02 & 6.82e-02 & 1.78e-01 & \textcolor{silver}{2.06e-01} & \textcolor{silver}{3.71e-01}\\
\multicolumn{1}{|c|}{\cellcolor{BlueViolet!20}} & M2 & \textcolor{silver}{8.95e-10} & \textcolor{gold}{5.17e-03} & \textcolor{silver}{9.27e-02} & 1.21e+11 & 2.78e+12\\
\multicolumn{1}{|c|}{\cellcolor{BlueViolet!20}} & M3 & 1.77e-01 & 4.73e-01 & 8.73e-01 & \textcolor{bronze}{6.03e-01} & \textcolor{bronze}{5.54e-01}\\
\multicolumn{1}{|c|}{\cellcolor{BlueViolet!20}} & M4 & \textcolor{bronze}{2.37e-07} & \textcolor{bronze}{1.54e-02} & \textcolor{gold}{4.43e-02} & \textcolor{gold}{5.18e-02} & \textcolor{gold}{1.84e-01}\\
\multicolumn{1}{|c|}{\cellcolor{BlueViolet!20}} & M5 & 6.25e-04 & 2.15e-02 & 9.01e-01 & 2.46e+02 & 1.64e+03\\
\multicolumn{1}{|c|}{\multirow{-6}{*}{\cellcolor{BlueViolet!20} \begin{turn}{90}$L^2$\end{turn}}} & M6 & \textcolor{gold}{1.49e-14} & \textcolor{silver}{1.49e-02} & \textcolor{bronze}{1.05e-01} & 1.28e+06 & 7.54e+06 \\
\hline
%MFE
\multicolumn{1}{|c|}{\cellcolor{BlueViolet!20}} & M1 & 3.71e-03 & 5.96e-03 & 1.05e-02 & \textcolor{silver}{9.19e-03} & \textcolor{gold}{9.42e-03}\\
\multicolumn{1}{|c|}{\cellcolor{BlueViolet!20}} & M2 & \textcolor{gold}{0.00} & \textcolor{gold}{1.94e-03} & \textcolor{gold}{5.11e-03} & 1.39e+10 & 3.51e+11\\
\multicolumn{1}{|c|}{\cellcolor{BlueViolet!20}} & M3 & 1.71e-02 & 4.49e-02 & 8.67e-02 & \textcolor{bronze}{6.57e-02} & \textcolor{bronze}{7.46e-02}\\
\multicolumn{1}{|c|}{\cellcolor{BlueViolet!20}} & M4 & \textcolor{bronze}{2.32e-03} & \textcolor{bronze}{3.59e-03} & \textcolor{silver}{5.87e-03} & \textcolor{gold}{5.82e-03} & \textcolor{silver}{1.43e-02}\\
\multicolumn{1}{|c|}{\cellcolor{BlueViolet!20}} & M5 & 3.45e-03 & 1.49e-02 & 1.70e-01 & 2.62e+00 & 4.46e+01\\
\multicolumn{1}{|c|}{\multirow{-6}{*}{\cellcolor{BlueViolet!20} \begin{turn}{90}MFE\end{turn}}} & M6 & \textcolor{silver}{2.04e-03} & \textcolor{gold}{3.43e-03} & \textcolor{bronze}{9.51e-03} & 2.56e+05 & 1.46e+06 \\
\hline
%Hausdorff
\multicolumn{1}{|c|}{\cellcolor{BlueViolet!20}} & M1 & 9.49e-02 & 1.67e-01 & 3.05e-01 & \textcolor{silver}{2.50e-01} & \textcolor{gold}{2.58e-01}\\
\multicolumn{1}{|c|}{\cellcolor{BlueViolet!20}} & M2 & \textcolor{bronze}{6.12e-02} & \textcolor{bronze}{9.25e-02} & \textcolor{bronze}{2.68e-01} & 1.07e+11 & 2.60e+12\\
\multicolumn{1}{|c|}{\cellcolor{BlueViolet!20}} & M3 & 3.72e-01 & 6.62e-01 & 1.09e+00 & \textcolor{bronze}{8.13e-01} & \textcolor{bronze}{6.34e-01}\\
\multicolumn{1}{|c|}{\cellcolor{BlueViolet!20}} & M4 & \textcolor{gold}{5.82e-02} & \textcolor{gold}{7.90e-02} & \textcolor{gold}{1.51e-01} & \textcolor{gold}{2.23e-01} & \textcolor{silver}{4.74e-01}\\
\multicolumn{1}{|c|}{\cellcolor{BlueViolet!20}} & M5 & 7.18e-02 & 2.98e-01 & 4.24e+00 & 3.87e+01 & 5.77e+02\\
\multicolumn{1}{|c|}{\multirow{-6}{*}{\cellcolor{BlueViolet!20} \begin{turn}{90}$d_\text{dHaus}$\end{turn}}} & M6 & \textcolor{silver}{5.85e-02} & \textcolor{silver}{8.95e-02} & \textcolor{silver}{2.53e-01} & 1.31e+06 & 7.63e+06\\
\hline

\end{tabular}
\caption{Statistics of the fitting errors for the whole test set. Here, gold, silver and bronze identify the first, second and third best performance.\label{table:fitting_measures_CWHOLE}}
\end{table}

As for classification measures, tables that specify recognition and fitting measures to the single point cloud artifacts are left to \ref{sec:appC} (see Tables \ref{table:fitting_measures_CLEAN}-\ref{table:fitting_measures_BUMPS}). 

\section{Concluding remarks}
Six methods were submitted for evaluation to this SHREC track on the fitting and recognition of simple geometric primitives on point clouds. Two approaches -- M1 and M2 -- base their whole pipeline on direct methods; M3 and M4 are fully based in neural networks; M5 and M6 exploits both methodologies, that is, neural architectures for the classification task and direct methods for the fitting and recognition tasks.

The dataset was specifically designed to allow the training of data-driven methods, with a training set of 46000 primitives and a test set of 925 models. Nine types of perturbations were randomly applied to each type of primitive in the ratio of 5000 training examples and 100 test samples, with the exception of local deformations that count 1000 training examples and 25 test samples.
The outcome of the methods shows that, on simple primitives, data-driven methods are reaching a satisfactory level of maturity, obtaining the best scores both in primitive classification and in the parameter estimation. Direct methods remain fiercely competitive, and seem to suffer more in classification than in fitting and recognition. We believe that a similar comparative analysis will be possible in the future on models with multiple primitives.

As a possible future development, to better assess the potential of these methods in a reverse engineering context, it would be interesting to create a dataset with real data, e.g., captured with a high quality industrial 3D scanner, and derive the ground truth coming from the specifications used in the fabrication of the targets. To determine which approach (or which paradigm, between direct and neural methods) has the best generalization capability, it would be certainly worthwhile to keep the training set introduced in this track and use real data just in the test set. This direction is however rather challenging, as it requires having sufficient diversity, in terms of models to be scanned, between the primitives and parameters used for training and testing sets.

\section*{Acknowledgments}
This work has been developed in the CNR IMATI research activities DIT.AD004.100, DIT.AD021.080.001 and DIT.AD021.125.
Ivan Sipiran was funded by the Agencia Nacional de Investigación y Desarrollo (ANID Chile), under grant number 11220211. Nguyen Quang Thuc was funded by Vingroup JSC and supported by the Master, Ph.D. Scholarship Programme of Vingroup Innovation Foundation (VINIF), Institute of Big Data, code VINIF.2021.ThS.JVN.06. The teams from University of Science, VNU-HCM, were funded by Gia Lam Urban Development and Investment Company Limited, Vingroup and supported by Vingroup Innovation Foundation (VINIF) under project code VINIF.2019.DA19. Yifan Qie's work benefited from the financial support of China Scholarship Council (Yifan QIE), under Grant NO.201806020187.

%%Vancouver style references.
\bibliographystyle{ieeetr}

\newpage\phantom{blabla}
\appendix
\section{Classification metrics\label{sec:appB}}

\begin{table}[h!]
\centering
\resizebox{0.5\textwidth }{!}{
\footnotesize	
\begin{tabular}{l|c|ccccc|c|}
%First row
\hhline{~-------}
& \cellcolor{ForestGreen!15}Method & \cellcolor{ForestGreen!15}T1
& \cellcolor{ForestGreen!15}T2 & \cellcolor{ForestGreen!15}T3 & \cellcolor{ForestGreen!15}T4 & 
\cellcolor{ForestGreen!15}T5 & 
\cellcolor{ForestGreen!15}Avg \\ \hline

%PPV
\multicolumn{1}{|l|}{\cellcolor{ForestGreen!15}} & M1 & \textcolor{gold}{1.0000} & \textcolor{silver}{0.9524} & \textcolor{silver}{0.9524} & \textcolor{silver}{0.9500} & \textcolor{gold}{1.0000} & \textcolor{bronze}{0.9710} \\
\multicolumn{1}{|l|}{\cellcolor{ForestGreen!15}} & M2 & \textcolor{silver}{0.9524} & 0.8333 & \textcolor{gold}{1.0000} & 0.7826 & \textcolor{silver}{0.9474} & 0.9031 \\
\multicolumn{1}{|l|}{\cellcolor{ForestGreen!15}} & M3 & \textcolor{gold}{1.0000} & \textcolor{gold}{1.0000} & \textcolor{silver}{0.9524} & \textcolor{silver}{0.9500} & \textcolor{gold}{1.0000} & \textcolor{silver}{0.9805} \\
\multicolumn{1}{|l|}{\cellcolor{ForestGreen!15}} & M4 & \textcolor{gold}{1.0000} & \textcolor{gold}{1.0000} & \textcolor{gold}{1.0000} & \textcolor{gold}{0.9524} & \textcolor{gold}{1.0000} & \textcolor{gold}{0.9905} \\
\multicolumn{1}{|l|}{\cellcolor{ForestGreen!15}} & M5 & \textcolor{gold}{1.0000} & \textcolor{bronze}{0.9048} & \textcolor{silver}{0.9524} & \textcolor{bronze}{0.8947} & \textcolor{silver}{0.9474} & 0.9398 \\
\multicolumn{1}{|l|}{\multirow{-6}{*}{\cellcolor{ForestGreen!15} \begin{turn}{90}PPV\end{turn}}} & M6 & \textcolor{gold}{1.0000} & 0.6897 & \textcolor{gold}{1.0000} & \textcolor{gold}{1.0000} & \textcolor{gold}{1.0000} & 0.9379 \\
\hline

%NPV
\multicolumn{1}{|l|}{\cellcolor{ForestGreen!15}} 
                       & M1 & \textcolor{gold}{1.0000} & \textcolor{gold}{1.0000} & \textcolor{gold}{1.0000} & \textcolor{silver}{0.9875} & \textcolor{silver}{0.9756} & \textcolor{bronze}{0.9926} \\
\hline

\end{tabular}
}
\caption{Classification metrics for clean segments.\label{table:classif_measures_CLEAN}}
\end{table}
\begin{table}[h!]
\centering
\resizebox{0.5\textwidth }{!}{
\footnotesize	
\begin{tabular}{l|c|ccccc|c|}
%First row
\hhline{~-------}
& \cellcolor{ForestGreen!15}Method & \cellcolor{ForestGreen!15}T1
& \cellcolor{ForestGreen!15}T2 & \cellcolor{ForestGreen!15}T3 & \cellcolor{ForestGreen!15}T4 & 
\cellcolor{ForestGreen!15}T5 & 
\cellcolor{ForestGreen!15}Avg \\ \hline

%PPV
\multicolumn{1}{|l|}{\cellcolor{ForestGreen!15}} 
                       & M1 & 0.8571 & \textcolor{bronze}{0.7941} & 0.8906 & \textcolor{silver}{0.9792} & \textcolor{gold}{1.0000} & \textcolor{bronze}{0.9042} \\
\multicolumn{1}{|l|}{\cellcolor{ForestGreen!15}} & M2 & 0.8955 & 0.7812 & 0.8596 & 0.8113 & \textcolor{bronze}{0.9153} & 0.8526 \\
\multicolumn{1}{|l|}{\cellcolor{ForestGreen!15}} & M3 & \textcolor{gold}{0.9836} & \textcolor{silver}{0.9483} & 0.9194 & \textcolor{gold}{1.0000} & 0.8939 & \textcolor{silver}{0.9490} \\
\multicolumn{1}{|l|}{\cellcolor{ForestGreen!15}} & M4 & \textcolor{gold}{0.9836} & \textcolor{gold}{1.0000} & \textcolor{gold}{1.0000} & \textcolor{bronze}{0.9677} & \textcolor{silver}{0.9836} & \textcolor{gold}{0.9870} \\
\multicolumn{1}{|l|}{\cellcolor{ForestGreen!15}} & M5 & \textcolor{silver}{0.9831} & 0.7761 & \textcolor{silver}{0.9464} & 0.8947 & 0.8852 & 0.8971 \\
\multicolumn{1}{|l|}{\multirow{-6}{*}{\cellcolor{ForestGreen!15}\begin{turn}{90}PPV\end{turn}}} & M6 & \textcolor{bronze}{0.9231} & 0.6667 & \textcolor{bronze}{0.9423} & 0.9455 & \textcolor{gold}{1.0000} & 0.8955 \\
\hline

%NPV
\multicolumn{1}{|l|}{\cellcolor{ForestGreen!15}} 
                       & M1 & \textcolor{gold}{1.0000} & \textcolor{bronze}{0.9741} & \textcolor{bronze}{0.9873} & 0.9484 & 0.9600 & \textcolor{bronze}{0.9740} \\
\hline

\end{tabular}
}
\caption{Classification metrics for segments suffering from uniform noise.\label{table:classif_measures_UNIF}}
\end{table}

\begin{table}[h!]
\centering
\resizebox{0.5\textwidth }{!}{
\footnotesize	
\begin{tabular}{l|c|ccccc|c|}
%First row
\hhline{~-------}
& \cellcolor{ForestGreen!15}Method & \cellcolor{ForestGreen!15}T1
& \cellcolor{ForestGreen!15}T2 & \cellcolor{ForestGreen!15}T3 & \cellcolor{ForestGreen!15}T4 & 
\cellcolor{ForestGreen!15}T5 & 
\cellcolor{ForestGreen!15}Avg \\ \hline

%PPV
\multicolumn{1}{|l|}{\cellcolor{ForestGreen!15}} 
                       & M1 & 0.8451 & 0.8333 & \textcolor{bronze}{0.9167} & \textcolor{bronze}{0.9600} & \textcolor{silver}{0.9623} & 0.9035 \\
\multicolumn{1}{|l|}{\cellcolor{ForestGreen!15}} & M2 & 0.9344 & 0.8254 & 0.8209 & 0.8750 & 0.8868 & 0.8685 \\
\multicolumn{1}{|l|}{\cellcolor{ForestGreen!15}} & M3 & \textcolor{gold}{1.0000} & \textcolor{silver}{0.9310} & 0.8939 & \textcolor{silver}{0.9661} & \textcolor{gold}{0.9828} & \textcolor{silver}{0.9548} \\
\multicolumn{1}{|l|}{\cellcolor{ForestGreen!15}} & M4 & \textcolor{silver}{0.9836} & \textcolor{gold}{0.9655} & \textcolor{gold}{0.9677} & \textcolor{gold}{0.9672} & \textcolor{gold}{0.9828} & \textcolor{gold}{0.9734} \\
\multicolumn{1}{|l|}{\cellcolor{ForestGreen!15}} & M5 & \textcolor{bronze}{0.9828} & \textcolor{bronze}{0.8413} & \textcolor{silver}{0.9444} & 0.9062 & 0.9180 & \textcolor{bronze}{0.9186} \\
\multicolumn{1}{|l|}{\multirow{-6}{*}{\cellcolor{ForestGreen!15}\begin{turn}{90}PPV\end{turn}}} & M6 & \textcolor{silver}{0.9836} & 0.6463 & 0.8871 & 0.9434 & \textcolor{bronze}{0.9524} & 0.8826 \\
\hline

%NPV
\multicolumn{1}{|l|}{\cellcolor{ForestGreen!15}} 
                       & M1 & \textcolor{gold}{1.0000} & \textcolor{silver}{0.9786} & \textcolor{bronze}{0.9792} & 0.9520 & \textcolor{bronze}{0.9636} & 0.9747 \\
\hline
\end{tabular}
}
\caption{Classification metrics for segments suffering from Gaussian noise.\label{table:classif_measures_GAUSS}}
\end{table}

\begin{table}[h!]
\centering
\resizebox{0.5\textwidth }{!}{
\footnotesize	
\begin{tabular}{l|c|ccccc|c|}
%First row
\hhline{~-------}
& \cellcolor{ForestGreen!15}Method & \cellcolor{ForestGreen!15}T1
& \cellcolor{ForestGreen!15}T2 & \cellcolor{ForestGreen!15}T3 & \cellcolor{ForestGreen!15}T4 & 
\cellcolor{ForestGreen!15}T5 & 
\cellcolor{ForestGreen!15}Avg \\ \hline

%PPV
\multicolumn{1}{|l|}{\cellcolor{ForestGreen!15}} 
                       & M1 & 0.8955 & 0.7910 & 0.8710 & \textcolor{silver}{0.9623} & \textcolor{bronze}{0.9608} & 0.8961 \\
\multicolumn{1}{|l|}{\cellcolor{ForestGreen!15}} & M2 & \textcolor{bronze}{0.9524} & 0.7361 & 0.8361 & 0.8431 & \textcolor{silver}{0.9811} & 0.8698 \\
\multicolumn{1}{|l|}{\cellcolor{ForestGreen!15}} & M3 & \textcolor{gold}{1.0000} & \textcolor{silver}{0.9355} & \textcolor{silver}{0.9500} & \textcolor{gold}{0.9828} & 0.9500 & \textcolor{silver}{0.9636} \\
\multicolumn{1}{|l|}{\cellcolor{ForestGreen!15}} & M4 & \textcolor{silver}{0.9836} & \textcolor{gold}{1.0000} & \textcolor{gold}{0.9836} & \textcolor{bronze}{0.9375} & \textcolor{gold}{0.9831} & \textcolor{gold}{0.9776} \\
\multicolumn{1}{|l|}{\cellcolor{ForestGreen!15}} & M5 & \textcolor{gold}{1.0000} & \textcolor{bronze}{0.8594} & \textcolor{bronze}{0.9444} & 0.9322 & 0.9062 & \textcolor{bronze}{0.9285} \\
\multicolumn{1}{|l|}{\multirow{-6}{*}{\cellcolor{ForestGreen!15}\begin{turn}{90}PPV\end{turn}}} & M6 & \textcolor{bronze}{0.9524} & 0.6463 & 0.9138 & \textcolor{silver}{0.9623} & 0.9545 & 0.8859 \\
\hline

%NPV
\multicolumn{1}{|l|}{\cellcolor{ForestGreen!15}} 
                       & M1 & \textcolor{gold}{1.0000} & 0.9700 & \textcolor{bronze}{0.9748} & 0.9636 & 0.9558 & 0.9728 \\
\hline

\end{tabular}
}
\caption{Classification metrics for segments suffering from undersampling.\label{table:classif_measures_UNDER}}
\end{table}

\begin{table}[h!]
\centering
\resizebox{0.5\textwidth }{!}{
\footnotesize	
\begin{tabular}{l|c|ccccc|c|}
%First row
\hhline{~-------}
& \cellcolor{ForestGreen!15}Method & \cellcolor{ForestGreen!15}T1
& \cellcolor{ForestGreen!15}T2 & \cellcolor{ForestGreen!15}T3 & \cellcolor{ForestGreen!15}T4 & 
\cellcolor{ForestGreen!15}T5 & 
\cellcolor{ForestGreen!15}Avg \\ \hline

%PPV
\multicolumn{1}{|l|}{\cellcolor{ForestGreen!15}} 
                       & M1 & 0.9231 & \textcolor{bronze}{0.8551} & \textcolor{gold}{1.0000} & \textcolor{gold}{1.0000} & \textcolor{gold}{1.0000} & \textcolor{bronze}{0.9556} \\
\multicolumn{1}{|l|}{\cellcolor{ForestGreen!15}} & M2 & 0.9077 & 0.8500 & 0.9455 & 0.8167 & 0.9167 & 0.8873 \\
\multicolumn{1}{|l|}{\cellcolor{ForestGreen!15}} & M3 & \textcolor{gold}{1.0000} & \textcolor{silver}{0.9344} & \textcolor{bronze}{0.9831} & \textcolor{silver}{0.9661} & \textcolor{bronze}{0.9672} & \textcolor{silver}{0.9702} \\
\multicolumn{1}{|l|}{\cellcolor{ForestGreen!15}} & M4 & \textcolor{gold}{1.0000} & \textcolor{gold}{0.9836} & \textcolor{silver}{0.9836} & \textcolor{gold}{1.0000} & \textcolor{gold}{1.0000} & \textcolor{gold}{0.9934} \\

\multicolumn{1}{|l|}{\cellcolor{ForestGreen!15}} & M5 & \textcolor{silver}{0.9828} & 0.7846 & 0.9194 & 0.8983 & 0.9464 & 0.9063 \\
\multicolumn{1}{|l|}{\multirow{-6}{*}{\cellcolor{ForestGreen!15}\begin{turn}{90}PPV\end{turn}}} & M6 & \textcolor{bronze}{0.9677} & 0.7308 & 0.9483 & \textcolor{bronze}{0.9298} & \textcolor{silver}{0.9778} & 0.9109 \\
\hline

%NPV
\multicolumn{1}{|l|}{\cellcolor{ForestGreen!15}} 
                       & M1 & \textcolor{gold}{1.0000} & \textcolor{silver}{0.9957} & \textcolor{gold}{1.0000} & 0.9600 & \textcolor{bronze}{0.9836} & \textcolor{bronze}{0.9879} \\
\hline

\end{tabular}
}
\caption{Classification metrics for segments suffering from missing data.\label{table:classif_measures_MISSING}}
\end{table}

\begin{table}[h!]
\centering
\resizebox{0.5\textwidth }{!}{
\footnotesize	
\begin{tabular}{l|c|ccccc|c|}
%First row
\hhline{~-------}
& \cellcolor{ForestGreen!15}Method & \cellcolor{ForestGreen!15}T1
& \cellcolor{ForestGreen!15}T2 & \cellcolor{ForestGreen!15}T3 & \cellcolor{ForestGreen!15}T4 & 
\cellcolor{ForestGreen!15}T5 & 
\cellcolor{ForestGreen!15}Avg \\ \hline

%PPV
\multicolumn{1}{|l|}{\cellcolor{ForestGreen!15}} & M1 & 0.8000 & 0.6957 & 0.8261 & \textcolor{gold}{1.0000} & \textcolor{gold}{1.0000} & 0.8643 \\
\multicolumn{1}{|l|}{\cellcolor{ForestGreen!15}} & M2 & \textcolor{silver}{0.9524} & 0.6400 & 0.7895 & 0.8235 & \textcolor{gold}{1.0000} & 0.8411 \\
\multicolumn{1}{|l|}{\cellcolor{ForestGreen!15}} & M3 & \textcolor{gold}{1.0000} & \textcolor{silver}{0.9500} & \textcolor{gold}{1.0000} & \textcolor{gold}{1.0000} & \textcolor{bronze}{0.8696} & \textcolor{silver}{0.9639} \\
\multicolumn{1}{|l|}{\cellcolor{ForestGreen!15}} & M4 & \textcolor{gold}{1.0000} & \textcolor{gold}{1.0000} & \textcolor{gold}{1.0000} & 0.9091 & \textcolor{gold}{1.0000} & \textcolor{gold}{0.9818} \\
\multicolumn{1}{|l|}{\cellcolor{ForestGreen!15}} & M5 & \textcolor{gold}{1.0000} & \textcolor{bronze}{0.7826} & \textcolor{silver}{0.9412} & \textcolor{silver}{0.9474} & \textcolor{silver}{0.9091} & \textcolor{bronze}{0.9160} \\
\multicolumn{1}{|l|}{\multirow{-6}{*}{\cellcolor{ForestGreen!15} \begin{turn}{90}PPV\end{turn}}} & M6 & \textcolor{bronze}{0.9091} & 0.5667 & \textcolor{bronze}{0.9375} & \textcolor{bronze}{0.9444} & \textcolor{gold}{1.0000} & 0.8715 \\
\hline

%NPV
\multicolumn{1}{|l|}{\cellcolor{ForestGreen!15}} & M1 & \textcolor{gold}{1.0000} & 0.9481 & \textcolor{bronze}{0.9870} & 0.9412 & \textcolor{bronze}{0.9302} & 0.9613 \\
\hline

\end{tabular}
}
\caption{Classification metrics for segments suffering from uniform noise and undersampling.\label{table:classif_measures_UNIFUNDER}}
\end{table}

\begin{table}[h!]
\centering
\resizebox{0.5\textwidth }{!}{
\footnotesize	
\begin{tabular}{l|c|ccccc|c|}
%First row
\hhline{~-------}
& \cellcolor{ForestGreen!15}Method & \cellcolor{ForestGreen!15}T1
& \cellcolor{ForestGreen!15}T2 & \cellcolor{ForestGreen!15}T3 & \cellcolor{ForestGreen!15}T4 & 
\cellcolor{ForestGreen!15}T5 & 
\cellcolor{ForestGreen!15}Avg \\ \hline

%PPV
\multicolumn{1}{|l|}{\cellcolor{ForestGreen!15}} & M1 & \textcolor{bronze}{0.9091} & 0.8095 & 0.8500 & 0.8947 & \textcolor{bronze}{0.9444} & 0.8816 \\
\multicolumn{1}{|l|}{\cellcolor{ForestGreen!15}} & M2 & \textcolor{silver}{0.9524} & 0.8571 & 0.7500 & 0.9412 & \textcolor{gold}{1.0000} & 0.9001 \\
\multicolumn{1}{|l|}{\cellcolor{ForestGreen!15}} & M3 & \textcolor{gold}{1.0000} & \textcolor{silver}{0.9500} & \textcolor{bronze}{0.8696} & \textcolor{gold}{1.0000} & \textcolor{gold}{1.0000} & \textcolor{gold}{0.9639} \\
\multicolumn{1}{|l|}{\cellcolor{ForestGreen!15}} & M4 & \textcolor{silver}{0.9524} & \textcolor{gold}{1.0000} & \textcolor{gold}{0.9524} & \textcolor{silver}{0.9524} & \textcolor{silver}{0.9500} & \textcolor{silver}{0.9614} \\
\multicolumn{1}{|l|}{\cellcolor{ForestGreen!15}} & M5 & \textcolor{gold}{1.0000} & \textcolor{bronze}{0.9048} & \textcolor{silver}{0.9444} & 0.9048 & 0.9000 & \textcolor{bronze}{0.9308} \\
\multicolumn{1}{|l|}{\multirow{-6}{*}{\cellcolor{ForestGreen!15} \begin{turn}{90}PPV\end{turn}}} & M6 & \textcolor{silver}{0.9524} & 0.6071 & 0.8571 & \textcolor{bronze}{0.9444} & 0.9167 & 0.8556 \\
\hline

%NPV
\multicolumn{1}{|l|}{\cellcolor{ForestGreen!15}} & M1 & \textcolor{gold}{1.0000} & 0.9620 & 0.9625 & 0.9630 & 0.9634 & 0.9702 \\
\hline

\end{tabular}
}
\caption{Classification metrics for segments suffering from Gaussian noise and undersampling.\label{table:classif_measures_GAUSSUNDER}}
\end{table}

\begin{table}[h!]
\centering
\resizebox{0.5\textwidth }{!}{
\footnotesize	
\begin{tabular}{l|c|ccccc|c|}
%First row
\hhline{~-------}
& \cellcolor{ForestGreen!15}Method & \cellcolor{ForestGreen!15}T1
& \cellcolor{ForestGreen!15}T2 & \cellcolor{ForestGreen!15}T3 & \cellcolor{ForestGreen!15}T4 & 
\cellcolor{ForestGreen!15}T5 & 
\cellcolor{ForestGreen!15}Avg \\ \hline

%PPV
\multicolumn{1}{|l|}{\cellcolor{ForestGreen!15}} & M1 & \textcolor{silver}{0.9091} & \textcolor{bronze}{0.8696} & \textcolor{gold}{1.0000} & \textcolor{gold}{1.0000} & \textcolor{gold}{1.0000} & \textcolor{bronze}{0.9557} \\
\multicolumn{1}{|l|}{\cellcolor{ForestGreen!15}} & M2 & \textcolor{bronze}{0.8696} & 0.8421 & 0.9412 & 0.7143 & \textcolor{bronze}{0.9000} & 0.8534 \\
\multicolumn{1}{|l|}{\cellcolor{ForestGreen!15}} & M3 & \textcolor{gold}{1.0000} & \textcolor{silver}{0.9091} & \textcolor{gold}{1.0000} & \textcolor{gold}{1.0000} & \textcolor{silver}{0.9524} & \textcolor{silver}{0.9723} \\
\multicolumn{1}{|l|}{\cellcolor{ForestGreen!15}} & M4 & \textcolor{gold}{1.0000} & \textcolor{gold}{1.0000} & \textcolor{gold}{1.0000} & \textcolor{gold}{1.0000} & \textcolor{gold}{1.0000} & \textcolor{gold}{1.0000} \\
\multicolumn{1}{|l|}{\cellcolor{ForestGreen!15}} & M5 & \textcolor{gold}{1.0000} & 0.6800 & \textcolor{silver}{0.9500} & \textcolor{bronze}{0.8421} & 0.8824 & 0.8709 \\
\multicolumn{1}{|l|}{\multirow{-6}{*}{\cellcolor{ForestGreen!15} \begin{turn}{90}PPV\end{turn}}} & M6 & \textcolor{silver}{0.9091} & 0.7692 & \textcolor{bronze}{0.9474} & \textcolor{silver}{0.9000} & \textcolor{gold}{1.0000} & 0.9051 \\
\hline

%NPV
\multicolumn{1}{|l|}{\cellcolor{ForestGreen!15}} & M1 & \textcolor{gold}{1.0000} & \textcolor{gold}{1.0000} & \textcolor{gold}{1.0000} & 0.9524 & \textcolor{silver}{0.9877} & \textcolor{bronze}{0.9880} \\
\hline

\end{tabular}
}
\caption{Classification metrics for segments suffering from uniform noise and missing data.\label{table:classif_measures_UNIFMISSING}}
\end{table}

\begin{table}[h]
\centering
\resizebox{0.5\textwidth }{!}{
\footnotesize	
\begin{tabular}{l|c|ccccc|c|}
%First row
\hhline{~-------}
& \cellcolor{ForestGreen!15}Method & \cellcolor{ForestGreen!15}T1
& \cellcolor{ForestGreen!15}T2 & \cellcolor{ForestGreen!15}T3 & \cellcolor{ForestGreen!15}T4 & 
\cellcolor{ForestGreen!15}T5 & 
\cellcolor{ForestGreen!15}Avg \\ \hline

%PPV
\multicolumn{1}{|l|}{\cellcolor{ForestGreen!15}} & M1 & 0.8696 & 0.8261 & \textcolor{gold}{1.0000} & \textcolor{gold}{1.0000} & \textcolor{gold}{1.0000} & \textcolor{bronze}{0.9391} \\
\multicolumn{1}{|l|}{\cellcolor{ForestGreen!15}} & M2 & \textcolor{bronze}{0.9048} & \textcolor{bronze}{0.9000} & 0.9000 & \textcolor{bronze}{0.8947} & 0.8500 & 0.8899 \\
\multicolumn{1}{|l|}{\cellcolor{ForestGreen!15}} & M3 & \textcolor{gold}{1.0000} & \textcolor{silver}{0.9500} & \textcolor{bronze}{0.9500} & \textcolor{gold}{1.0000} & \textcolor{silver}{0.9500} & \textcolor{silver}{0.9700} \\
\multicolumn{1}{|l|}{\cellcolor{ForestGreen!15}} & M4 & \textcolor{gold}{1.0000} & \textcolor{gold}{1.0000} & \textcolor{silver}{0.9524} & \textcolor{gold}{1.0000} & \textcolor{gold}{1.0000} & \textcolor{gold}{0.9905} \\
\multicolumn{1}{|l|}{\cellcolor{ForestGreen!15}} & M5 & \textcolor{silver}{0.9474} & 0.8421 & 0.9000 & \textcolor{silver}{0.9091} & \textcolor{silver}{0.9500} & 0.9097 \\
\multicolumn{1}{|l|}{\multirow{-6}{*}{\cellcolor{ForestGreen!15} \begin{turn}{90}PPV\end{turn}}} & M6 & \textcolor{gold}{1.0000} & 0.6429 & 0.9474 & \textcolor{gold}{1.0000} & \textcolor{bronze}{0.9375} & 0.9055 \\
\hline

%NPV
\multicolumn{1}{|l|}{\cellcolor{ForestGreen!15}} & M1 & \textcolor{gold}{1.0000} & \textcolor{bronze}{0.9870} & \textcolor{gold}{1.0000} & 0.9524 & \textcolor{bronze}{0.9756} & \textcolor{bronze}{0.9830} \\
\hline

\end{tabular}
}
\caption{Classification metrics for segments suffering from Gaussian noise and missing data.\label{table:classif_measures_GAUSSMISSING}}
\end{table}

\begin{table}[h!]
\centering
\resizebox{0.5\textwidth }{!}{
\footnotesize	
\begin{tabular}{l|c|ccccc|c|}
%First row
\hhline{~-------}
& \cellcolor{ForestGreen!15}Method & \cellcolor{ForestGreen!15}T1
& \cellcolor{ForestGreen!15}T2 & \cellcolor{ForestGreen!15}T3 & \cellcolor{ForestGreen!15}T4 & 
\cellcolor{ForestGreen!15}T5 & 
\cellcolor{ForestGreen!15}Avg \\ \hline

%PPV
\multicolumn{1}{|l|}{\cellcolor{ForestGreen!15}} & M1 & \textcolor{gold}{1.0000} & \textcolor{gold}{1.0000} & \textcolor{silver}{0.8333} & \textcolor{gold}{1.0000} & \textcolor{gold}{1.0000} & \textcolor{silver}{0.9667} \\
\multicolumn{1}{|l|}{\cellcolor{ForestGreen!15}} & M2 & \textcolor{gold}{1.0000} & \textcolor{silver}{0.8333} & \textcolor{silver}{0.8333} & \textcolor{gold}{1.0000} & \textcolor{gold}{1.0000} & \textcolor{bronze}{0.9333} \\
\multicolumn{1}{|l|}{\cellcolor{ForestGreen!15}} & M3 & \textcolor{gold}{1.0000} & \textcolor{gold}{1.0000} & \textcolor{silver}{0.8333} & \textcolor{gold}{1.0000} & \textcolor{gold}{1.0000} & \textcolor{silver}{0.9667} \\
\multicolumn{1}{|l|}{\cellcolor{ForestGreen!15}} & M4 & \textcolor{gold}{1.0000} & \textcolor{gold}{1.0000} & \textcolor{gold}{1.0000} & \textcolor{gold}{1.0000} & \textcolor{gold}{1.0000} & \textcolor{gold}{1.0000} \\
\multicolumn{1}{|l|}{\cellcolor{ForestGreen!15}} & M5 & \textcolor{gold}{1.0000} & \textcolor{gold}{1.0000} & \textcolor{silver}{0.8333} & \textcolor{silver}{0.7143} & \textcolor{gold}{1.0000} & 0.9095 \\
\multicolumn{1}{|l|}{\multirow{-6}{*}{\cellcolor{ForestGreen!15} \begin{turn}{90}PPV\end{turn}}} & M6 & \textcolor{gold}{1.0000} & \textcolor{silver}{0.8333} & \textcolor{gold}{1.0000} & \textcolor{gold}{1.0000} & \textcolor{gold}{1.0000} & \textcolor{silver}{0.9667} \\
\hline

%NPV
\multicolumn{1}{|l|}{\cellcolor{ForestGreen!15}} & M1 & \textcolor{gold}{1.0000} & \textcolor{gold}{1.0000} & \textcolor{gold}{1.0000} & \textcolor{gold}{1.0000} & \textcolor{silver}{0.9524} & \textcolor{silver}{0.9905} \\
\hline

\end{tabular}
}
\caption{Classification metrics for segments suffering from local deformations.\label{table:classif_measures_DEFORM}}
\end{table}

\newpage\phantom{blabla}
\newpage\phantom{blabla}
\newpage\phantom{blabla}
\newpage\phantom{blabla}
\newpage\phantom{blabla}
\section{Fitting metrics\label{sec:appC}}

\begin{table}[h!]
\centering
\resizebox{0.5\textwidth }{!}{
\footnotesize	
\begin{tabular}{c|>{\centering}m{0.723cm}|>{\centering}m{1.2cm} >{\centering}m{1.2cm} >{\centering}m{1.2cm}  >{\centering}m{1.2cm} c|}
%First row
\hhline{~------}
& \cellcolor{BlueViolet!20}Method & \cellcolor{BlueViolet!20}Q1
& \cellcolor{BlueViolet!20}Q2 & \cellcolor{BlueViolet!20}Q3 & \cellcolor{BlueViolet!20}mean & 
\cellcolor{BlueViolet!20}std \\ \hline

%L2
\multicolumn{1}{|l|}{\cellcolor{BlueViolet!20}} & M1 & 1.73e-02 & 4.10e-02 & 1.23e-01 & \textcolor{silver}{1.24e-01} & \textcolor{bronze}{2.10e-01}\\
\multicolumn{1}{|l|}{\cellcolor{BlueViolet!20}} & M2 & \textcolor{bronze}{4.11e-10} & \textcolor{bronze}{5.80e-10} & \textcolor{gold}{8.86e-10} & 1.46e+01 & 1.35e+02\\
\multicolumn{1}{|l|}{\cellcolor{BlueViolet!20}} & M3 & 1.24e-01 & 4.14e-01 & 6.92e-01 & \textcolor{bronze}{5.19e-01} & 5.11e-01\\
\multicolumn{1}{|l|}{\cellcolor{BlueViolet!20}} & M4 & 8.20e-09 & 2.86e-08 & \textcolor{silver}{1.55e-07} & 8.72e-03 & \textcolor{gold}{2.07e-02}\\
\multicolumn{1}{|l|}{\cellcolor{BlueViolet!20}} & M5 & \textcolor{gold}{4.41e-15} & \textcolor{gold}{1.36e-13} & 1.81e-02 & 9.58e+01 & 4.12e+02\\
\multicolumn{1}{|l|}{\multirow{-6}{*}{\cellcolor{BlueViolet!20} \begin{turn}{90}$L^2$\end{turn}}} & M6 & \textcolor{silver}{6.30e-15} & \textcolor{silver}{2.86e-13} & \textcolor{bronze}{1.64e-05} & \textcolor{gold}{1.26e-02} & \textcolor{silver}{3.79e-02} \\
\hline
%MFE
\multicolumn{1}{|l|}{\cellcolor{BlueViolet!20}} & M1 & 3.22e-03 & 4.99e-03 & 8.03e-03 & \textcolor{silver}{6.35e-03} & \textcolor{gold}{5.29e-03}\\
\hline
\end{tabular}
}
\caption{Statistics of the fitting errors for clean segments.\label{table:fitting_measures_CLEAN}}
\end{table}

\begin{table}[h!]
\centering
\resizebox{0.5\textwidth }{!}{
\footnotesize	
\begin{tabular}{c|>{\centering}m{0.723cm}|>{\centering}m{1.2cm} >{\centering}m{1.2cm} >{\centering}m{1.2cm}  >{\centering}m{1.2cm} c|}
%First row
\hhline{~------}
& \cellcolor{BlueViolet!20}Method & \cellcolor{BlueViolet!20}Q1
& \cellcolor{BlueViolet!20}Q2 & \cellcolor{BlueViolet!20}Q3 & \cellcolor{BlueViolet!20}mean & 
\cellcolor{BlueViolet!20}std \\ \hline
\multicolumn{1}{|l|}{\cellcolor{BlueViolet!20}} & M1 & 2.40e-02 & 8.30e-02 & 2.21e-01 & \textcolor{silver}{2.38e-01} & \textcolor{silver}{4.07e-01} \\ 
\multicolumn{1}{|l|}{\cellcolor{BlueViolet!20}} & M2 & \textcolor{silver}{2.67e-03} & \textcolor{silver}{1.74e-02} & \textcolor{bronze}{1.73e-01} & 4.40e+10 & 3.45e+11\\
\multicolumn{1}{|l|}{\cellcolor{BlueViolet!20}} & M3 & 1.68e-01 & 4.81e-01 & 8.83e-01 & \textcolor{bronze}{6.03e-01} & \textcolor{bronze}{5.43e-01}\\
\multicolumn{1}{|l|}{\cellcolor{BlueViolet!20}} & M4 & \textcolor{bronze}{4.39e-03} & \textcolor{gold}{1.50e-02} & \textcolor{gold}{4.02e-02} & \textcolor{gold}{5.20e-02} & \textcolor{gold}{1.92e-01}\\
\multicolumn{1}{|l|}{\cellcolor{BlueViolet!20}} & M5 & 4.66e-03 & \textcolor{bronze}{2.14e-02} & 9.07e-01 & 3.59e+02 & 2.16e+03\\
\multicolumn{1}{|l|}{\multirow{-6}{*}{\cellcolor{BlueViolet!20} \begin{turn}{90}$L^2$\end{turn}}} & M6 & \textcolor{gold}{2.08e-03} & 2.33e-02 & \textcolor{silver}{1.53e-01} & 2.09e+06 & 9.54e+06 \\
\hline
%MFE
\multicolumn{1}{|l|}{\cellcolor{BlueViolet!20}} & M1 & 4.02e-03 & 6.78e-03 & \textcolor{bronze}{1.17e-02} & \textcolor{silver}{1.06e-02} & \textcolor{gold}{1.11e-02}\\
\hline
\end{tabular}
}
\caption{Statistics of the fitting errors for segments suffering from uniform noise.\label{table:fitting_measures_UNIF}}
\end{table}

\begin{table}[h!]
\centering
\resizebox{0.5\textwidth }{!}{
\footnotesize	
\begin{tabular}{c|>{\centering}m{0.723cm}|>{\centering}m{1.2cm} >{\centering}m{1.2cm} >{\centering}m{1.2cm}  >{\centering}m{1.2cm} c|}
%First row
\hhline{~------}
& \cellcolor{BlueViolet!20}Method & \cellcolor{BlueViolet!20}Q1
& \cellcolor{BlueViolet!20}Q2 & \cellcolor{BlueViolet!20}Q3 & \cellcolor{BlueViolet!20}mean & 
\cellcolor{BlueViolet!20}std \\ \hline
\multicolumn{1}{|l|}{\cellcolor{BlueViolet!20}} & M1 & 2.86e-02 & 8.42e-02 & \textcolor{silver}{2.00e-01} & \textcolor{silver}{2.38e-01} & \textcolor{silver}{4.15e-01}\\
\multicolumn{1}{|l|}{\cellcolor{BlueViolet!20}} & M2 & \textcolor{gold}{7.30e-03} & \textcolor{silver}{5.34e-02} & \textcolor{bronze}{2.47e-01} & 3.32e+11 & 4.89e+12\\
\multicolumn{1}{|l|}{\cellcolor{BlueViolet!20}} & M3 & 2.04e-01 & 5.19e-01 & 8.97e-01 & \textcolor{bronze}{6.48e-01} & \textcolor{bronze}{6.09e-01}\\
\multicolumn{1}{|l|}{\cellcolor{BlueViolet!20}} & M4 & 2.43e-02 & \textcolor{gold}{4.17e-02} & \textcolor{gold}{8.15e-02} & \textcolor{gold}{9.64e-02} & \textcolor{gold}{2.53e-01}\\
\multicolumn{1}{|l|}{\cellcolor{BlueViolet!20}} & M5 & \textcolor{bronze}{1.98e-02} & 7.73e-02 & 1.57e+00 & 1.41e+02 & 6.27e+02\\
\multicolumn{1}{|l|}{\multirow{-6}{*}{\cellcolor{BlueViolet!20} \begin{turn}{90}$L^2$\end{turn}}} & M6 & \textcolor{silver}{1.84e-02} & \textcolor{bronze}{7.39e-02} & 2.48e-01 & 1.95e+06 & 9.27e+06 \\
\hline
%MFE
\multicolumn{1}{|l|}{\cellcolor{BlueViolet!20}} & M1 & 4.22e-03 & 6.39e-03 & \textcolor{bronze}{1.28e-02} & \textcolor{silver}{1.04e-02} & \textcolor{gold}{9.74e-03}\\
\hline
\end{tabular}
}
\caption{Statistics of the fitting errors for segments suffering from Gaussian noise.\label{table:fitting_measures_GAUS}}
\end{table}

\begin{table}[h!]
\centering
\resizebox{0.5\textwidth }{!}{
\footnotesize	
\begin{tabular}{c|>{\centering}m{0.723cm}|>{\centering}m{1.2cm} >{\centering}m{1.2cm} >{\centering}m{1.2cm}  >{\centering}m{1.2cm} c|}
%First row
\hhline{~------}
& \cellcolor{BlueViolet!20}Method & \cellcolor{BlueViolet!20}Q1
& \cellcolor{BlueViolet!20}Q2 & \cellcolor{BlueViolet!20}Q3 & \cellcolor{BlueViolet!20}mean & 
\cellcolor{BlueViolet!20}std \\ \hline
\multicolumn{1}{|l|}{\cellcolor{BlueViolet!20}} & M1 & 2.24e-02 & 6.95e-02 & 2.28e-01 & \textcolor{silver}{2.25e-01} & \textcolor{silver}{3.89e-01}\\
\multicolumn{1}{|l|}{\cellcolor{BlueViolet!20}} & M2 & \textcolor{silver}{9.75e-10} & \textcolor{gold}{7.64e-03} & \textcolor{bronze}{1.37e-01} & 3.59e+11 & 4.90e+12\\
\multicolumn{1}{|l|}{\cellcolor{BlueViolet!20}} & M3 & 2.03e-01 & 5.48e-01 & 9.30e-01 & \textcolor{bronze}{6.06e-01} & \textcolor{bronze}{5.02e-01}\\
\multicolumn{1}{|l|}{\cellcolor{BlueViolet!20}} & M4 & \textcolor{bronze}{2.11e-07} & \textcolor{bronze}{1.96e-02} & \textcolor{gold}{5.05e-02} & \textcolor{gold}{6.98e-02} & \textcolor{gold}{2.50e-01} \\
\multicolumn{1}{|l|}{\cellcolor{BlueViolet!20}} & M5 & 8.45e-04 & 2.47e-02 & 1.70e+00 & 3.18e+02 & 1.94e+03\\
\multicolumn{1}{|l|}{\multirow{-6}{*}{\cellcolor{BlueViolet!20} \begin{turn}{90}$L^2$\end{turn}}} & M6 & \textcolor{gold}{1.23e-14} & \textcolor{silver}{1.74e-02} & \textcolor{silver}{1.11e-01} & 1.46e+06 & 8.36e+06 \\
\hline
%MFE
\multicolumn{1}{|l|}{\cellcolor{BlueViolet!20}} & M1 & 3.68e-03 & 6.38e-03 & 1.18e-02 & \textcolor{gold}{9.75e-03} & \textcolor{gold}{9.54e-03}\\
\hline

\end{tabular}
}
\caption{Statistics of the fitting errors for segments suffering from under sampling.\label{table:fitting_measures_UNDER}}
\end{table}

\begin{table}[h!]
\centering
\resizebox{0.5\textwidth }{!}{
\footnotesize	
\begin{tabular}{c|>{\centering}m{0.723cm}|>{\centering}m{1.2cm} >{\centering}m{1.2cm} >{\centering}m{1.2cm}  >{\centering}m{1.2cm} c|}
%First row
\hhline{~------}
& \cellcolor{BlueViolet!20}Method & \cellcolor{BlueViolet!20}Q1
& \cellcolor{BlueViolet!20}Q2 & \cellcolor{BlueViolet!20}Q3 & \cellcolor{BlueViolet!20}mean & 
\cellcolor{BlueViolet!20}std \\ \hline
\multicolumn{1}{|l|}{\cellcolor{BlueViolet!20}} & M1 & 2.23e-02 & 6.94e-02 & 1.68e-01 & \textcolor{silver}{2.33e-01} & \textcolor{silver}{4.39e-01}\\
\multicolumn{1}{|l|}{\cellcolor{BlueViolet!20}} & M2 & \textcolor{silver}{7.42e-10} & \textcolor{gold}{4.80e-03} & \textcolor{silver}{8.95e-02} & 1.71e+10 & 2.44e+11\\
\multicolumn{1}{|l|}{\cellcolor{BlueViolet!20}} & M3 & 1.51e-01 & 4.65e-01 & 8.34e-01 & \textcolor{bronze}{5.86e-01} & \textcolor{bronze}{5.44e-01}\\
\multicolumn{1}{|l|}{\cellcolor{BlueViolet!20}} & M4 & \textcolor{silver}{1.70e-07} & \textcolor{silver}{1.36e-02} & \textcolor{gold}{4.19e-02} & \textcolor{gold}{4.82e-02} & \textcolor{gold}{1.65e-01}\\
\multicolumn{1}{|l|}{\cellcolor{BlueViolet!20}} & M5 & 1.75e-11 & \textcolor{bronze}{1.87e-02} & 8.00e-01 & 2.18e+02 & 1.44e+03\\
\multicolumn{1}{|l|}{\multirow{-6}{*}{\cellcolor{BlueViolet!20} \begin{turn}{90}$L^2$\end{turn}}} & M6 & \textcolor{gold}{1.64e-14} & \textcolor{bronze}{1.87e-02} & \textcolor{bronze}{1.05e-01} & 1.67e+06 & 8.42e+06 \\
\hline
%MFE
\multicolumn{1}{|l|}{\cellcolor{BlueViolet!20}} & M1 & 3.68e-03 & 5.76e-03 & 1.06e-02 & \textcolor{silver}{9.54e-03} & \textcolor{silver}{1.05e-02}\\
\hline
\end{tabular}
}
\caption{Statistics of the fitting errors for segments suffering from missing data.\label{table:fitting_measures_MISSING}}
\end{table}

\begin{table}[h!]
\centering
\resizebox{0.5\textwidth }{!}{
\footnotesize	
\begin{tabular}{c|>{\centering}m{0.723cm}|>{\centering}m{1.2cm} >{\centering}m{1.2cm} >{\centering}m{1.2cm}  >{\centering}m{1.2cm} c|}
%First row
\hhline{~------}
& \cellcolor{BlueViolet!20}Method & \cellcolor{BlueViolet!20}Q1
& \cellcolor{BlueViolet!20}Q2 & \cellcolor{BlueViolet!20}Q3 & \cellcolor{BlueViolet!20}mean & 
\cellcolor{BlueViolet!20}std \\ \hline
\multicolumn{1}{|l|}{\cellcolor{BlueViolet!20}} & M1 & 2.66e-02 & 9.42e-02 & 2.21e-01 & \textcolor{silver}{1.95e-01} & \textcolor{silver}{3.17e-01}\\
\multicolumn{1}{|l|}{\cellcolor{BlueViolet!20}} & M2 & \textcolor{gold}{3.78e-03} & \textcolor{gold}{1.86e-02} & \textcolor{bronze}{2.17e-01} & 8.50e+10 & 4.19e+11\\
\multicolumn{1}{|l|}{\cellcolor{BlueViolet!20}} & M3 & 2.51e-01 & 5.46e-01 & 9.40e-01 & \textcolor{bronze}{6.11e-01} & \textcolor{bronze}{5.09e-01}\\
\multicolumn{1}{|l|}{\cellcolor{BlueViolet!20}} & M4 & 6.35e-03 & \textcolor{silver}{1.87e-02} & \textcolor{gold}{4.66e-02} & \textcolor{gold}{8.46e-02} & \textcolor{gold}{3.12e-01}\\
\multicolumn{1}{|l|}{\cellcolor{BlueViolet!20}} & M5 & \textcolor{bronze}{5.98e-03} & \textcolor{bronze}{2.14e-02} & 1.84e+00 & 3.59e+02 & 2.06e+03\\
\multicolumn{1}{|l|}{\multirow{-6}{*}{\cellcolor{BlueViolet!20} \begin{turn}{90}$L^2$\end{turn}}} & M6 & \textcolor{silver}{4.40e-03} & 2.26e-02 & \textcolor{silver}{1.75e-01} & 2.30e+06 & 1.11e+07 \\
\hline
%MFE
\multicolumn{1}{|l|}{\cellcolor{BlueViolet!20}} & M1 & 3.86e-03 & 7.25e-03 & \textcolor{bronze}{1.42e-02} & \textcolor{silver}{1.12e-02} & \textcolor{gold}{1.10e-02}\\
\hline

\end{tabular}
}
\caption{Statistics of the fitting errors for segments suffering from uniform noise and undersampling.\label{table:fitting_measures_UNIFUNDER}}
\end{table}

\begin{table}[h!]
\centering
\resizebox{0.5\textwidth }{!}{
\footnotesize	
\begin{tabular}{c|>{\centering}m{0.723cm}|>{\centering}m{1.2cm} >{\centering}m{1.2cm} >{\centering}m{1.2cm}  >{\centering}m{1.2cm} c|}
%First row
\hhline{~------}
& \cellcolor{BlueViolet!20}Method & \cellcolor{BlueViolet!20}Q1
& \cellcolor{BlueViolet!20}Q2 & \cellcolor{BlueViolet!20}Q3 & \cellcolor{BlueViolet!20}mean & 
\cellcolor{BlueViolet!20}std \\ \hline
\multicolumn{1}{|l|}{\cellcolor{BlueViolet!20}} & M1 & 3.05e-02 & 1.14e-01 & 3.75e-01 & \textcolor{silver}{3.33e-01} & \textcolor{silver}{5.11e-01}\\
\multicolumn{1}{|l|}{\cellcolor{BlueViolet!20}} & M2 & \textcolor{gold}{8.27e-03} & \textcolor{silver}{4.91e-02} & \textcolor{silver}{2.76e-01} & 9.65e+11 & 8.35e+12\\
\multicolumn{1}{|l|}{\cellcolor{BlueViolet!20}} & M3 & 2.21e-01 & 5.99e-01 & 1.00e+00 & \textcolor{bronze}{6.48e-01} & \textcolor{bronze}{5.44e-01}\\
\multicolumn{1}{|l|}{\cellcolor{BlueViolet!20}} & M4 & 2.80e-02 & \textcolor{gold}{4.17e-02} & \textcolor{gold}{9.50e-02} & \textcolor{gold}{1.13e-01} & \textcolor{gold}{2.95e-01}\\
\multicolumn{1}{|l|}{\cellcolor{BlueViolet!20}} & M5 & \textcolor{bronze}{2.32e-02} & 8.68e-02 & 1.71e+00 & 1.78e+02 & 6.26e+02\\
\multicolumn{1}{|l|}{\multirow{-6}{*}{\cellcolor{BlueViolet!20} \begin{turn}{90}$L^2$\end{turn}}} & M6 & \textcolor{silver}{1.26e-02} & \textcolor{bronze}{7.35e-02} & \textcolor{bronze}{2.80e-01} & 2.26e+06 & 9.63e+06 \\
\hline
\multicolumn{1}{|l|}{\cellcolor{BlueViolet!20}} & M1 & 4.60e-03 & 7.55e-03 & \textcolor{bronze}{1.31e-02} & \textcolor{silver}{1.11e-02} & \textcolor{gold}{1.00e-02}\\
\hline
\end{tabular}
}
\caption{Statistics of the fitting errors for segments suffering from Gaussian noise and undersampling.\label{table:fitting_measures_GAUSSUNDER}}
\end{table}

\begin{table}[h!]
\centering
\resizebox{0.5\textwidth }{!}{
\footnotesize	
\begin{tabular}{c|>{\centering}m{0.723cm}|>{\centering}m{1.2cm} >{\centering}m{1.2cm} >{\centering}m{1.2cm}  >{\centering}m{1.2cm} c|}
%First row
\hhline{~------}
& \cellcolor{BlueViolet!20}Method & \cellcolor{BlueViolet!20}Q1
& \cellcolor{BlueViolet!20}Q2 & \cellcolor{BlueViolet!20}Q3 & \cellcolor{BlueViolet!20}mean & 
\cellcolor{BlueViolet!20}std \\ \hline
\multicolumn{1}{|l|}{\cellcolor{BlueViolet!20}} & M1 & 2.49e-02 & 8.86e-02 & 2.51e-01 & \textcolor{silver}{3.08e-01} & \textcolor{silver}{5.28e-01}\\
\multicolumn{1}{|l|}{\cellcolor{BlueViolet!20}} & M2 & \textcolor{silver}{2.78e-03} & \textcolor{silver}{1.89e-02} & \textcolor{bronze}{1.72e-01} & 4.94e+10 & 4.30e+11\\
\multicolumn{1}{|l|}{\cellcolor{BlueViolet!20}} & M3 & 1.17e-01 & 4.84e-01 & 8.34e-01 &\textcolor{bronze}{5.71e-01} & \textcolor{bronze}{5.60e-01}\\
\multicolumn{1}{|l|}{\cellcolor{BlueViolet!20}} & M4 & \textcolor{bronze}{3.81e-03} & \textcolor{gold}{1.48e-02} & \textcolor{gold}{3.46e-02} & \textcolor{gold}{3.40e-02} & \textcolor{gold}{8.47e-02}\\
\multicolumn{1}{|l|}{\cellcolor{BlueViolet!20}} & M5 & 5.08e-03 & \textcolor{bronze}{2.30e-02} & 2.81e-01 & 3.07e+02 & 1.70e+03\\
\multicolumn{1}{|l|}{\multirow{-6}{*}{\cellcolor{BlueViolet!20} \begin{turn}{90}$L^2$\end{turn}}} & M6 & \textcolor{gold}{2.43e-03} & 2.58e-02 & \textcolor{silver}{1.10e-01} & 2.69e+06 & 1.08e+07 \\
\hline
\multicolumn{1}{|l|}{\cellcolor{BlueViolet!20}} & M1 & 4.50e-03 & 7.60e-03 & \textcolor{bronze}{1.20e-02} & \textcolor{silver}{1.14e-02} & \textcolor{silver}{1.26e-02}\\
\hline
\end{tabular}
}
\caption{Statistics of the fitting errors for segments suffering from uniform noise and missing data.\label{table:fitting_measures_UNIFMISS}}
\end{table}

\begin{table}[h!]
\centering
\resizebox{0.5\textwidth }{!}{
\footnotesize	
\begin{tabular}{c|>{\centering}m{0.723cm}|>{\centering}m{1.2cm} >{\centering}m{1.2cm} >{\centering}m{1.2cm}  >{\centering}m{1.2cm} c|}
%First row
\hhline{~------}
& \cellcolor{BlueViolet!20}Method & \cellcolor{BlueViolet!20}Q1
& \cellcolor{BlueViolet!20}Q2 & \cellcolor{BlueViolet!20}Q3 & \cellcolor{BlueViolet!20}mean & 
\cellcolor{BlueViolet!20}std \\ \hline
\multicolumn{1}{|l|}{\cellcolor{BlueViolet!20}} & M1 & 2.27e-02 & \textcolor{bronze}{7.24e-02} & \textcolor{silver}{1.49e-01} & \textcolor{silver}{2.06e-01} & \textcolor{silver}{4.09e-01}\\
\multicolumn{1}{|l|}{\cellcolor{BlueViolet!20}} & M2 & \textcolor{gold}{6.60e-03} & \textcolor{silver}{4.49e-02} & \textcolor{bronze}{1.50e-01} & 3.81e+09 & 3.59e+10\\
\multicolumn{1}{|l|}{\cellcolor{BlueViolet!20}} & M3 & 2.08e-01 & 4.70e-01 & 8.34e-01 & \textcolor{bronze}{6.25e-01} & \textcolor{bronze}{5.82e-01}\\
\multicolumn{1}{|l|}{\cellcolor{BlueViolet!20}} & M4 & \textcolor{bronze}{2.15e-02} & \textcolor{gold}{4.11e-02} & \textcolor{gold}{7.82e-02} & \textcolor{gold}{1.03e-01} & \textcolor{gold}{2.64e-01}\\
\multicolumn{1}{|l|}{\cellcolor{BlueViolet!20}} & M5 & \textcolor{silver}{1.77e-02} & 1.09e-01 & 1.68e+00 & 1.57e+02 & 6.74e+02\\
\multicolumn{1}{|l|}{\multirow{-6}{*}{\cellcolor{BlueViolet!20} \begin{turn}{90}$L^2$\end{turn}}} & M6 & 2.23e-02 & 8.31e-02 & 2.25e-01 & 2.38e+06 & 9.84e+06 \\
\hline
\multicolumn{1}{|l|}{\cellcolor{BlueViolet!20}} & M1 & 3.81e-03 & 5.96e-03 & \textcolor{bronze}{1.24e-02} & \textcolor{silver}{1.00e-02}& \textcolor{silver}{1.01e-02}\\
\hline
\end{tabular}
}
\caption{Statistics of the fitting errors for segments suffering from Gaussian noise and missing data.\label{table:fitting_measures_GAUSSMISS}}
\end{table}

\begin{table}[h!]
\centering
\resizebox{0.5\textwidth }{!}{
\footnotesize	
\begin{tabular}{c|>{\centering}m{0.723cm}|>{\centering}m{1.2cm} >{\centering}m{1.2cm} >{\centering}m{1.2cm}  >{\centering}m{1.2cm} c|}
%First row
\hhline{~------}
& \cellcolor{BlueViolet!20}Method & \cellcolor{BlueViolet!20}Q1
& \cellcolor{BlueViolet!20}Q2 & \cellcolor{BlueViolet!20}Q3 & \cellcolor{BlueViolet!20}mean & 
\cellcolor{BlueViolet!20}std \\ \hline
\multicolumn{1}{|l|}{\cellcolor{BlueViolet!20}} & M1 & \textcolor{silver}{1.63e-02} & 4.11e-02 & 1.26e-01 & \textcolor{bronze}{1.23e-01} & \textcolor{bronze}{1.94e-01}\\
\multicolumn{1}{|l|}{\cellcolor{BlueViolet!20}} & M2 & \textcolor{gold}{1.38e-03} & \textcolor{gold}{3.16e-03} & \textcolor{gold}{9.06e-03} & 2.37e-01 & 7.69e-01\\
\multicolumn{1}{|l|}{\cellcolor{BlueViolet!20}} & M3 & 1.89e-01 & 6.40e-01 & 1.18e+00 & 7.56e-01 & 7.14e-01\\
\multicolumn{1}{|l|}{\cellcolor{BlueViolet!20}} & M4 & 5.23e-03 & 1.82e-02 & \textcolor{bronze}{2.39e-02} & \textcolor{silver}{2.36e-02} & \textcolor{silver}{2.55e-02}\\
\multicolumn{1}{|l|}{\cellcolor{BlueViolet!20}} & M5 & \textcolor{bronze}{2.44e-03} & \textcolor{bronze}{1.19e-02} & 1.96e+00 & 3.02e+02 & 7.86e+02\\
\multicolumn{1}{|l|}{\multirow{-6}{*}{\cellcolor{BlueViolet!20} \begin{turn}{90}$L^2$\end{turn}}} & M6 & 2.67e-03 & \textcolor{silver}{1.15e-02} & \textcolor{silver}{2.37e-02} & \textcolor{gold}{1.87e-02} & \textcolor{gold}{2.19e-02} \\
\hline
\multicolumn{1}{|l|}{\cellcolor{BlueViolet!20}} & M1 & 3.67e-03 & 4.76e-03 & 7.53e-03 & \textcolor{bronze}{6.53e-03} & \textcolor{bronze}{5.46e-03}\\
\hline
\end{tabular}
}
\caption{Statistics of the fitting errors for segments suffering from small deformations.\label{table:fitting_measures_BUMPS}}
\end{table}

\end{document}